\documentclass[12pt,a4,english]{article}
\usepackage[latin9]{inputenc}
\usepackage{geometry}
\usepackage{verbatim}
\usepackage{amsmath}
\usepackage{graphicx}
\usepackage{longtable}
\usepackage{setspace}
\usepackage[round,sort,authoryear]{natbib}
\usepackage{amsfonts}
\usepackage{pdflscape}
\usepackage{upgreek}
\usepackage{dsfont}
\usepackage[colorlinks=true,linkcolor=blue, citecolor=blue, urlcolor = blue]{hyperref}
\usepackage{chngcntr}
\usepackage{comment}

\makeatletter
\usepackage{graphicx}
\makeatother
\usepackage{babel}
\usepackage{amsfonts}
\usepackage{amsthm}
\usepackage[toc,page]{appendix}
\usepackage{booktabs,caption}
\usepackage{multirow}
\usepackage[labelfont=bf,textfont=md]{caption}
\usepackage[labelsep=space]{caption}
\usepackage{hyperref}
\usepackage{titlesec}
\usepackage{lmodern}
\usepackage[T1]{fontenc}

\usepackage{mathtools}

\makeatletter
\newcommand{\nextverbatimspread}[1]{%
  \def\verbatim@font{%
    \linespread{#1}\normalfont\ttfamily
    \gdef\verbatim@font{\normalfont\ttfamily}}
}
\makeatother

\DeclarePairedDelimiter{\floor}{\lfloor}{\rfloor}
\usepackage[T1]{fontenc}

\usepackage{natbib}
\usepackage{bibentry}
\nobibliography*

\usepackage{fancyhdr}
\numberwithin{equation}{section}

\theoremstyle{definition}
\newtheorem{theorem}{Theorem}
\newtheorem{definition}{Definition}
\newtheorem{assumption}{Assumption}
\newtheorem{lemma}{Lemma}

\newtheorem{proposition}{Proposition}
\newtheorem{remark}{Remark}
\newtheorem{corollary}{Corollary}

\usepackage{amsmath}
\newcommand\norm[1]{\left\lVert#1\right\rVert}

\newif\ifshow 
\showfalse 

\ifshow
  \includecomment{wrap}
\else
  \excludecomment{wrap} 
\fi

\titleformat*{\section}{\large \bfseries}
\titleformat*{\subsection}{\normalsize \bfseries}
\titleformat*{\subsubsection}{\normalsize \bfseries}

\begin{document}
\pagenumbering{roman}

\title{ {\Large \textbf{Structural Break Detection in Quantile Predictive Regression Models with Persistent Covariates}\thanks{I would like to thank \textit{Jose Olmo, Tassos Magdalinos} and \textit{Jean-Yves Pitarakis} for guidance, support and continuous encouragement throughout the PhD programme. Furthermore, this study has been inspired by the work of \textit{Peter C. B. Phillips} in time series econometrics. The author acknowledge the use of the Iridis 5 HPC Facility and associated support services at the University of Southampton in the completion of this work. Financial support from the VC PhD scholarship of the University of Southampton is also gratefully acknowledged. All the usual disclaimers apply.

Parts of this paper are derived from my Ph.D. thesis at the University of Southampton with title: \textcolor{blue}{\textit{"Aspects of Estimation and Inference for Predictive Regression Models"}}. } \\
}
}

\author{\textbf{Christis Katsouris}\footnote{Ph.D Candidate, Department of Economics, University of Southampton, Southampton, SO17 1BJ, UK. \textit{E-mail Address}: \textcolor{blue}{C.Katsouris@soton.ac.uk}. \textit{Website}: \textcolor{blue}{sites.google.com/view/christiskatsouris} } 
\\ \textit{University of Southampton} 
\\
\\   
\small \textcolor{blue}{JOB MARKET PAPER} 
\\
\\
\large  First version: November 13, 2021
\\
\large This version: November 13, 2021
}

\date{}

\maketitle



\begin{abstract}
\vspace*{-0.3em}
We propose an econometric environment for structural break detection in nonstationary quantile predictive regressions. We establish the limit distributions for a class of Wald and fluctuation type statistics based on both the ordinary least squares estimator and the endogenous instrumental regression estimator proposed by \cite{PhillipsMagdal2009econometric}. Although the asymptotic distribution of these test statistics appears to depend on the chosen estimator, the IVX based tests are shown to be asymptotically nuisance parameter-free regardless of the degree of persistence and consistent under local alternatives. The finite-sample performance of both tests is evaluated via simulation experiments. An empirical application to house pricing index returns demonstrates the practicality of the proposed break tests for regression quantiles of nonstationary time series data.    
\vspace*{0.3 em}

\vspace*{0.3 em}

\end{abstract}

\newpage

\begin{spacing}{0.20}
  \tableofcontents
\end{spacing}

\newpage 

\color{blue}

\begin{center}
\textbf{Notation and Abbreviations}
\end{center}

\medskip
\begin{footnotesize}

\begin{tabular}{cp{0.95\textwidth}}
  $x \vee y$   & $\text{max} \left\{ x, y \right\}$ \\
  $x \wedge y$ & $\text{min} \left\{ x, y \right\}$ \\
  $:=$      & equality by definition \\
  $\equiv$  & equivalent statement \\ 
  $\floor{ \ . \ }$ & integer part of the argument \\
  $| \ . \ |$ & absolute value of the argument \\
  $\mathds{1} \big\{ . \big\}$ & indicator function \\
  $\mathbb{E} \left[ \ . \ \right]$ & expectation operator \\
  $\mathbb{P} \left( \ . \ \right)$ & probability operator \\
  $\text{Var} \left( \ . \ \right)$ & variance operator \\ 
  \\
  
  $\mathcal{O}_{\mathbb{P}} \left( \ . \ \right)$ & Order of convergence \\ 
  $\overset{ \mathcal{D} }{ \to }$ & Convergence in distribution      \\
  $\overset{ \mathbb{P} }{ \to }$  & Convergence in probability       \\
  $\Rightarrow$            & Weakly convergence argument \\
  $\mathcal{D} \left( [0,1] \right)$ &  Skorokhod topology \\
  \\
  
  $\mathcal{SQ}^{ols}_n ( \lambda, \uptau )$ & sup-$\mathcal{Q}$ test based on OLS estimator \\
  \\
  $\mathcal{SQ}^{ivx}_n ( \lambda, \uptau )$ & sup-$\mathcal{Q}$ test based on IVX estimator \\ 
  \\
  $\mathcal{SW}^{ols}_n ( \lambda, \uptau )$ &  sup-Wald test based on OLS estimator \\
  \\
  $\mathcal{SW}^{ivx}_n ( \lambda, \uptau )$ &  sup-Wald test based on IVX estimator \\ 
  \\
  $\mathcal{W}^{ols}_n ( \lambda, \uptau )$ & (self-normalized) sup-Wald test based on OLS estimator \\
  \\
  $\mathcal{W}^{ivx}_n ( \lambda, \uptau )$ & (self-normalized) sup-Wald test based on IVX estimator \\
  \\

   $\mathcal{DQ}^{ols}_n ( \lambda, \uptau )$ & double sup-$\mathcal{Q}$ test based on OLS estimator \\
  \\
  $\mathcal{DQ}^{ivx}_n ( \lambda, \uptau )$ & double sup-$\mathcal{Q}$ test based on IVX estimator \\ 
  \\
  $\mathcal{DW}^{ols}_n ( \lambda, \uptau )$ &  double sup-Wald test based on OLS estimator \\
  \\
  $\mathcal{DW}^{ivx}_n ( \lambda, \uptau )$ & double sup-Wald test based on IVX estimator \\ 
  \\
\end{tabular}\\

\end{footnotesize}

\color{black}

\newpage 

\bigskip

\color{black}

\begin{itemize}

\item \textbf{Summary:} We propose a set of structural break tests for testing the null hypothesis of no parameter instability in nonstationary quantile predictive regressions models. More specifically, we derive the asymptotic distribution of Wald type and fluctuation type statistics based on both the OLS and IVX estimators. We show that the limiting distribution of both the fluctuation and Wald test statistics under high persistence weakly convergent into a nonstandard and nonpivotal limiting distributions. In those cases, the underline stochastic processes  depend on the innovation distribution and cannot be identified with conventional processes commonly employed in the structural break literature such as a Brownian motion or Brownian bridge in the topological space.    

\medskip

\item \textbf{Related Literature:} The relevant literature to the econometric environment studied in the paper is the framework of predictive regression models proposed by  \cite{PhillipsMagdal2009econometric}, \cite{kostakis2015Robust} and \cite{lee2016predictive} who extend the asymptotic theory to the case of quantile predictive regression models. Further relevant studies include the framework proposed by \cite{xiao2009quantile} for the quantile cointegrating regression model as well as the paper of \cite{qu2008testing} who propose a framework for break testing in regression quantiles. 

\medskip

\item \textbf{Research Objective:} Our main research objective is to investigate the implementation of structural break tests in quantile predictive regressions under the assumption of nonstationarity, that is, regressors being generated as near unit root processes. To do this, we study the asymptotic distribution of the testing procedures and evaluate their statistical performance via Monte Carlo simulations.  

\medskip

\item \textbf{Research Contributions:} To the best of our knowledge the proposed framework in this paper that unifies the persistence properties of regressors in quantile predictive regression models and structural break testing is a novel aspect not previously examined in the literature. Our contributions in the literature is a statistical framework for testing for structural breaks in the relation between the regressand and the predictors based on a conditional quantile functional form.  

\medskip

\item \textbf{Research Findings:} The main research findings of our study are summarized. We derive the limit distributions of structural break tests suitable for quantile predictive regression models with regressors being either high persistent or mildly integrated. We focus on fluctuation type statistics as well as Wald type statistics which are constructed based on either the OLS or IVX estimators. Our tests can be used as diagnostic tools for break detection when estimating predictive regressions under nonstationarity based on a conditional quantile functional form.







\end{itemize}

\newpage 

\begin{wrap}

\color{blue}

\textbf{Reference:} Hypothesis testing near singularities and boundaries. 

For instance, one may find that an asymptotic distribution is $\chi^2_p$ with $p$ degrees of freedom at almost all points of the exponent rate of persistence within the unit circle but at the boundary, such that $\gamma_x = 1$, it discontinuously jumps to a different distribution. For instance a mixture of $\chi^2_p$ or something more complicated. Therefore, one could derive a general asymptotic distribution which encompasses all these different points of the exponent rate covering for instance the singularity case as well but then it simplifies to a standard $\chi^2_p$ limiting distribution when we are not at the boundary of the parameter space that the exponent rate of persistence belongs to.

On the other hand, for the true non-asymptotic distribution, for any fixed sample size actually it appears to be the case that due to the presence of the nuisance parameter of persistence the limiting distribution can be different from the corresponding limit for large samples. Furthermore, one might conclude that when considering the asymptotics at the singularity or boundary could be relevant to testing even when the true parameter value is near that point, for fixed sample sizes. However, as it appears from some preliminary simulation experiments and from the standard local-to-unity asymptotics, as the sample size is increased, the region on which the asymptotics give poor approximations shrinks, but no matter how large a sample is, the discontinuous behaviour of the asymptotic distribution indicates there is some parameter region on which it is inappropriate for empirical use. More specifically, an interesting aspect for future research it would be to propose a different approximation that the one obtained from standard local-to-unity asymptotics. In particular such an approximate distribution could be dependent on both sample size and parameter value, but has no discontinuous jump near the boundaries or singularities. In other words, for hypothesis testing in nonstationary time series models, complications arise when the asymptotic distributions depend on nuisance parameters. 

\color{red}
\textbf{The finite Sample Inference Procedure} Using the conditions discussed in the previous sections, we are able to provide the key results on finite sample inference. Thus, consider the following GMM function for estimating $\theta_0$
\begin{align}
L_n(\theta) = \frac{1}{2} \left[ \frac{1}{\sqrt{n}} \sum_{t=1}^n m_t (\theta) \right]^{\prime} \boldsymbol{W}_n \left[ \frac{1}{\sqrt{n}} \sum_{t=1}^n m_t (\theta) \right], 
\end{align}
where $m_t(\theta) = \big[ \uptau - \mathds{1} \left\{ y_t \leq q( D, \theta, \uptau ) \right\} \big] g(Z_t)$. Then, a convenient and natural choice of $W_n$ is given by 
\begin{align}
W_n = \frac{1}{ \uptau( 1 - \uptau) } \left[ \frac{1}{n} \sum_{t=1}^n g(Z_t) g(Z_t)^{\prime} \right]^{-1}
\end{align}
which equals to the inverse of the variance of $n^{-1/2} \sum_{t=1}^n m_t( \theta_0 )$ conditional on $Z_1,...,Z_n$. Therefore, since this conditional variance does not depend on $\theta_0$, the GMM function with $W_n$ defined above also corresponds to the continuous-updating estimator of Hansen.

\end{wrap}

\color{black}

\newpage 

\setcounter{page}{1}
\pagenumbering{arabic}

\section{Introduction}
\label{Section1}

Time series predictability is a research question that sparked the development of various methodologies for estimation and inference in predictive regression models. Related studies include \cite{jansson2006optimal}, \cite{mikusheva2007uniform}, \cite{Phillips2014Confidence}, \cite{cai2014testing}, \cite{kostakis2015Robust}, \cite{Kasparis2015nonparametric}, \cite{gonzalo2012regime, gonzalo2017inferring}, \cite{ren2019balanced},  \cite{demetrescu2020testing}, \cite{yang2020testing}, \cite{georgiev2021extensions},  \cite{andersen2021consistent}, \cite{harvey2021simpleA} as well as \cite{dou2021generalized} among others\footnote{The problem of distorted statistical inference in predictive regression models under nearly integrated predictors has been also reported in previous studies such as   \cite{elliott1994inference}, \cite{elliott2011control}, \cite{campbell2006efficient} and  references therein.}. The aforementioned frameworks operate under the general assumption of a stable relation between the predictant and the predictors of the model. However, the possible presence of parameter instability under these conditions require a different treatment  (e.g., see \cite{pitarakis2017simple} and  \cite{georgiev2018testing}). Thus, to derive limits for structural break tests suitable for predictive regressions, necessitate to employ certain regularity conditions and invariance principles of partial sum processes as in \cite{Phillips2007limit, PhillipsMagdal2009econometric}, to obtain stochastic integral approximations.    

Most of the current literature focuses on detecting structural break in the conditional mean of the distribution of $f ( y_t | \boldsymbol{x}_t)$ or $f ( y_t | \boldsymbol{x}_{t-1})$ where $\boldsymbol{x}_{t}$ is assumed to follow a near unit root process such as in \cite{cai2015testing}, \cite{georgiev2018testing} and \cite{Katsouris2021breaks}, leaving the conditional quantile distribution vastly unexplored. Consequently, this paper addresses these issues by developing a framework suitable for structural break testing for quantile regressions under nonstationarity which has not seen much attention. Specifically, we build on related studies such as \cite{lee2016predictive} and \cite{fan2019predictive} who propose a framework for estimation and inference in quantile predictive regressions as well as the study of \cite{qu2008testing} who focus on break testing procedures for regression quantiles. More precisely, the latter approach studies a linear quantile model with stationary covariates, while we consider the quantile predictive regression model with possibly nonstationary predictors as in \cite{lee2016predictive}. Another related study is presented by \cite{aue2017piecewise} in the context of piecewise quantile autoregressions. Therefore, proposing tests which bridge the gap between these two approaches is a further development of the current toolkit while the local-to-unity theory proposed by the seminal work of \cite{Phillips1987time, Phillips1987towards, Phillips1988testing} can facilitate the asymptotic theory.   

Consider the $\uptau-$th conditional quantile of $y_t$ which is defined as following   
\begin{align}
\label{quantile}
\mathsf{Q}_{y } \left( \uptau | x \right) = F^{-1}_{y | x }  ( \uptau | x ) := \mathsf{inf} \big\{ \mathsf{s} : F_{y | x } ( \mathsf{s} | x ) \geq \uptau \big\}
\end{align} 
Specifically, using the conditional quantile function \eqref{quantile} (see, \cite{kiefer1967bahadur}) to be the underline functional form for the predictive regression model then the induced econometric environment permits to investigate the presence of quantile predictability.  

\newpage

As a result, the parameter vector of the predictive regression model is quantile dependent for some fixed quantile level within a compact set. In practise, the parameter vector of a quantile regression can be estimated as the unique solution of the following  unconstrained optimization problem (see, \cite{koenker1987estimation} and \cite{portnoy1991asymptotic})
\begin{align}
\underset{ \boldsymbol{b} \in \mathbb{R}^p }{ \mathsf{arg \ min}  } \ \sum_{t=1}^n \uprho_{\uptau} \big( y_t -  \boldsymbol{x}_{t-1}^{\prime}  \boldsymbol{b}  \big)
\end{align}
where $\uprho_{\uptau} ( \mathsf{u} ) \equiv \mathsf{u} \left( \uptau - \mathds{1} \left\{ \mathsf{u} < 0 \right\} \right)$, is the check function as defined by \cite{Koenker1978regression}, $\boldsymbol{b}$ is some parameter vector and $\boldsymbol{x}_{t-1}$ is the lagged regressor of the model. Our research interest is on the asymptotic behaviour of the parameter estimators of $\boldsymbol{b}$ under both the null hypothesis of no parameter instability as well as under the alternative hypothesis of structural break at an unknown break-point location within the full sample. 

The structural break literature has various applications for different modelling conditions, however there are currently limited studies related to break testing in quantile time series models even under the assumption of stationarity and ergodicity such as in the seminal paper of \cite{andrews1993tests} for linear models. Furthermore, when considering predictive regression models inference regarding structural break involves deriving nonstandard asymptotic theory due to the presence of the nuisance coefficient of persistence. More precisely, testing based on conventional estimation methods (such as OLS-based tests) has been proved to have size distortions for persistent regressors (e.g., see \cite{georgiev2018testing} and \cite{Katsouris2021breaks}). Our study is considered as a unified framework for structural break detection in quantile predictive regressions (see, \cite{lee2016predictive} and \cite{fan2019predictive}) which encompasses regressor properties such as high persistent, mildly integrated or stationary under certain parameter space restrictions on the persistence parameters. In particular, the proposed tests can statistically evaluate for breaks in the model coefficients at both fixed and multiple quantiles of the underline conditional quantile distribution. We investigate the asymptotic theory and implementation of both Wald type statistics as in \cite{andrews1993tests} as well as  fluctuation type statistics as in \cite{qu2008testing} but within the setting of nonstationary quantile predictive regressions.

Our contributions are threefold. Firstly, we study the estimation problem of quantile predictive regressions with multiple predictors assumed to be generated as either near unit root or mildly integrated processes. We construct quantile predictability Wald statistics under the null of no predictability and derive the asymptotic distributions (as in \cite{lee2016predictive}). Secondly, we propose a structural break testing procedure for quantile predictive regressions which permits testing for the presence of breaks at the tails. Thirdly, we examine the statistical performance of these tests for detecting parameter instability in the coefficients of quantile predictive regressions with extensive simulation experiments of empirical size. 

\newpage

Under the null hypothesis of no structural break the specific weakly convergence argument permits to establish convergence to stochastic integral approximations defined with respect to a two-parameter process; limit results employed to derive the asymptotic distributions of fluctuation type statistics. Similarly, for the Wald type statistics we employ the asymptotic theory developed by \cite{PhillipsMagdal2009econometric} and \cite{lee2016predictive}. Thus, we introduce invariance principles for partial sum processes of matrix moments based on these functionals to obtain asymptotic results for the proposed econometric environment.  

Throughout the paper, we assume that all random elements are defined within a probability space denoted with the triple $\left( \Omega, \mathcal{F}, \mathbb{P} \right)$. All limits are taken as $n \to \infty$, where $n$ is the sample size. Denote with $\mathcal{D} \left( [0,1] \right)$ to be the set of functions on $[0,1]$ that are right continuous and have left limits, equipped with the Skorokhod metric. Then, the symbol $"\Rightarrow"$ is used to denote the weak convergence of the associated probability measures as $n \to \infty$. The symbol $\overset{ \mathcal{D} }{\to}$ and $\overset{ \mathbb{P} }{ \to }$  are employed to denote convergence in distribution and convergence in probability respectively. Moreover,  we denote with $\mathcal{O}_{ \mathbb{P} }(.)$ and $o_{ \mathbb{P} }(.)$ the stochastic order of convergence in probability (see, \cite{billingsley1968convergence}). In terms of vector and matrix notation, for any random matrix $X$, $\norm{ X }_p$ denotes the $L_p-$norm, that is, $\norm{ X }_p = \left( \mathbb{E} \norm{ X }^p \right)^{ 1 / p}$ where $p$ is some positive constant. The proposed test statistics are constructed based on the Euclidean norm. Moreover, we employ the term $\textit{stationary regressor}$ meaning a regressor of the predictive regression model which is generated from the local-to-unity specification that satisfies $\upgamma_x = 0$ and $|1 +  c_i | < 1 \ \forall i$.   

The paper is organized as follows. Section \ref{Section2} discusses the econometric environment of the quantile predictive regression model along with main assumptions, the estimation methodology as well as the testing hypotheses of interest. Section \ref{Section3} presents the testing procedure and asymptotic theory of structural break testing for a fixed quantile level. Section \ref{Section5} investigates the finite sample performance of the proposed tests via Monte Carlo experiments. Section \ref{Section6} illustrates the implementation of of the proposed testing procedures with an empirical application. Section \ref{Section7} concludes.  Proofs of main limit results can be found in the Appendix (see Section \ref{Section8}).

\subsection{Literature Review}


Estimation and inference with quantile models has been proposed to the literature with the seminal work of \cite{Koenker1978regression} and \cite{Koenker1982Robust}. In particular, the uniform Bahadur-type representation established by \cite{koenker1987estimation} as well as the representation of quantiles proposed by \cite{knight1989limit} (see also \cite{koul1995autoregression}) are commonly used to derive limit distributions for estimators and test statistics. Several papers in the literature follow these methodologies that examine related aspects to the proposed estimation environment and testing procedures.

\newpage

Firstly, related literature to the problem of structural change for regression quantiles include the studies of \cite{su2008testing} and 
\cite{qu2008testing}. Both frameworks develop diagnostic tools for break detection in quantile regression models under the assumption of stationary covariates. In the former case the alternative hypothesis of a single break is formulated with respect to the magnitude of the break-point. In the latter case, the author also propose a multiple break point testing procedure.  Additionally, \cite{furno2014quantile} implements these quantile regression based statistics using the methodology proposed by \cite{chow1960tests} which implies testing for break at a known location in the sample.

A different perspective is presented by 
\cite{hoga2018structural} who consider detecting for breaks using tail dependence measures (see also \cite{hoga2017change}\footnote{The framework proposed by \cite{hoga2017change} considers a set of change point tests for the tail index of random variables and contributes to the literature of nonparametric modelling methods.}) based on an empirical estimator of extremal dependence. An extension of the method to an alternative hypothesis with multiple breaks is also examined. All aforementioned procedures correspond to structural break tests suitable for quantile models\footnote{Notice that the literature of break tests presented here differs from the literature of slope heterogeneity in quantile models. Related aspects to inference are \cite{koenker2001quantile}, \cite{chernozhukov2005extremal}, \cite{portnoy2012nearly} and \cite{escanciano2018quantile} which are of independent interest.} while in our study we focus on implementing testing procedures specifically for quantile predictive regression models in which the time series properties of predictors are modelled with the nuisance parameter of persistence. In particular, a separate autoregressive model with a local unit root coefficient matrix is used to model the unknown degree of persistence which implies that conventional approaches for deriving large sample theory are no longer valid. However, the fluctuation based tests implemented by \cite{qu2008testing} for break detection in quantile models with stationary covariates provides a suitable econometric environment for investigating the effect of nonstationarity to the asymptotic distribution of the tests. 

Secondly, asymptotic theory for quantile time series regressions has been examined by \cite{Koenker2002inference, Koenker2004unit, koenker2006quantile} in the context of autoregressive model specifications and unit root testing\footnote{A unified framework for econometric inference with nearly integrated regressors and unit roots is proposed by the seminal work of \cite{Phillips1988Regression, Phillips1987towards, Phillips1987time}. Further relevant literature includes the study of \cite{stock1994unit} who discuss aspects related to testing with unit roots and structural breaks in time series models as well as  Chapter 14 in \cite{davidson2000econometric} that has additional derivations and examples.}. Further applications in the time series literature include testing procedures for threshold effects under the assumption of a conditional quantile function as in the studies of \cite{Galvao2011threshold, galvao2014testing} and the case of nonstationary (nonlinear) quantile regressions by
\cite{xiao2009quantile}, \cite{cho2015quantile}, \cite{li2016estimation} and \cite{uematsu2019nonstationary}. Moreover, \cite{kato2009asymptotics} derives related limit results which are useful when considering the asymptotic behaviour of quantile estimators for a wide class of modelling approaches and econometric conditions. All these studies cover both stationary and nonstationary models however they operate under the null hypothesis of no parameter instability in model parameters. Therefore, our proposed testing procedure, is considered to be a novel contribution to the time series econometrics literature.  

\newpage

Thirdly, the proposed structural break testing methodology is closely related to the problem of unidentified parameters under the null hypothesis which is  well-known in the econometrics and statistics literature as the Davies problem (see, \cite{davies1977hypothesis}). The particular aspect which is relevant to parameter admissibility has been investigated in problems of estimation and testing such as in the studies of \cite{hansen1996inference}, \cite{andrews1994optimal}, \cite{pitarakis2004least} and \cite{elliott2015nearly}. A more recent approach to the unidentified parameter problem under the null is presented by \cite{mccloskey2017bonferroni}\footnote{Specifically, \cite{mccloskey2017bonferroni} proposes a framework for a set of flexible size-corrected critical values construction methods that lead to tests with correct asymptotic size and desirable power properties in testing problems with nuisance parameter under the null hypothesis.}. Therefore, to overcome this challenge, we employ the supremum operator when constructing Wald type statistics. Furthermore, a nonparametric estimation approach for quantile regressions such in the study of \cite{qu2015nonparametric}, may require different moment conditions for estimation and inference which is beyond the scope of this paper. 

Lastly, in terms of the model structure of the proposed modelling approach of the paper, we impose standard econometric assumptions and conditions in the literature of predictive regressions. More precisely, by imposing a standard martingale difference condition on the equation innovations $u_t$, this implies an orthogonality condition between the innovations and the model regressors, such that $\mathsf{Cov} \big( u_t, \boldsymbol{x}_{t} \big) = \mathbb{E} \big[ \boldsymbol{x}_{t} \mathbb{E} \left( u_t | \mathcal{F}_{t-1} \right) \big] = 0$. Furthermore, the model structure we follow does not allow for endogeneity even though the IVX instrumentation method implies the construction of endogenous instruments. Specifically, as explained by \cite{PhillipsMagdal2009econometric} and \cite{kostakis2015Robust}, the IVX filtration implies the construction of instrumental covariates based on information obtained only from the regressors of the model. According to \cite[p.~731]{WangPhillips2012specification} the model structure would permit for the presence of endogeneity when the equation error could be serially dependent and cross-correlated under certain moment restrictions (see also \cite{yang2020testing}\footnote{Specifically, \cite{yang2020testing} propose a unified IVX-AR Wald statistic that accounts for serial correlation in the error terms of the linear predictive regression model. The IVX-AR estimator corrects the size distortions arising from serially correlated error under the presence of high persistence.}). However, the particular aspect can complicate the derivation of the limit distributions as additional considerations will be needed, such as to incorporate conditional heteroscedasticity, which is beyond the scope of our study and we leave as future research related to the proposed framework.      

Based on the aforementioned challenges in the literature, we are motivated to develop an econometric framework for structural break testing in quantile predictive regressions with persistent covariates. Overall our research objectives are concentrated around two main pillars: \textit{(i)} to introduce a new set of weak convergence results for certain partial sums of functionals which are based on either the OLS or the IVX estimators and involve nonstationary time series. Specifically, these functionals are instrumental for deriving the asymptotic behaviour of the proposed test statistics; and \textit{(ii)} to demonstrate the practicality of a new set of structural break tests for detecting parameter instability.


\newpage

\section{Econometric Environment and Testing Problem}
\label{Section2}

\subsection{Econometric Model and Assumptions}
\label{Section2.1}

Consider the (linear) predictive regression model
\begin{align}
\label{model1}
y_t &= \alpha + \boldsymbol{\beta}^{\prime} \boldsymbol{x}_{t-1} + u_t, \ \ \ \ 1 \leq t \leq n, 
\\ 
\label{model2}
\boldsymbol{x}_t &= \boldsymbol{\Phi}_n \boldsymbol{x}_{t-1} + \boldsymbol{v}_t
\end{align}
where $y_t \in \mathbb{R}^{n \times 1}$ is a scalar dependent variable and  $\boldsymbol{x}_{t-1} \in \mathbb{R}^{n \times p}$ is a $p-$dimensional vector of predictors such that $\boldsymbol{x}_{t}$ is generated as a near unit root process (or local unit root process) with an autocorrelation coefficient matrix as defined by the studies of \cite{PhillipsMagdal2009econometric} and \cite{kostakis2015Robust} expressed as below
\begin{align}
\label{autoregressive.matrix}
\boldsymbol{\Phi}_n = \left( \boldsymbol{I}_p + \frac{ \boldsymbol{C}_p  }{ n^{ \upgamma_x } } \right), \ \ \text{for some} \ \upgamma_x > 0.
\end{align}
where $n$ is the sample size, and $\boldsymbol{C}_p  = \mathsf{diag} \left\{ c_1,...,c_p \right\}$  is a $p \times p $ diagonal matrix with the coefficients of persistence $c_i$ for $i = 1,...,p$. We consider that the predictors of the model are allowed to belong only to one of the two degree of persistence as specified below
\begin{itemize}
\item \textit{Local Unit Root} (\textit{LUR}): $\upgamma_x = 1$ and $c_i \in ( - \infty, 0  ),  \ \forall \ i = 1,...,p$.
\item \textit{Mildly Integrated} (\textit{MI}): $\upgamma_x \in (0,1)$ and $c_i \in ( - \infty, 0), \ \forall \ i = 1,...,p$.
\end{itemize} 
Let $\mathcal{F}_t$ denote the natural filtration, then for the error term of the predictive regression we assume that $\mathbb{E} \left( u_t | \mathcal{F}_{t-1} \right) = 0$ and $\mathbb{E} \left( u^2_t | \mathcal{F}_{t-1} \right) = \sigma^2_{uu}$. Specifically, the innovation structure of the predictive regression model allows to impose a linear process dependence for $\boldsymbol{v}_t $, with a conditionally homoscedastic \textit{martingale difference sequence} condition such that
\begin{align*}
\boldsymbol{v}_t = \sum_{j=1}^{\infty} \boldsymbol{\varphi}_j \boldsymbol{\varepsilon}_{t-j}, \ \ \boldsymbol{\varepsilon}_t \sim \textit{mds} \left( \boldsymbol{0}, \boldsymbol{\Sigma} \right), \ 
\end{align*}  
with necessary conditions for the linear process representation to hold given as below
\begin{align}
\boldsymbol{\Sigma} > 0, \ \ \ \ \ \sum_{j=0}^{\infty}  j \norm{ \boldsymbol{\varphi}_j } < \infty \ \ \text{such that} \ \  \boldsymbol{\varphi}_o ( \mathsf{z} )  = \sum_{j=0}^{\infty} \mathsf{z}^j \boldsymbol{\varphi}_j.
\end{align}
Denote with $\boldsymbol{e}_t = \left( u_t, \boldsymbol{v}_t^{\prime} \right)^{\prime}$, then under regularity conditions, the following invariance principle (\textit{FCLT}) holds (see  \cite{Phillips1992asymptotics})
\begin{small} 
\begin{align}
\label{fclt1}
\frac{1}{\sqrt{n}} \sum_{t=1}^{ \floor{ nr } }
\boldsymbol{e}_t :=
\frac{1}{\sqrt{n}}
\sum_{j=1}^{ \floor{ nr } }
\begin{bmatrix}
u_{t} \\
\boldsymbol{v}_{t}
\end{bmatrix}
\equiv
\begin{bmatrix}
B_{un} (r) \\
\boldsymbol{B}_{vn} (r)
\end{bmatrix}
\Rightarrow
\begin{bmatrix}
B_{u} (s) \\
\boldsymbol{B}_{v} (s)
\end{bmatrix}
:= 
\mathcal{BM}
\begin{bmatrix}
\sigma^2_{uu}  &  \boldsymbol{\sigma}^{\prime}_{uv} 
\\
\boldsymbol{\sigma}_{vu} &  \boldsymbol{\Sigma}_{vv}   
\end{bmatrix}_{ \textcolor{red}{ (p + 1) \times (p + 1) } }
\end{align}
\end{small}
where $\boldsymbol{\Sigma}_{vv} \in \mathbb{R}^{ p \times p}$ is a positive definite covariance matrix and $0 < r < 1$.

\newpage

Specifically, the individual components of the vector sequence $\boldsymbol{e}_t = \left( u_t, \boldsymbol{v}_t^{\prime} \right)^{\prime}$ have partial sums processes that weakly converge into their Brownian motion counterparts as below 
\begin{align}
B_{un} (r) 
:= 
\frac{1}{\sqrt{n}} \sum_{t=1}^{ \floor{ nr } } u_{t} 
&\Rightarrow  
B_{u} (r) := \mathcal{N} \big( 0, r \sigma_{uu}^2 \big)
\\
\boldsymbol{B}_{vn} (r) 
:= 
\frac{1}{\sqrt{n}} \sum_{t=1}^{ \floor{ nr } } \boldsymbol{v}_{t}
&\Rightarrow 
\boldsymbol{B}_{v} (r) := \mathcal{N} \big( \boldsymbol{0}, r \boldsymbol{\Sigma}_{vv} \big)
\end{align}
where $\boldsymbol{B}(r) = \big( B_{u} (r),  \boldsymbol{B}_{v} (r)^{\prime} \big)^{\prime}$ being a $(p \times 1)$ Brownian motion with long-run covariance matrix $\boldsymbol{\Sigma}_{ee}$, that is, a Gaussian vector process with almost surely continuous sample paths. More precisely, since $\boldsymbol{x}_t$ is an adapted process to $\mathcal{F}_t$ then in practise there exists a correlated vector Brownian motion $\boldsymbol{B}_n(r) = \big( B_{un} (r), \boldsymbol{B}_{vn} (r)^{\prime} \big)^{\prime}$ such that 
\begin{align}
\left( \frac{1}{\sqrt{n}  } \sigma_{uu}^{-1} \sum_{t=1}^{ \floor{nr} } u_t,  \frac{1}{ \sqrt{n} } \boldsymbol{\Sigma}_{vv}^{-1/2} \sum_{t=1}^{ \floor{nr} } \boldsymbol{v}_t \right)^{\prime} \Rightarrow \boldsymbol{B}(r) = \big( B_{u} (r), \boldsymbol{B}_{v}(r)^{\prime} \big)^{\prime}, \ \ 0 < r < 1
\end{align}
on $\mathcal{D}\left( [0,1] \right)^2$ as $n \to \infty$, with covariance matrix as in \eqref{fclt1} implying joint convergence. 
Assuming the above conditions hold, then the following local to unity principle applies (as proposed by   \cite{Phillips1987towards, Phillips1987time})
\begin{align}
\label{fclt2}
\frac{ \boldsymbol{x}_{[nr]} }{  \sqrt{n} } \Rightarrow \boldsymbol{J}_c(r),\ \ \ \text{where} \ \ \boldsymbol{J}_c(r) =  \int_0^r e^{(r-s) \boldsymbol{C}_p } d \boldsymbol{B}_v(s).
\end{align}

The functional $\boldsymbol{J}_c(r)$ represents the \textit{Ornstein-Uhlenbeck} process\footnote{The OU process is a stationary Gaussian process with an autocorrelation function that decays exponentially over time. Moreover, the continuous time OU diffusion process has a unique solution and this property allows to approximate asymptotic terms for estimators and corresponding test statistics as a function of the nuisance parameter of persistence (see, \cite{perron1991continuous}).} which is employed to derive stochastic integral approximations for the nonstationary predictive regression model \eqref{model1}-\eqref{model2}. Denote with $\boldsymbol{K}_c(r) := \boldsymbol{\Sigma}_{vv} \boldsymbol{J}_c(r)$, where $\boldsymbol{K}_c(r)$ is a $p-$dimensional Gaussian process defined as $\boldsymbol{K}_c(r) = \displaystyle \int_0^r e^{(r-s) \boldsymbol{C}_p } d \boldsymbol{B}_v(s)$ is the solution of \textit{Black-Scholes} differential equation $d \boldsymbol{K}_c(r) \equiv c \boldsymbol{K}_c(r) + d \boldsymbol{B}_v(r)$, with $\boldsymbol{K}_c(r) =0$ as the initial condition. 

\medskip

The specific autoregression matrix specification \eqref{autoregressive.matrix} allows to examine other persistence properties such as unit root processes, when $c_i = 0$ for all $i  \in \left\{ 1,...,p \right\}$, or explosive processes when $c_i > 0$. In this paper, we consider two types of nonstationarity, that is, the near unit or high persistent regressors and the mildly integrated regressors. Therefore, in both cases the coefficient of persistent, $c_i$, lies below the unit root boundary. Thus, the main difference between these two persistence classes is that the mildly integrated regressors have an exponent rate below the unit boundary, which implies that these regressors are less persistent  than regressors generated from near unit root processes. Furthermore, one can consider extending our estimation and testing framework to the case of explosive regressors; we leave this aspect for future research.


\newpage

\subsubsection{Quantile Predictive Regression Model}

Our main goal is to investigate the asymptotic theory and empirical implementation of structural break tests suitable for quantile predictive regression models. Therefore, we consider suitable modifications of the  innovation structure that corresponds to the linear predictive regression model, following standard conditions and assumptions employed for quantile time series models, currently presented in literature. More precisely, the conditional quantile function of $y_t$ denoted with $\mathsf{Q}_{y_t} \left( \uptau | \mathcal{F}_{t-1} \right)$, replaces the conditional mean function of the predictive regression which implies the following model specification
\begin{align}
\mathsf{Q}_{y_t} \left( \uptau | \mathcal{F}_{t-1} \right) :=  F^{-1}_{ y_t | \boldsymbol{x}_{t-1} } (\uptau) \equiv \alpha (\uptau) + \boldsymbol{\beta} (\uptau)^{\prime} \boldsymbol{x}_{t-1}. 
\end{align}
such that $F_{ y_t | \boldsymbol{x}_{t-1} } (\uptau) := \mathbb{P} \big( y_t \leq \mathsf{Q}_{y_t} \left( \uptau | \mathcal{F}_{t-1} \right) \big| \mathcal{F}_{t-1} \big) \equiv \uptau$, where $\uptau \in (0,1)$ is some quantile level in the compact set $(0,1)$. Therefore, in order to define the innovation structure that corresponds to the quantile predictive regression, we employ the piecewise derivative of the loss function such that $\psi_{\uptau} ( \mathsf{u} ) = \big[ \uptau - \mathds{1} \left\{ \mathsf{u} < 0 \right\} \big]$. Consequently, this implies that $u_t (\uptau) := u_{t} - F^{-1}_{ u } (\uptau)$ where $F^{-1}_{ u } (\tau)$  denotes the unconditional $\uptau-$quantile of the error term $u_{t}$. Then, the corresponding invariance principle for the nonstationary quantile predictive regression model is formulated as below  
\begin{align}
\frac{1}{\sqrt{n}}  \sum_{t=1}^{ \floor{ nr } } 
\begin{bmatrix}
\psi_{\uptau} \big( u_t (\uptau) \big) 
\\
\boldsymbol{v}_{t}
\end{bmatrix}
\Rightarrow
\begin{pmatrix}
B_{ \psi_{\uptau} } ( r )_{ \textcolor{red}{ ( 1 \times n ) }  } 
\\
\boldsymbol{B}_v ( r )_{ \textcolor{red}{ ( p \times n ) }  }
\end{pmatrix}
\equiv
\mathcal{BM} 
\begin{bmatrix}
\uptau (1 - \uptau) & \boldsymbol{\sigma}_{\psi_{\uptau} v}^{\prime}
\\
\boldsymbol{\sigma}_{v \psi_{\uptau}} & \boldsymbol{\Omega}_{vv}  
\end{bmatrix}
\end{align}   
\begin{assumption}
\label{assumption1} 
The following conditions for the innovation sequence hold:
\begin{itemize}
\item[\textbf{(\textit{i})}] The sequence of stationary conditional \textit{probability distribution functions (pdf)} denoted with $\big\{ f_{ u_t (\uptau), t-1}(.) \big\}$ evaluated at zero with a non-degenerate mean function such that $f_{ u_t (\uptau)  }(0) := \mathbb{E} \left[  f_{ u_t (\uptau), t-1}(0) \right] > 0$ satisfies a $\textit{FCLT}$ given as below
\begin{align}
\frac{1}{ \sqrt{n} } \sum_{t=1}^{ \floor{nr} } \big( f_{  u_t (\uptau), t-1}(0) - \mathbb{E} \left[  f_{ u_t (\uptau), t-1}(0) \right] \big) \Rightarrow B_{ f_{  u_t (\uptau) } } (r).
\end{align}
\item[ \textbf{(\textit{ii})} ]  For each $t$ and $\uptau \in (0,1)$, $f_{ u_t (\uptau), t-1}(.)$ is uniformly bounded away from zero with a corresponding conditional distribution function $F_t(.)$ which is absolutely  continuous with respect to Lebesgue measure on $\mathbb{R}$ (see, \cite{neocleous2008monotonicity}, \cite{goh2009nonstandard} and \cite{lee2016predictive}).
\end{itemize}
\end{assumption}
Assumption \ref{assumption1} \textbf{(\textit{i})} provides a standard weak convergence argument to a Brownian motion process that corresponds to the underline distribution generating the innovation sequence of the quantile predictive regression model (see, \cite{lee2016predictive} and \cite{fan2019predictive}). Furthermore, Assumption \ref{assumption1} \textbf{(\textit{ii})} provides the weak convergence argument for the sparsity function of the model to its  Brownian motion counterpart for some $0 < r < 1$. 

\newpage

\subsection{Estimation Methodology}

We investigate the statistical properties of the estimation methodology in relation to the handling of the nuisance parameter of persistence. We derive the asymptotic distributions of the associated test statistics based on two optimization methods which are known to have different convergence rates in the time series predictability literature.

\subsubsection{OLS based estimation}

We consider the OLS based estimation using the check function $\uprho_{\uptau} ( . )$, which is common practise for optimization problems of quantile series. Specifically, the asymptotic behaviour of the quantity $\boldsymbol{\mathcal{E}}_n(\uptau) \equiv \sqrt{n} \big( \boldsymbol{\beta}_n(\uptau) - \boldsymbol{\beta}_0(\uptau) \big)$ is of interest. The traditional approach to asymptotics for $\hat{\boldsymbol{\beta}}(\uptau)$ is to employ a Bahadur representation which allows to decompose the expression into a Brownian bridge component and an error term (see, \cite{portnoy2012nearly} and \cite{kato2009asymptotics}). Furthermore, various studies are concerned with the determination of sharp error bounds for the specific error term. However in our setting, $\sqrt{n}-$consistent asymptotics do not always apply due to the presence of  nonstationarity. Additionally, the chosen estimator affects the stochastic rates of convergence. 

Similar to the linear predictive regression (a model with a conditional mean functional form), under the assumption of persistent regressors, the OLS estimator has been proved to be biased due to the presence of nuisance parameters (e.g., see \cite{campbell2006efficient}), resulting to distorted statistical inference\footnote{Relevant studies in the literature which examine the asymptotic behaviour of standard $t-$tests under these conditions include \cite{Phillips2013predictive, Phillips2016robust}, \cite{lee2016predictive}, \cite{fan2019predictive},  \cite{kostakis2015Robust} and \cite{Kasparis2015nonparametric} among others.}. Nevertheless, focusing on the two persistence properties we introduced previously, we derive its limit distribution which is useful when considering structural break tests under nonstationarity.      

Denote with $\boldsymbol{ \theta } ( \uptau ) = \big[ \alpha (\uptau), \boldsymbol{\beta} (\uptau)^{\prime} \big]^{\prime} \in \mathbb{R}^{ (p+1) \times 1 }$ and $\boldsymbol{X}_{t-1} = \left( \boldsymbol{1}, \boldsymbol{x}_{t-1}^{\prime} \right)^{\prime} \in \mathbb{R}^{ (p+1) \times n }$, then the OLS based estimator is obtained by solving the following optimization problem  
\begin{align}
\widehat{\boldsymbol{ \theta }}_n^{qr} \left( \uptau \right) := \underset{ \boldsymbol{ \theta } \in \mathbb{R}^{p+1} }{ \mathsf{arg \ min} } \ \sum_{t=1}^n \uprho_{\tau} \big( y_{t} - \boldsymbol{X}_{t-1}^{\prime} \boldsymbol{\theta}  \big)
\end{align}
where $\uprho_{\uptau}( \mathsf{u} ) = \mathsf{u} \big( \uptau - \mathds{1} \left\{ \mathsf{u} < 0 \right\} \big)$ with $\uptau \in (0,1)$, represents the asymmetric quantile regression function. Following \cite{lee2016predictive}, we use the normalization matrices below which are different according to the persistence properties of predictors such that 
\begin{align}
\label{normalization}
\boldsymbol{D}_n := 
\left\{
\begin{array}{ll}
      \mathsf{diag} \big( \sqrt{n}, n \boldsymbol{I}_p  \big)    & \text{for} \  \textit{LUR},  \\
      \mathsf{diag} \left( \sqrt{n}, n^{\frac{ 1 + \upgamma_x }{2}} \boldsymbol{I}_p \right)    & \text{for} \  \textit{MI} . 
\end{array} 
\right. 
\end{align}
Then, Corollary \ref{corollary1} summarizes the asymptotic distribution of the OLS-QR estimator (see also Theorem 2.1 \cite{lee2016predictive}) for mildly integrated and high persistent regressors.

\newpage

\begin{corollary}
\label{corollary1}
Under Assumption  \ref{assumption1} and \textit{FCLT} \eqref{fclt1}-\eqref{fclt1} it follows that: $\boldsymbol{D}_n \left( \widehat{\boldsymbol{ \theta }}_n^{qr} \left( \uptau \right) - \boldsymbol{ \theta } \left( \uptau \right)  \right)$
\begin{small}
\begin{align*}
\Rightarrow
\left\{
\begin{array}{ll}
    f_{ u_t (\uptau) }(0)^{-1} 
    \begin{bmatrix}
    1 & \displaystyle \int_0^1 \boldsymbol{J}_c(r)^{\prime} dr \\
    \displaystyle \int_0^1 \boldsymbol{J}_c(r) dr & \displaystyle \int_0^1 \boldsymbol{J}_c(r) \boldsymbol{J}_c(r)^{\prime} dr 
\end{bmatrix}^{-1}_{ \textcolor{red}{ ( p + 1 ) \times (p + 1) } } 
\begin{bmatrix}
   B_{\psi_{\uptau}}(1)_{ \textcolor{red}{ (1 \times n) } }  
    \\
   \displaystyle \int_0^1 \boldsymbol{J}_c(r) dB_{\psi_{\uptau}} dr_{ \textcolor{red}{ ( p \times n ) } }
\end{bmatrix}                 & \textit{LUR},  
\\
     \mathcal{N} \displaystyle  \left( 0, \frac{ \uptau (1 - \uptau) }{  f_{ u_t (\uptau) }(0)^2  }   
     \begin{bmatrix}
        1  &  \boldsymbol{0}^{\prime} \\
        \boldsymbol{0}  & \boldsymbol{V}_{xx}^{-1}
     \end{bmatrix}_{ \textcolor{red}{(p + 1) \times (p + 1)} } 
     \right)         
         & \textit{MI}.  
\end{array} 
\right. 
\end{align*}
\end{small}
where the stochastic matrix $\boldsymbol{V}_{xx}$ is defined by the following expression 
\begin{small}
\begin{align*}
\boldsymbol{V}_{xx} := \int_0^{\infty} e^{r \boldsymbol{C}_p } \boldsymbol{\Omega}_{xx}  e^{r\boldsymbol{C}_p} dr, \ \text{where} \ \boldsymbol{\Omega}_{xx} := \sum_{m=-\infty}^{\infty} \mathbb{E} \left( \boldsymbol{v}_{t} \boldsymbol{v}_{t-m}^{\prime} \right) = \boldsymbol{\varphi}_{o} (1) \boldsymbol{\Sigma} \boldsymbol{\varphi}_{o} (1)^{\prime}. 
\end{align*}
\end{small}
\end{corollary}
Consequently, the limiting joint distribution of the model intercept and slopes for the nonstationary quantile predictive regression model under the assumption of mildly integrated regressors, is a mixed normal of the form $\mathcal{MN} \big( 0, \boldsymbol{\Sigma}^{\star} \big)$, where
\begin{small}
\begin{align}
\boldsymbol{\Sigma}^{\star} 
:=
\frac{ \uptau ( 1 - \uptau ) }{  f_{ u_t (\uptau_0) }(0)^2  }   
\begin{bmatrix}
  1  &  \boldsymbol{0}^{\prime} \\
  \boldsymbol{0}  & \boldsymbol{V}_{xx}^{-1}
\end{bmatrix}, \ \ \text{for some} \ \uptau \in (0,1).
\end{align} 
\end{small}
while under high persistence the asymptotic behaviour of $\boldsymbol{D}_n \left( \widehat{\boldsymbol{ \theta }}_n^{qr} \left( \uptau \right) - \boldsymbol{ \theta } \left( \uptau \right)  \right)$ depends on functionals of OU processes which are more challenging to approximate, especially if one is interested to obtain sharp error bounds\footnote{In particular, the study of  \cite{portnoy2012nearly} obtains a near $\sqrt{n}-$consistent error bound by employing the "Hungarian construction" which requires to approximate the quantity $\boldsymbol{\mathcal{E}}_n(\uptau)$ using a Brownian bridge limit which converges to this non-zero Gaussian process with an appropriate rate of convergence.  } as in \cite{portnoy2012nearly}.   

Furthermore, we assume that the sparsity coefficient can be consistently estimated with a fixed unbiased estimator for all $t$. This a strong assumption which can be regarded as a trade-off relation between the complicated objective function (i.e., non-differentiable and nonstationary) and the tractable error term of the model \citep{uematsu2019nonstationary}. On the other hand, it permits to consider the limiting distributions when testing for a set of parameter restrictions. Thus, to overcome the problem of nonstandard statistical inference due to the presence of the nuisance coefficient of persistence, we employ the instrumental variable regression approach proposed by \cite{PhillipsMagdal2009econometric}. 

The estimator of \cite{PhillipsMagdal2009econometric} performs reasonably well in finite samples and even performs better that the OLS counterpart (see \cite{georgiev2021extensions}), demonstrating the robustness of the method in filtering out abstract degree of peristence when testing for linear restrictions in predictive regressions. In addition, the suggested instrumental variable approach is by definition neither spurious, since it is always correlated with the corresponding regressor, not a poor instrument, because the correlation of unit root processes tends to one asymptotically. 

\newpage

\subsubsection{IVX based estimation}

The endogenous instrumentation (IVX) procedure for predictive regression models\footnote{Notice that the related asymptotic theory which is robust to abstract degree of persistence and results to nuisance-parameter free inference was pioneered by \cite{Magdalinos2009limit} in the context of cointegration models.} proposed by \cite{PhillipsMagdal2009econometric} implies the use of a mildly integrated instrumental variable. The instrumented variable is  constructed as below 
\begin{align}
\label{IVX.instrument}
\widetilde{ \boldsymbol{z} }_{t} = \sum_{ j = 0 }^{ t - 1} \left( \boldsymbol{I}_p + \frac{ \boldsymbol{C}_z }{ n^{ \upgamma_z } }   \right) \big( \boldsymbol{x}_{t-j} - \boldsymbol{x}_{t-j-1} \big), 
\end{align}
where $\boldsymbol{C}_{z} = \mathsf{diag} \{ c_{z1},...,c_{zp} \}$ is a $p \times p$ diagonal matrix such that $c_{zj} < 0 \ \forall \ j \in \left\{ 1,..., p \right\}$ with $0 < \upgamma_z < 1$, where $\upgamma_z$ is the exponent rate of the persistence  coefficient of the instrumental variable, such that $\upgamma_z \neq \upgamma_x$. The IVX filtering methodology  transforms a possibly nonstationary autoregressive process that generates the set of predictors, $\boldsymbol{x}_t$, which encompasses both stable or unstable processes based on the behaviour of the local unit root coefficient, into a mildly integrated process which is less persistent than the endogenous variables. Another statistical property is the choice of the exponent rate for the coefficient of persistence that corresponds to the instrumental variable $c_{zj}$. Specifically, the econometric literature has documented a choice of $\upgamma_z$ close to 0.95 as a reasonable value with desirable finite-sample properties when constructing predictability tests (see, \cite{lee2016predictive}, \cite{Phillips2016robust} and \cite{kostakis2015Robust}). Thus, to account for the different convergence rates due to nonstationarity and obtain the asymptotic distribution of the IVX-QR estimator we employ the following normalization matrices 
\begin{align}
\tilde{ \boldsymbol{Z} }_{t-1,n} := \tilde{ \boldsymbol{D} }_n^{-1} \tilde{ \boldsymbol{z} }_{t-1} \ \ \text{and} \ \ \tilde{ \boldsymbol{X} }_{t-1,n} := \tilde{\boldsymbol{D}}_n^{-1} \tilde{ \boldsymbol{x} }_{t-1} 
\end{align} 
where $\tilde{\boldsymbol{D} }_n = n^{\frac{ 1 + \upgamma_x \wedge \upgamma_z }{2}} \boldsymbol{I}_p$, such that $\upgamma_x  \wedge \upgamma_z \equiv \mathsf{min} \left( \upgamma_x, \upgamma_z \right)$ which is identical for both the case of \textit{local unit root} and \textit{mildly integrated} regressors. Furthermore, we denote with $y_t (\uptau) := y_t - \alpha( \uptau ) + \mathcal{O}_{ \mathbb{P} } ( n^{- 1/ 2} )$ to be the zero-intercept \textit{QR} dependent variable. The particular dequantiling procedure permits to reformulate the quantile model as $y_t (\uptau) = \boldsymbol{x}_{t-1}^{\prime} \beta( \uptau ) + u_t (\uptau )$ that simplifies the derivations for the asymptotics of the model estimator which is known to have different convergence rates when an intercept is included (e.g., see \cite{gonzalo2012regime, gonzalo2017inferring}).  
Then, the IVX-QR estimator for the quantile regression, is defined by the following unconstrained optimization problem
\begin{align}
\label{min.function}
\widehat{\boldsymbol{\beta}}_n^{ivx-qr} ( \uptau ) 
:= \underset{ \boldsymbol{\beta} \in \mathbb{R}^p }{ \mathsf{ arg \ inf } } \ \frac{1}{2} \ \left\{ \left( \sum_{t=1}^n h_t \left( \boldsymbol{\beta} \right) \right)^{\prime} \left( \sum_{t=1}^n h_t \left( \boldsymbol{\beta} \right) \right) \right\},   
\end{align}
where $h_t (.)$ is defined below such that $\psi_{ \uptau } ( \mathsf{u} ) := \big[ \mathds{1} \left\{ \mathsf{u} \leq 0 \right\} - \uptau \big]$ and $\psi_{ \uptau } ( \mathsf{u} ) \overset{  \mathsf{maps \ to} }{ \mapsto } \uprho_{\uptau}^{-1} ( \mathsf{u} )$
\begin{align}
h_t \left( \boldsymbol{\beta} \right) 
:= 
\tilde{ \boldsymbol{z} }_{t-1} \times \psi_{\uptau} \big( u_t (  \uptau ) \big) 
\equiv 
\tilde{\boldsymbol{z}}_{t-1} \times \big[ \uptau - \mathds{1} \big\{ y_t ( \uptau ) < \boldsymbol{x}_{t-1} ^{\prime} \boldsymbol{\beta} \big\} \big]
\end{align}

\newpage

The minimization of expression \eqref{min.function} leads to the following first order condition:
\begin{align}
\sum_{t=1}^{n} \tilde{ \boldsymbol{Z} }_{t-1,n} \times \big[ \uptau - \mathds{1} \big\{ y_t ( \uptau ) < \boldsymbol{x}_{t-1} ^{\prime} \boldsymbol{\beta}_n^{ivx-qr} (\uptau) \big\} \big] = o_{ \mathbb{P} }(1).
\end{align}
Furthermore, it can be proved that for both the cases of high persistent and mildly integrated predictors (\textit{LUR} or \textit{MI}) the asymptotic distribution of the IVX-QR estimator is identical as presented by Corollary \ref{corollary2}. Notice that for the interested reader the limit distribution for other classes of persistence (e.g., such as predictors which exhibit near stationary or mildy explosive persistence) can be found in Theorem 3.1 of \cite{lee2016predictive}. 

\medskip

\begin{corollary}(IVX-QR Limit Theory) 
\label{corollary2}
Under Assumption \ref{assumption1} it follows that 
\begin{align}
\tilde{\boldsymbol{D}}_n \left( \widehat{\boldsymbol{\beta}}_n^{ivx-qr} \left( \uptau \right) - \boldsymbol{\beta} \left( \uptau  \right) \right) 
\Rightarrow
\mathcal{N} \displaystyle  \left( 0, \frac{  \uptau  (1 -  \uptau ) }{ f_{u_t ( \uptau) }(0)^2  } \big( \boldsymbol{\Gamma}_{cxz} \boldsymbol{V}_{cxz}^{-1}  \boldsymbol{\Gamma}_{cxz}^{\prime} \big)^{-1} \right) 
\end{align}
which is a mixed Gaussian distribution due to the stochastic covariance matrix. 
\end{corollary} 
Analytic definitions of the covariance matrices $\boldsymbol{\Gamma}_{cxz}$ and $\boldsymbol{V}_{cxz}$ can be found in expressions (3.4) and (3.5) in \cite{lee2016predictive}. Lastly,  Lemma \ref{lemma1} presents the asymptotic behaviour of the self-normalized Wald statistic based on the IVX-QR estimator (see, Proposition 3.1 in \cite{lee2016predictive}) for both the cases of \textit{LUR} and \textit{MI} regressors. 

\medskip

\begin{lemma}
\label{lemma1}
(Self-normalized IVX-QR) Under Assumption \ref{assumption1} it holds that, 
\begin{align}
\frac{ \widehat{f_{ u_t ( \uptau )} }(0)^2   }{  \uptau ( 1- \uptau)  } \left( \widehat{\boldsymbol{\beta}}^{ivx-qr}_n ( \uptau ) - \boldsymbol{\beta} ( \uptau ) \right)^{\prime}  \big( \boldsymbol{X}^{\prime} \boldsymbol{P}_{ \tilde{ \boldsymbol{Z} }  } \boldsymbol{X} \big) \left( \widehat{\boldsymbol{\beta}}^{ivx-qr}_n ( \uptau ) - \boldsymbol{\beta} ( \uptau ) \right) \Rightarrow \chi^2_p 
\end{align}
where
\begin{align*}
\big( \boldsymbol{X}^{\prime} \boldsymbol{P}_{ \tilde{ \boldsymbol{Z} }  } \boldsymbol{X} \big) 
:= \left( \boldsymbol{X}^{\prime} \tilde{\boldsymbol{Z}} \right) \left( \tilde{\boldsymbol{Z}}^{\prime} \tilde{\boldsymbol{Z}} \right)^{-1} \left( \tilde{\boldsymbol{Z}} ^{\prime} \boldsymbol{X} \right) 
\equiv 
\left( \sum_{t=1}^n \boldsymbol{x}_{t-1} \tilde{\boldsymbol{z} }_{t-1}^{\prime} \right) \left( \sum_{t=1}^n \tilde{\boldsymbol{z}}_{t-1} \tilde{\boldsymbol{z} }^{\prime}_{t-1} \right)^{-1}  \left( \sum_{t=1}^n  \tilde{\boldsymbol{z} }_{t-1} \boldsymbol{x}_{t-1}^{\prime} \right)
\end{align*}
such that $\widehat{f_{ u_t ( \uptau )} }(0)^2$ is a consistent estimator of $f_{ u_t ( \uptau )}(0)^2$. 
\end{lemma}
Furthermore, the above result can be generalized when testing for a set of linear restrictions under the null hypothesis, $\mathcal{H}_0 : \boldsymbol{R} \boldsymbol{\beta} ( \uptau ) = \boldsymbol{q} ( \uptau )$ where $\boldsymbol{R}$ is a $r \times p$ known matrix and $\boldsymbol{q} ( \uptau )$ is a prespecified vector. The corresponding asymptotic distribution for the IVX-Wald statistic for the quantile predictive regression is given by the following expression
\begin{align*}
\frac{ \widehat{f_{ u_t ( \uptau )} }(0)^2   }{  \uptau ( 1- \uptau)  } \left( \boldsymbol{R} \widehat{\boldsymbol{\beta}}^{ivx-qr}_n ( \uptau ) - \boldsymbol{q} ( \uptau ) \right)^{\prime}  \bigg[ \boldsymbol{R} \big( \boldsymbol{X}^{\prime} \boldsymbol{P}_{ \tilde{ \boldsymbol{Z} }  } \boldsymbol{X} \big)^{-1} \boldsymbol{R}^{\prime} \bigg]^{-1} \left( \boldsymbol{R} \widehat{\boldsymbol{\beta}}^{ivx-qr}_n ( \uptau ) - \boldsymbol{q} ( \uptau ) \right) \Rightarrow \chi^2_r 
\end{align*}
where $\chi^2_r$ denotes the chi-square random variate with $r$ degrees of freedom such that $\mathbb{P} \left( \chi^2 \geq \chi^2_{ r ; \upalpha } \right) = \upalpha$, where $0 < \upalpha < 1$ denotes the fixed significance level. 

\newpage

\subsection{Testing Hypotheses}
\label{Section2.3}

In this section we present the testing problem of interest in this paper. More precisely, we consider two type of testing hypotheses, that is: \textit{(i)} testing for structural break for a fixed quantile level $\uptau \in (0,1)$ and \textit{(ii)} testing for a structural break multiple quantile levels. For both testing hypotheses we operate under the assumption of a single structural break at an unknown location. Notice, that similar formulations can be found in the study of \cite{qu2008testing}, however the focus of our study in the quantile predictive regression model with possibly nonstationary predictors in the context of return predictability literature. 

\paragraph{Testing Hypothesis A.} 

The first testing hypothesis of interest is concerned with testing for structural break in a pre-specified quantile with the null and alternative hypothesis given as below
\begin{align}
\label{hypothesisA}
\mathcal{H}^{(A)}_0&: 
\boldsymbol{\beta}_t ( \uptau ) = \boldsymbol{\beta}_0 ( \uptau ) \ \ \ \text{for all} \ \ 1 \leq t \leq n, \ \textcolor{blue}{\text{for a fixed} \ \uptau \in (0,1)}, 
\nonumber
\\
\\
\mathcal{H}^{(A)}_1&:
\boldsymbol{\beta}_t ( \uptau ) =
\begin{cases}
\boldsymbol{\beta}_1 ( \uptau )  & \text{where} \ 1 \leq t \leq \kappa
\nonumber
\\
\boldsymbol{\beta}_2 ( \uptau )  & \text{where} \ \kappa + 1 \leq t \leq n 
\end{cases}
\end{align}
\textcolor{blue}{for a fixed $\uptau \in (0,1)$}, where $\kappa = \floor{ \lambda n }$ the unknown break-point with $\lambda \in (0,1)$.

\paragraph{Testing Hypothesis B.}

The second testing hypothesis of interest is concerned with testing for structural break across multiple quantiles, that is, quantiles contained in a set $\mathcal{T}_{\iota}$, with the null and alternative hypothesis given as below
\begin{align}
\label{hypothesisB}
\mathcal{H}^{(B)}_0&: 
\boldsymbol{\beta}_t ( \uptau ) = \boldsymbol{\beta}_0 ( \uptau ) \ \ \ \text{for all} \ \ 1 \leq t \leq n, \ \textcolor{blue}{\text{and for all} \ \ \uptau \in \mathcal{T}_{\iota}}, 
\nonumber
\\
\\
\mathcal{H}^{(B)}_1&:
\boldsymbol{\beta}_t ( \uptau ) =
\begin{cases}
\boldsymbol{\beta}_1 ( \uptau )  & \text{where} \ 1 \leq t \leq \kappa
\nonumber
\\
\boldsymbol{\beta}_2 ( \uptau )  & \text{where} \ \kappa + 1 \leq t \leq n 
\end{cases}
\end{align}
\textcolor{blue}{for some $\uptau \in \mathcal{T}_{\iota}$}, where $\kappa = \floor{ \lambda n }$ the unknown break-point with $\lambda \in (0,1)$.  

where $\boldsymbol{\beta}_0 ( \uptau )$ is the value of the true population parameter under the null hypothesis of no parameter instability. 

\medskip

\begin{remark}
\label{remark2}
Notice that the statistical problem given by \textit{Testing Hypothesis A} allow us to focus on a particular quantile of interest, e.g., any fixed quantile level $\uptau \equiv \uptau_0 \in (0,1)$. On the other hand, the inference problem given by \textit{Testing Hypothesis B} permits testing for structural break in the coefficients of the quantile predictive regression model by investigating the presence of breaks in the conditional distribution, that is, at any possible quantile level within the compact set $\mathcal{T}_{\iota } := [ \iota , 1 - \iota  ]$ where $0 < \iota  < 1/2$.  
\end{remark}

\newpage

Both \textit{Testing Hypothesis A} and \textit{B} summarize the modelling environment under the null as well as under the alternative hypothesis. More precisely, under the null hypothesis the parameter vector is taken to be constant throughout the sample such that $\boldsymbol{\beta}_t(\uptau) \equiv \boldsymbol{\beta}_0(\uptau)$ for $t = 1,...,n$ where  $\boldsymbol{\beta}_0(\uptau)$ is the unknown quantile dependent regression parameter. Therefore, in practise we are interested in testing the null hypothesis that $\boldsymbol{\beta}_t(\uptau)$ remains constant, that is,  $\boldsymbol{\beta}_t(\uptau) = \boldsymbol{\beta}_0(\uptau)$ for all $t$ against the alternative that the quantile dependent parameter vector $\boldsymbol{\beta}_t(\uptau)$ has a single structural break at an unknown location within the full sample, resulting to two regimes\footnote{Notice that the two regimes we refer to here, are not equivalent to testing methodologies proposed in studies such as \cite{gonzalo2012regime, gonzalo2017inferring} and \cite{galvao2014testing} in which emphasis is given to testing the null hypothesis of linearity based on the presence of no threshold effect.}. Under the alternative hypothesis:
\begin{align}
\mathsf{Q}_{y_t} \left( \uptau | \mathcal{F}_{t-1} \right) = \boldsymbol{\beta}_1(\uptau) \boldsymbol{x}_{t-1} \mathds{1} \big\{ t \leq \kappa \big\} + \boldsymbol{\beta}_2(\uptau) \boldsymbol{x}_{t-1} \mathds{1} \big\{ t > \kappa \big\} + u_t
\end{align} 
where $\mathcal{F}_t$ denotes the $\sigma-$field generated by $\left\{ \boldsymbol{x}_{t-1}, \boldsymbol{x}_{t-2},..., \right\}$. Therefore, it is convenient to write the hypotheses with a different formulation. Denote with $\boldsymbol{\upbeta}_{(1)}(\uptau) = \boldsymbol{\beta}_1(\uptau)$ and $\boldsymbol{\upbeta}_{(2)}(\uptau) = \boldsymbol{\beta}_2(\uptau) - \boldsymbol{\beta}_1(\uptau)$. Furthermore, to construct the model so that it can capture the magnitude of the structural break we denote with $\boldsymbol{\mathcal{X}}_{t-1} = \big( \boldsymbol{x}_{t-1}^{\prime}, \boldsymbol{x}_{t-1}^{\prime} \mathds{1} \left\{ t > \kappa \right\} \big)^{\prime}$ and $\boldsymbol{\vartheta}(\uptau) = \big( \boldsymbol{\upbeta}_{(1)}(\uptau)^{\prime}, \boldsymbol{\upbeta}_{(2)}(\uptau)^{\prime} \big)^{\prime}$. Thus, we express the null hypothesis as following
\begin{align}
\label{form.null}
\mathcal{H}_0^{(A)}: \mathsf{Q}_{y_t} \left( \uptau | \mathcal{F}_{t-1} \right) =
\boldsymbol{\mathcal{X}}_{t-1}^{\prime} \boldsymbol{\vartheta}(\uptau), \ \text{with} \ \boldsymbol{\upbeta}_{(2)}(\uptau) = \boldsymbol{0} \ \text{for some fixed} \ \uptau_0 \in \mathcal{T}_{\iota} 
\end{align} 
Then, the alternative hypothesis can be formulated as below
\begin{align}
\label{form.alter}
\mathcal{H}_1^{(A)}: \mathsf{Q}_{y_t} \left( \uptau | \mathcal{F}_{t-1} \right) =
\boldsymbol{\mathcal{X}}_{t-1}^{\prime} \boldsymbol{\vartheta}(\uptau), \ \text{with} \ \boldsymbol{\upbeta}_{(2)}(\uptau) \neq \boldsymbol{0} \ \text{for some fixed} \ \uptau \in \mathcal{T}_{\iota} 
\end{align}
Furthermore, since we consider the nonstationary quantile predictive regression model without an intercept, following the formulations given by expressions \eqref{form.null} and \eqref{form.alter} for the null and alternative hypotheses respectively, then we define the quantile dependent estimator as the optimization problem below 
\begin{align}
\widehat{\boldsymbol{ \vartheta }}_n \big( \lambda , \uptau \big) := \underset{ \boldsymbol{ \vartheta } \in \mathbb{R}^{2p} }{ \mathsf{arg \ min} } \ \sum_{t=1}^n \uprho_{\tau} \big( y_{t} - \boldsymbol{\mathcal{X}}_{t-1}^{\prime} \boldsymbol{b}  \big),
\end{align}
Therefore, with the above formulation of the estimator $\widehat{\boldsymbol{ \vartheta }}_n ( \uptau, \lambda )$ is the quantile dependent regression estimator when we employ $\boldsymbol{\mathcal{X}}_{t-1}$ to be the model predictor variables. Specifically, when the $\mathcal{H}_0^{(A)}$ is true, under suitable regularity conditions, $\widehat{\boldsymbol{ \vartheta }}_2 ( \lambda, \uptau )$ converges in probability to $\boldsymbol{0}$ for each $( \lambda, \uptau ) \in \Lambda_{\eta} \times \mathcal{T}_{\iota}$. On the other hand, when $\mathcal{H}_1^{(A)}$ is true, $\widehat{\boldsymbol{ \vartheta }}_2 ( \lambda ; \uptau_0 )$ converges in probability to $\boldsymbol{\upbeta}_{(2)}(\uptau_0) = \big( \boldsymbol{\beta}_2(\uptau_0) - \boldsymbol{\beta}_1(\uptau_0) \big) \neq \boldsymbol{0}$. In summary, since the quantile level $\uptau_0$ especially for \textit{Testing Hypothesis B} is unknown a prior, then it is reasonable to reject $\mathcal{H}_0$ when the magnitude of $\widehat{\boldsymbol{ \vartheta }}_2 ( \lambda, \uptau )$ is suitable large for some $( \lambda, \uptau ) \in \Lambda_{\eta} \times \mathcal{T}_{\iota}$. Thus, an example of a suitable test statistic to test whether $\mathcal{H}_0$ against the alternative hypothesis $\mathcal{H}_1$ is to employ the supremum of the Wald process.

\newpage

In general, a Wald type statistic has the following form
\begin{align}
\mathcal{SW}_n \big( \uptau, \lambda \big) := \underset{ ( \lambda, \uptau ) \in \Lambda_{\eta} \times \mathcal{T}_{\iota} }{ \mathsf{sup} } n \hat{\boldsymbol{\upbeta}}_{(2)}(\uptau_0) \bigg[   \boldsymbol{V}_n ( \lambda; \uptau_0 ) \bigg] \hat{\boldsymbol{\upbeta}}_{(2)}(\uptau_0)
\end{align}
where $\boldsymbol{V}_n ( \lambda; \uptau_0 )$ is the asymptotic covariance matrix of the stochastic process $\sqrt{n} \hat{\boldsymbol{\upbeta}}_{(2)}(\uptau_0)$, under the null hypothesis. However, since the covariance matrix that corresponds to the population regression parameters is in practise unknown, is replaced by a suitable consistent estimate that holds under the null hypothesis of no structural break in the quantile predictive regression model. Obviously within the aforementioned structural break setting which is our main research focus in the paper, the break-point location is not identified under the null hypothesis as we explained in the introduction.  

Moreover, additionally to Remark \ref{remark2}, the Testing Hypotheses of interest \textit{A} and \textit{B} correspond to two different test functions such that  Testing Hypothesis \textit{A} requires to formulate test statistics by employing the supremum funcitonal while Testing Hypothesis \textit{B} requires to construct test statistics with the use of the double supremum functional. Intuitively, we are interested to examine both hypotheses since it might be the case that \textit{A} is not rejected, that is, there are no statistical evidence of the presence of a structural break for a given quantile level $\uptau_0 \in (0,1)$, while when employing Testing Hypothesis \textit{B}, it could be the case that the test statistic provides statistical evidence of rejecting the null hypothesis, implying that a structural break still exist at some other quantile level not the one which is kept fixed, within the set (0,1). 

\section{Testing for structural break for a fixed quantile level}
\label{Section3}

Next, we focus on the structural break testing procedures for the quantile regression model with regressors generated as near unit root processes based on two test statistics.

\subsection{Preliminary Setting}

We consider structural break tests for a fixed quantile level, say $\uptau_0 \in (0,1)$. Consider the subgradient\footnote{The required convexity arguments for obtaining estimators based on the conditional quantile function are based on the convexity lemma result presented by \cite{pollard1991asymptotics}. Furthermore, related results are presented by \cite{koenker1987estimation}.} $S_n \big( \lambda, \uptau_0 , \boldsymbol{b} \big)$, based on the subsample $1 \leq t \leq \kappa$
\begin{align}
\label{functional1}
S_n \big( \lambda, \uptau_0 , \boldsymbol{b} \big) = n^{- 1 / 2} \sum_{t=1}^{\floor{ \lambda n} } \boldsymbol{x}_{t-1} \psi_{ \uptau } \big( y_t - \boldsymbol{x}_{t-1}^{\prime} \boldsymbol{b} \big), 
\end{align}
where $\boldsymbol{b}$ corresponds to an estimator of the parameter vector $\boldsymbol{\beta} ( \uptau_0 )$ which encompasses both the OLS and IVX estimators under suitable parametrizations.

\newpage

The continuous function $\psi_{ \uptau } ( . )$ is defined as $\psi_{ \uptau } ( \mathsf{u} ) = \big[ \uptau_0 - \mathds{1} \left\{ \mathsf{u} \leq 0 \right\} \big]$ and $\kappa = \floor{ \lambda n }$ denotes the unknown break-point location implying a break fraction $\lambda \equiv \underset{ n \to \infty }{ \mathsf{lim} } \kappa/n$ such that $\lambda \in \Lambda_{\eta}:= [ \eta, 1 - \eta]$ is a compact set. Thus, under the null hypothesis of no structural break with  stationary and ergodic regressors, the quantity  $\psi_{ \uptau } \big( y_t - \boldsymbol{x}_{t-1}^{\prime} \boldsymbol{\beta} ( \uptau_0 ) \big)$ is a pivotal statistic. In particular, since $\boldsymbol{x}_{t-1}^{\prime} \boldsymbol{\beta} ( \uptau_0 )$ is equal to the conditional $\uptau-$quantile of $y_t$ given $\boldsymbol{x}_{t-1}$, then the random variables $\mathds{1} \left\{ y_1 \leq \boldsymbol{x}_{t-1}^{\prime} \boldsymbol{\beta} ( \uptau_0 ) \right\},..., \mathds{1} \left\{ y_n \leq \boldsymbol{x}_{t-1}^{\prime} \boldsymbol{\beta} ( \uptau_0 ) \right\}$ are independent Bernoulli trials with success probability $\uptau$ (see, \cite{galvao2014testing}), which implies a sequence of  random variables with mean zero and variance $\uptau_0 ( 1 - \uptau_0 )$. A similar result should hold in our setting of possibly nonstationary regressors. Furthermore, denote with $\boldsymbol{X} = \left( x_1^{\prime} ,..., x_n^{\prime} \right)^{\prime}$ and define the following auxiliary quantity
\begin{align}
\label{functional2}
\mathcal{J}_n \big( \lambda, \uptau_0 , \boldsymbol{\beta}_0 ( \uptau ) \big) 
:= 
\left( n^{-1} \boldsymbol{X}^{\prime} \boldsymbol{X} \right)^{- 1 / 2} S_n \big( \lambda, \uptau_0, \boldsymbol{\beta}_0 ( \uptau )  \big).  
\end{align}
Furthermore, replacing the unknown parameter vector $\boldsymbol{\beta}_0 ( \uptau_0 )$ with the quantile dependent regression estimator based on the full sample, under the null hypothesis of no structural break in the model, the quantity given by \eqref{functional2} can be formulated as below
\begin{align}
\label{random}
\hat{\mathcal{J}}_n \big( \lambda, \uptau_0, \widehat{ \boldsymbol{ \beta} }_n( \uptau_0 ) \big) = \big( \boldsymbol{X}^{\prime} \boldsymbol{X} \big)^{- 1 / 2} \sum_{t=1}^{\floor{ \lambda n} } \boldsymbol{x}_{t-1} \psi_{ \uptau } \left( y_t -  \boldsymbol{x}_{t-1}^{\prime} \widehat{ \boldsymbol{ \beta} }_n( \uptau_0 ) \right).
\end{align}
for some $0 < \lambda < 1$ and $\uptau_0 \in (0,1)$. 

\medskip
The use of fluctuation type statistics provide a way for statistical inference regarding the presence of structural breaks in model coefficients (\cite{leisch2000monitoring}). Intuitively for these class of tests we consider the asymptotic behaviour of the corresponding empirical processes to decide whether to accept or reject the null. In particular, the fluctuation type test converges to a nondegenerate limiting distribution under the null hypothesis, since the random quantity given by expression \eqref{random} is essentially governed by the invariance principle under the null. In practise, when the quantile dependent parameters exhibit no structural break in the sample, then $\widehat{ \boldsymbol{ \beta} }_n( \uptau_0 )$ is a consistent estimator  and as a result, $\hat{\mathcal{J}}_n \big( \lambda, \uptau_0, \widehat{ \boldsymbol{ \beta} }_n( \uptau_0 ) \big)$ has the same stochastic order as its population counterpart. On the other hand, when the null hypothesis is false, the underline stochastic process exhibit excessive fluctuations. Specifically, under the alternative hypothesis, model parameters have a break at some unknown location in the sample, which implies that $\widehat{ \boldsymbol{\beta} }_n( \uptau_0 )$  will differ significantly from the true value for some sub-sample and the estimated residuals will have high fluctuations (beyond the usual increments of a Wiener process) resulting to falsely rejecting the null due to a large value of the statistic \cite[p.~172]{qu2008testing}. 

\newpage

The focus of the proposed econometric environment in this paper is the structural break detection in the model parameters of quantile predictive regression with possibly nonstationary regressors, under the assumption that these are generated as near unit root processes. Intuitively, the two persistence classes (mildly integrated and near unit root) we consider encompasses moderate deviations from the unit boundary similar to the case of near integrated (see \cite{Phillips1988Regression}). Therefore, both the value (and sign) of the coefficient of persistence as well as its exponent rate\footnote{Practically, these are nuisance parameters however via Monte Carlo simulations we can choose suitable values for $c_i$ and $\upgamma_x$ in order to simulate these experimental conditions and thus evaluate the finite-sample performance of the test statistics with high persistence or mildly integrated regressors.} (a tuning parameter), determine the asymptotic behaviour of functionals based on these near unit root process. As a result, the chosen estimator can affect the asymptotic theory of the proposed test statistics as well as the corresponding functionals which we examine separately below. 

\subsubsection{OLS based functionals}

Within the proposed econometric environment which corresponds to the modelling of nonstationary quantile time series models, regressors are assumed to follow a local unit root process. 
Thus, we expect that OLS based functionals will dependent on the nuisance coefficient of persistence\footnote{Notice that this is the standard inference problem in the predictability literature. Further details regarding the bias (nonstandard distortion) occurred in predictability tests (i.e., $t-$tests) in quantile predictive regression models can be found in the study of \cite{lee2016predictive}. In our study, we aim to compare both the OLS as well as the instrumental variable approach of \cite{PhillipsMagdal2009econometric}.}. Furthermore, due to the presence of both a model intercept and the set of nonstationary regressors, we also need to modify the functionals given by expressions \eqref{functional1}-\eqref{functional2} in order to account for the different convergence rates. 

Therefore, to obtain equivalent representations to the quantity  $S_n \big( \lambda, \uptau_0, \boldsymbol{b} \big)$, we consider the corresponding partial sum process of the functional $\boldsymbol{K}_{nx}  \big( \uptau_0, \boldsymbol{\theta}^{ols}_n( \uptau_0 ) \big)$ as given by Definition \ref{definition1}. We  obtain the limit result for these functionals based on the full sample and then focus on deriving invariance principles for the corresponding  partial sum processes for the two estimators under examination (see, Section \ref{Section3.2}). More precisely, the functionals given by Definition \ref{definition1} correspond to a quantile regression ordinary least squares estimator and are employed when the asymptotic behaviour of the OLS based test statistics is concerned (see also Lemma A1 in \cite{lee2016predictive}). 

\begin{definition}
\label{definition1}
\begin{align}
\boldsymbol{K}_{nx}  \big( \uptau_0, \boldsymbol{\theta}^{ols}_n( \uptau_0 ) \big)
&:= 
\boldsymbol{D}_n^{-1} \sum_{t=1}^n \boldsymbol{X}_{t-1} \psi_{\uptau} \big(  u_t ( \uptau_0 ) \big) 
\\
\boldsymbol{L}_{nx} \big( \uptau_0, \boldsymbol{\theta}^{ols}_n(\uptau_0) \big)  
&:= 
\boldsymbol{D}_n^{-1} \left[ \sum_{t=1}^n f_{ u_t ( \uptau ), t-1 } (0) \boldsymbol{X}_{t-1} \boldsymbol{X}_{t-1}^{\prime} \right] \boldsymbol{D}_n^{-1} 
\end{align}
for some $\uptau_0 \in (0,1)$ where $\psi_{\uptau} \big(  u_t ( \uptau_0 ) \big) = \big[ \uptau_0 - \mathds{1} \big\{ y_t - \boldsymbol{X}_{t-1} ^{\prime}\boldsymbol{\theta}^{ols}_n(\uptau_0)  \leq 0 \big\} \big]$. 
\end{definition}

\newpage

A key observation is that under the null hypothesis of no parameter instability these functional converge to a nondegenerate limit distribution. Corollary \ref{corollary3A} demonstrates the asymptotic distributions of the functionals given by Definition \ref{definition1}. 

\medskip

\begin{corollary}
\label{corollary3A}
Under the assumption that the pair $\left\{ y_t, \boldsymbol{x}_{t-1} \right\}_{t=1}^n$ is generated by the model \eqref{model1}-\eqref{model2} then for both \textit{LUR} and \textit{MI} regressors it holds that
\begin{itemize}

\item[\textit{(i)}] $\boldsymbol{K}_{nx} \big( \uptau_0,  \boldsymbol{\theta}^{ols}_n(\uptau_0) \big)
\Rightarrow 
\boldsymbol{K}_{x} \big( \uptau_0,  \boldsymbol{\theta}_0(\uptau_0) \big)$, for some $\uptau_0 \in (0,1)$ as $n \to \infty$,

\item[\textit{(ii)}] $\boldsymbol{L}_{nx} \big( \uptau_0, \boldsymbol{\theta}^{ols}_n(\uptau_0) \big) 
\Rightarrow 
\boldsymbol{L}_{x} \big( \uptau_0, \boldsymbol{\theta}_0(\uptau_0) \big)$, for some $\uptau_0 \in (0,1)$ as $n \to \infty$,
\end{itemize}
where 
\begin{small}
\begin{align}
\boldsymbol{K}_{x} \big( \uptau_0, \boldsymbol{\theta}_0(\uptau_0) \big)
&\equiv 
\begin{cases}
\begin{bmatrix}
  B_{\psi_{\uptau}}(1)_{ \textcolor{red}{ ( 1 \times n) }  }  
    \\
  \displaystyle \int_0^1 \boldsymbol{J}_c(r) dB_{ \psi_{ \uptau} } 
\end{bmatrix}_{ \textcolor{red}{ ( p + 1 ) \times n }  }  & \ \ \ \ \ \ \ \ \ \ \ \ \textit{LUR},   
\\
\\
\mathcal{N} \displaystyle  \left( \boldsymbol{0}, \uptau_0 (1 - \uptau_0) \times   
     \begin{bmatrix}
        1  &  \boldsymbol{0}^{\prime} \\
        \boldsymbol{0}  & \boldsymbol{V}_{xx}
     \end{bmatrix}_{ \textcolor{red}{ ( p + 1 ) \times ( p + 1 ) }  }  
     \right)_{ \textcolor{red}{ ( p + 1 ) \times n }  }         
         & \ \ \ \ \ \ \ \ \ \ \ \ \textit{MI}.  
\end{cases}
\\
\nonumber
\\
\boldsymbol{L}_{x} \big( \uptau_0, \boldsymbol{\theta}_0(\uptau_0) \big)
&\equiv 
\begin{cases}
f_{ u_t (\uptau)} (0) \times 
\begin{bmatrix}
  1  &  \displaystyle \int_0^1 \boldsymbol{J}_c(r)^{\prime} \\
  \displaystyle \int_0^1 \boldsymbol{J}_c(r)  &   \displaystyle \int_0^1 \boldsymbol{J}_c(r) \boldsymbol{J}_c(r)^{\prime}
\end{bmatrix}_{ \textcolor{red}{ ( p + 1 ) \times ( p + 1 ) }  }       & \ \ \ \textit{LUR},   
\\
\\
f_{ u_t ( \uptau )} (0) \times 
\begin{bmatrix}
1  &  \boldsymbol{0}^{\prime} \\
\boldsymbol{0}  & \boldsymbol{V}_{xx}
\end{bmatrix}_{ \textcolor{red}{ ( p + 1 ) \times ( p + 1 ) }  }     & \ \ \ \textit{MI}.  
\end{cases}
\end{align}
\end{small}
where the stochastic matrix $\boldsymbol{V}_{xx}$ is defined by \cite{PhillipsMagdal2009econometric} as
\begin{align*}
\boldsymbol{V}_{xx} := \int_0^{\infty} e^{r \boldsymbol{C}_p } \boldsymbol{\Omega}_{xx}  e^{r\boldsymbol{C}_p} dr, \ \text{where} \ \boldsymbol{\Omega}_{xx} := \sum_{m=-\infty}^{\infty} \mathbb{E} \left( \boldsymbol{v}_{t} \boldsymbol{v}_{t-m}^{\prime} \right) = \boldsymbol{\varphi}_{o} (1) \boldsymbol{\Sigma} \boldsymbol{\varphi}_{o} (1)^{\prime}. 
\end{align*}
\end{corollary}

\medskip

\begin{remark}
Notice that an important aspect for robust inference in quantile regressions\footnote{In some studies presented in the literature the use of the check function is defined to be the difference of the indicator function from the quantile level, as in \cite{zhou1998statistical}; however both expressions are equivalent due to the monotonicity property of the check function.} is the consistent estimation of the sparsity coefficient (see, discussion presented in \cite{koenker1999goodness}) and also conditions proposed by \cite{koltchinskii1997m}, especially in finite samples. In our setting the self-normalized property of Wald type tests ensures that the sparsity coefficient does not affect the estimation accuracy.  
\end{remark}

\newpage

\subsubsection{IVX based functionals}

In this Section, we derive the asymptotic distribution of the IVX based functionals which are useful to obtain the asymptotic behaviour of the proposed structural break tests under the assumption of nonstationary regressors in the model. We employ the embedded normalization version of the instruments such that $\tilde{ \boldsymbol{Z} }_{t-1,n} := \tilde{ \boldsymbol{D} }_n^{-1} \tilde{ \boldsymbol{z} }_{t-1}$. 

\medskip

\begin{definition}
\label{definition2}
\begin{align}
\boldsymbol{K}_{nz} \big( \uptau_0, \boldsymbol{\beta}_n^{ivx} ( \uptau_0 ) \big) 
&:= 
\sum_{t=1}^n \tilde{\boldsymbol{Z}}_{t-1,n} \psi_{\uptau} \big(  u_t ( \uptau_0 ) \big) 
\\
\boldsymbol{L}_{nz} \big( \uptau_0, \boldsymbol{\beta}_n^{ivz} ( \uptau_0 ) \big) 
&:= 
\left[ \sum_{t=1}^n f_{ u_t ( \uptau ), t-1 } (0) \tilde{\boldsymbol{Z}}_{t-1,n} \tilde{\boldsymbol{Z}}_{t-1,n}^{\prime} \right] 
\\
\boldsymbol{M}_{nz} \big( \uptau_0, \boldsymbol{\beta}^{ivx}_n ( \uptau_0 ) \big) 
&:= 
\left[ \sum_{t=1}^n f_{ u_t \left( \uptau  \right), t-1 } (0) \tilde{\boldsymbol{Z}}_{t-1,n} \boldsymbol{X}^{\prime}_{t-1,n} \right] 
\end{align}
for some $\uptau_0 \in (0,1)$ where $\psi_{\uptau} \big(  u_t ( \uptau_0 ) \big) = \big[ \uptau_0 - \mathds{1} \big\{ y_t - \boldsymbol{\beta}^{ivx}_n(\uptau_0)^{\prime} \boldsymbol{x}_{t-1} \leq 0 \big\} \big]$.   
\end{definition}

\medskip

\begin{corollary}
\label{corollary4A}
Under the assumption that the pair $\left\{ y_t, \boldsymbol{x}_{t-1} \right\}_{t=1}^n$ is generated by the model \eqref{model1}-\eqref{model2} then for both \textit{LUR} and \textit{MI} regressors it holds that
\begin{itemize}

\item[\textit{(i)}] $\boldsymbol{K}_{nz} \big( \uptau_0, \boldsymbol{\beta}_n^{ivx} ( \uptau_0 ) \big) \Rightarrow \boldsymbol{K}_{z} \big( \uptau_0, \boldsymbol{\beta}_0 ( \uptau_0 ) \big) \equiv 
\mathcal{N} \big( \boldsymbol{0}, \uptau_0 ( 1 - \uptau_0) \boldsymbol{V}_{cxz} \big)$,

\item[\textit{(ii)}] $\boldsymbol{L}_{nz} \big( \uptau_0, \boldsymbol{\beta}_n^{ivz} (\uptau_0) \big) \Rightarrow \boldsymbol{L}_{z} \big( \uptau_0, \boldsymbol{\beta}_0 ( \uptau_0 ) \big) \equiv 
f_{ u_t (\uptau)} (0) \times \boldsymbol{V}_{cxz}$, 

\item[\textit{(iii)}] $\boldsymbol{M}_{nz} \big( \uptau_0, \boldsymbol{\beta}_n^{ivx} (\uptau_0) \big) \Rightarrow \boldsymbol{M}_{z} \big( \uptau_0, \boldsymbol{\beta}_0 ( \uptau_0 ) \big) \equiv f_{ u_t \left( \uptau  \right) } (0) \times \boldsymbol{\Gamma}_{cxz}$, 
\end{itemize}
where the definition of the asymptotic matrix $\boldsymbol{V}_{cxz}$ depends on the stochastic dominance of the two exponent rates (see, \cite{PhillipsMagdal2009econometric} and \cite{lee2016predictive}) such as 
\begin{align}
\boldsymbol{V}_{cxz} \equiv 
\begin{cases}
\boldsymbol{V}_{zz} = 
\displaystyle \int_{0}^{\infty} e^{r \boldsymbol{C}_z} \boldsymbol{\Omega}_{xx} e^{r \boldsymbol{C}_z} dr,  & \  \text{when} \ 0 < \upgamma_z < \upgamma_x  < 1,
\\
\\
\boldsymbol{V}_{xx} = 
\displaystyle \int_{0}^{\infty} e^{r \boldsymbol{C}_p} \boldsymbol{\Omega}_{xx} e^{r \boldsymbol{C}_p} dr,  & \ \text{when} \ 0 < \upgamma_x < \upgamma_z < 1.
\end{cases}
\end{align}
\end{corollary}
Moreover, the definition of the moment matrix $\boldsymbol{\Gamma}_{cxz}$ is presented by \cite{lee2016predictive} via expression (3.4) which is the corresponding asymptotic limit given by expression (20) in \cite{PhillipsMagdal2009econometric} as given below 
\begin{align}
\boldsymbol{\Gamma}_{cxz} 
:=
\begin{cases}
\displaystyle  - \boldsymbol{C}_z^{-1} \left( \boldsymbol{\Omega}_{xx} + \int_0^1 \boldsymbol{J}_c (r) d\boldsymbol{J}_c^{\prime} \right), & \ \text{when} \ \upgamma_x = 1,
\\
- \boldsymbol{C}_z^{-1} \bigg( \boldsymbol{\Omega}_{xx} +    \boldsymbol{C}_p \boldsymbol{V}_{xx} \bigg), &  \ \text{when} \ 0 < \upgamma_z < \upgamma_x < 1,
\\
\boldsymbol{V}_{xx}, & \ \text{when} \ 0 < \upgamma_x < \upgamma_z  < 1.
\end{cases}
\end{align}

\newpage

The proofs of Corollary \ref{corollary3A} and \ref{corollary4A} can be found in the Appendix of the paper. Note that the stochastic convergence of these functional holds for large sample size, $n \to \infty$, and the existence of well-defined moment matrices with negligible higher-order terms. Therefore, to facilitate the development of the asymptotic theory we define the following empirical process for some parameter vector $\boldsymbol{b} \in \mathbb{R}^p$ such that
\begin{align*}
\boldsymbol{G}_n \left( \uptau, \boldsymbol{b} \right) := n^{ - ( 1 + \upgamma_x ) / 2 } \sum_{t=1}^n \boldsymbol{z}_{t-1} \times \left\{ \psi_{\uptau} \big( u_t \left( \uptau \right) - \boldsymbol{x}_{t-1}^{\prime} \boldsymbol{b} \big) - \mathbb{E}_{ \mathcal{F}_{t - 1} } \left[ \psi_{\uptau} \big( u_t \left( \uptau \right) - \boldsymbol{x}_{t-1}^{\prime} \boldsymbol{b}  \big) \right] \right\} 
\end{align*}
where $\uptau \in (0,1)$ and $0 < \upgamma_x < 1$.  In particular, the empirical process $\boldsymbol{G}_n \left( \uptau, \boldsymbol{b} \right)$ is consider stochastically $\varrho-$equicontinuous over $\mathcal{T}_{\iota} \times B$, such that for any $\epsilon > 0$, 
\begin{align}
\underset{ \delta \to 0 }{ \mathsf{lim} } \ \underset{ n \to \infty }{ \mathsf{lim \ sup} } \  \mathbb{P} \left( \underset{ [ \delta ] }{ \mathsf{sup} } \ \big| G_n \big( \uptau_1 , \boldsymbol{b}_1 \big) - G_n \big( \uptau_2, \boldsymbol{b}_2 \big) \big| > \epsilon \right) = 0,
\end{align}
where $[ \mathcal{\delta} ] := \big\{ ( \uptau_1, \boldsymbol{b}_1 ), ( \uptau_2, \boldsymbol{b}_2 ) \in \left( \mathcal{T} \times B \right)^2 : \ \varrho \big( ( \uptau_1, \boldsymbol{b}_1 ), ( \uptau_2, \boldsymbol{b}_2 ) \big) < \delta \big\}$.

\medskip

\begin{remark}
The above expression is often employed to derive asymptotics for quantile regression models (with stationary regressors). Specifically, one can consider the validity of the stochastic equicontinuity proof of \cite{bickel1975one} under nonstationarity. Practically, since the regressors employed when estimating the inverse of the quantile function $\psi_{ \uptau } ( . )$, that is, $\tilde{\boldsymbol{z}}_{t-1}$ is mildly integrated, inducing a nearly stationary process, then the conditions given by \cite{bickel1975one} are valid and the proof follows with modifications to accommodate the nonstationary quantile predictive regression (\cite{lee2016predictive}).    
\end{remark}
An additional condition for convergence in probability for the empirical process is imposed by Lemma \ref{lemma3}, which can be employed to derive the convergence rate of the IVX estimator for the nonstationary quantile predictive regression model.  

\medskip

\begin{lemma}
\label{lemma3}
For a generic constant $\mathcal{C}_1 > 0$
\begin{align}
\mathsf{sup} \big\{ \big\| \boldsymbol{G}_n ( \uptau, \boldsymbol{b} ) - \boldsymbol{G}_n (  \uptau, \boldsymbol{0} ) \big\| : \norm{ \boldsymbol{b} } \leq n^{(1+ \delta) / 2} \mathcal{C}_1 \big\} = o_{ \mathbb{P} } (1). 
\end{align}
where $\boldsymbol{b}$ is some estimator of the model parameter vector. 
\end{lemma}  
More precisely, Lemma \ref{lemma3} provides a simplified way to derive the convergence limit for the IVX-QR estimator (see, also \cite{lee2016predictive}) that ensures consistent estimation of the model parameters for the quantile predictive regression model. A related study to our setting with detailed derivations for nonstandard inference problems, (Wald type statistics), for nonstationary quantile regressions is presented in the study of \cite{goh2009nonstandard}. Overall, the asymptotic theory of this paper aims to combine unit root asymptotics with empirical process methods. Specifically, we employ a two-parameter empirical process that converges weakly to a two-parameter Brownian motion. Therefore, our asymptotic distributions involve stochastic integrals with respect to this two-parameter process.

\newpage

\subsection{Invariance principles for partial sum processes}
\label{Section3.2}

To obtain the asymptotic distributions of the test statistics, we consider the asymptotic behaviour of the partial sum processes of the functionals defined in the previous section. We focus in the case of nonstationary regressors which are either high persistent or mildly integrated (see, Section \ref{Section2} for definitions and \cite{kostakis2015Robust}). Moreover, since we derive and compare the limit distributions of structural break tests based on the chose estimation methodology, we derive invariance principles that correspond to each of these two estimators. Therefore, we define with
\begin{align}
S_{nx}^{ols} \big( \lambda, \uptau_0 , \boldsymbol{\theta}^{ols}_n ( \uptau_0 ) \big) 
:=  
\boldsymbol{D}_n^{-1} \sum_{t=1}^{\floor{ \lambda n} } \boldsymbol{X}_{t-1} \psi_{ \uptau } \big( u_t ( \uptau_0 ) \big), \ \ \text{for some} \ \ 0 < \lambda < 1,
\end{align}
where $u_t ( \uptau_0 ) = \big( y_t -  \boldsymbol{X}_{t-1}^{\prime} \boldsymbol{\theta}^{ols}_n( \uptau_0 ) \big)$ for $\uptau_0 \in (0,1)$, which can be determined uniquely, making the mapping $\psi_{ \uptau } ( \mathsf{u} ) \mapsto \uprho_{\uptau}^{-1} ( \mathsf{u} )$ one-to-one and well-defined. Moreover, we denote with $\boldsymbol{X}_{t-1} = \big( \boldsymbol{1}, \boldsymbol{x}^{\prime}_{t-1} \big)^{\prime}$ the regressors and $\boldsymbol{\theta} (\uptau_0)= \big( \alpha(\uptau_0), \boldsymbol{\beta}^{\prime}(\uptau_0)   \big)^{\prime}$ the parameters. 

Recall that for mildly integrated regressors it holds that (see, Corollary \ref{corollary3A}) 
\begin{align}
\boldsymbol{K}_{nx}  \big( \uptau_0, \boldsymbol{\theta}^{ols}_n(\uptau_0) \big)
&:= 
\boldsymbol{D}_n^{-1} \sum_{t=1}^n \boldsymbol{X}_{t-1} \psi_{\uptau} \big(  u_t ( \uptau_0 ) \big) 
\Rightarrow  \mathcal{N} \displaystyle  \left( \boldsymbol{0}, \uptau_0 (1 - \uptau_0)  
  \begin{bmatrix}
  1  &  \boldsymbol{0}^{\prime} \\
  \boldsymbol{0}  & \boldsymbol{V}_{xx}
  \end{bmatrix} 
\right)     
\end{align}
Similarly, we can show that $S_{nx}^{ols} \big( \lambda, \uptau , \boldsymbol{\theta}^{ols}_n ( \uptau_0 ) \big)  \Rightarrow S_{x} \big( \lambda, \uptau_0 , \boldsymbol{\theta}_0( \uptau_0 ) \big)$ as $n \to \infty$, where
\begin{align}
S_{x} \big( \lambda, \uptau_0 , \boldsymbol{\theta_0} ( \uptau_0 ) \big) 
\equiv
\mathcal{N} \displaystyle  \left( \boldsymbol{0}, \uptau_0 (1 - \uptau_0) 
 \lambda  
  \begin{bmatrix}
  1  &  \boldsymbol{0}^{\prime} \\
  \boldsymbol{0}  & \boldsymbol{V}_{xx}
  \end{bmatrix} 
\right)     
\end{align}
\color{black}
for some $0 < \lambda < 1$. Then, for the corresponding IVX based functional it holds that 
\begin{align}
S_{nz}^{ivx} \big( \lambda, \uptau_0 , \boldsymbol{\beta}^{ivx}_n ( \uptau_0 ) \big) 
:=  
\sum_{t=1}^{\floor{ \lambda n} } \tilde{ \boldsymbol{Z} }_{t-1,n} \psi_{ \uptau } \big( u_t ( \uptau_0 ) \big)
\Rightarrow \mathcal{N} \big(  \boldsymbol{0}, \uptau_0 ( 1 - \uptau_0) \lambda \boldsymbol{V}_{cxz} \big)
\end{align}
\color{black}
where $\tilde{ \boldsymbol{Z} }_{t-1,n} := \tilde{ \boldsymbol{D} }_n^{-1} \tilde{ \boldsymbol{z} }_{t-1}$ since we employ the corresponding dequantiled model. 

\begin{definition}
\label{definition3}
\begin{align}
\hat{\mathcal{J}}^{ols}_{nx} \big( \lambda, \uptau_0, \widehat{ \boldsymbol{\theta} }^{ols}_n( \uptau_0 ) \big) 
&:= 
\big( \boldsymbol{X}^{\prime} \boldsymbol{X} \big)^{- 1 / 2} \sum_{t=1}^{\floor{ \lambda n} } \boldsymbol{X}_{t-1} \psi_{ \uptau } \left( y_t -  \boldsymbol{X}_{t-1}^{\prime}\widehat{ \boldsymbol{\theta} }^{ols}_n( \uptau_0 ) \right),
\\
\hat{\mathcal{J}}^{ivx}_{nx} \big( \lambda, \uptau_0, \widehat{ \boldsymbol{ \beta} }^{ivx}_n( \uptau_0 ) \big) 
&:= 
\big( \boldsymbol{X}^{\prime} \tilde{\boldsymbol{Z}} \big)^{- 1 / 2} \sum_{t=1}^{\floor{ \lambda n} } \tilde{\boldsymbol{Z}}_{t-1,n} \psi_{ \uptau } \left( y_t - \boldsymbol{X}_{t-1,n} ^{\prime} \widehat{ \boldsymbol{ \beta} }^{ivx}_n( \uptau_0 )\right),
\\
\hat{\mathcal{J}}^{ivz}_{nx} \big( \lambda, \uptau_0, \widehat{ \boldsymbol{ \beta} }^{ivz}_n( \uptau_0 ) \big)
&:= 
\big( \tilde{\boldsymbol{Z}}^{\prime} \tilde{\boldsymbol{Z}} \big)^{- 1 / 2} \sum_{t=1}^{\floor{ \lambda n} } \tilde{\boldsymbol{Z}}_{t-1,n} \psi_{ \uptau } \left( y_t -  \tilde{\boldsymbol{Z}}_{t-1,n}^{\prime} \widehat{ \boldsymbol{ \beta} }^{ivz}_n( \uptau_0 ) \right).
\end{align}
for some $0 < \lambda < 1$ and $\uptau_0 \in (0,1)$. 
\end{definition}

\newpage

Consider the functionals given by Definition \ref{definition3}, then when we employ the OLS estimator for a model with mildly integrated regressors, $\upgamma_x \in (0,1)$, the following limit result holds
\begin{align}
\hat{\mathcal{J}}^{ols}_{nx} \big( \lambda, \uptau_0, \widehat{ \boldsymbol{\theta} }^{ols}_n( \uptau_0 ) \big) 
\nonumber
&= 
\big( \boldsymbol{X}^{\prime} \boldsymbol{X} \big)^{- 1 / 2} \sum_{t=1}^{\floor{ \lambda n} } \boldsymbol{X}_{t-1} \psi_{ \uptau } \left( y_t - \boldsymbol{X}_{t-1}^{\prime} \widehat{ \boldsymbol{\theta} }^{ols}_n( \uptau_0 ) \right)
\\
&\equiv
\left( \boldsymbol{D}_n^{-1} \left[ \sum_{t=1}^n \boldsymbol{X}^{\prime}_{t-1} \boldsymbol{X}_{t-1} \right] \boldsymbol{D}_n^{-1} \right)^{- 1 / 2} \left\{ \boldsymbol{D}_n^{-1} \sum_{t=1}^{ \floor{\lambda n} } \boldsymbol{X}_{t-1} \psi_{\uptau} \big( u_t (\uptau_0) \big) \right\}
\nonumber
\\
&\Rightarrow
\left\{ 
\begin{bmatrix}
  1  &  \boldsymbol{0}^{\prime} \\
  \boldsymbol{0}  & \boldsymbol{V}_{xx}
  \end{bmatrix}  \right\}^{-1 / 2}
\times
\mathcal{N} \displaystyle  \left( \boldsymbol{0}, \uptau_0 (1 - \uptau_0)   \lambda 
\begin{bmatrix}
  1  &  \boldsymbol{0}^{\prime} \\
  \boldsymbol{0}  & \boldsymbol{V}_{xx}
  \end{bmatrix} 
\right) 
\nonumber
\\
&=
\sqrt{\uptau_0 ( 1 - \uptau_0 ) } \times  \mathcal{N} \big( \boldsymbol{0}, \lambda \boldsymbol{I}_p \big).
\end{align} 
since the term $\frac{1}{ n^{ 1 + \upgamma_x } } \sum_{t=1}^{ \floor{\lambda n} } \boldsymbol{x}_{t-1} \boldsymbol{x}_{t-1}^{\prime} \overset{ \mathbb{P} }{ \to } \lambda \boldsymbol{V}_{xx}$ converges in probability. A similar limit result holds for the IVZ based functional such that $\hat{\mathcal{J}}^{ivz}_{nx} \big( \lambda, \uptau_0, \widehat{ \boldsymbol{\theta} }^{ivz}_n( \uptau_0 ) \big) \Rightarrow \sqrt{\uptau_0 ( 1 - \uptau_0 ) } \times  \mathcal{N} \big( \boldsymbol{0}, \lambda \boldsymbol{I}_p \big)$, regardless of whether the regressors exhibit high persistence. On the other hand, the OLS based functional has a nonstandard limit distribution with regressors of high persistence. In the case of the IVX functional one needs to consider the limit result for these two classes of persistence separately. These conjectures are summarized 
and proved by Proposition \ref{proposition1A} in the next section where we formalize the test statistics.  

\subsection{Test Statistics}
\label{Section3.3}

We consider as detectors two types of test statistics commonly employed in the literature related to structural break testing methodologies. The first type of test corresponds to the fluctuation type statistic studied by \cite{qu2008testing} specifically for a quantile regression model, while the second type of test corresponds to the Wald statistic proposed by the seminal paper of \cite{andrews1993tests} for the linear regression model. Both test statistics utilize the supremum functional since the underline assumptions allow for a structural break for the coefficients of the nonstationary quantile predictive regression model at an unknown break-point location within the full sample. 

Therefore, the null hypothesis of interest (e.g., see \eqref{hypothesisA}) is formulated as below
\begin{align}
\mathcal{H}^{(A)}_0: \boldsymbol{\theta}^{(1)}_n \big( \lambda ; \uptau_0 \big) 
= 
\boldsymbol{\theta}^{(2)}_n \big( \lambda ; \uptau_0 \big) \ \ \ \text{versus} \ \ \ \mathcal{H}^{(A)}_1: \boldsymbol{\theta}^{(1)}_n \big( \lambda ; \uptau_0 \big) \neq \boldsymbol{\theta}^{(2)}_n \big( \lambda ; \uptau_0 \big)
\end{align}
where $\boldsymbol{\theta}^{(j)}_n ( \lambda ; \uptau_0 ) = \big( \alpha^{(j)}_n(\lambda ; \uptau_0), \boldsymbol{\beta}^{(j)}_n(\lambda ; \uptau_0)^{\prime} \big)^{\prime}$, for $j \in \left\{ 1, 2 \right\}$ and the location of the break-point is denoted with $\kappa = \floor{ \lambda n }$ for some $0 < \lambda < 1$. Specifically, the implementation of structural break tests for the purpose of detecting parameter instability in nonstationary quantile predictive regressions is a novel aspect in the literature. To facilitate for the development of large sample theory, Assumption \ref{assumption2} presents necessary conditions relating the matrix moments to the quantile structure of the model. 

\newpage

\begin{assumption}
\label{assumption2}
The regressors of the nonstationary quantile predictive regression model which follow a near unit process, are assumed to satisfy the following conditions: 

\begin{itemize}

\item[(a)] $\displaystyle \underset{ n \to \infty }{ \mathsf{plim} } \ \frac{1}{ n^{ 1 + \upgamma_x } } \sum_{t=1}^{ \floor{ \lambda n } } f_{ u_t ( \uptau ), t-1 } (0)  \boldsymbol{x}_{t-1} \boldsymbol{x}_{t-1}^{\prime}  = \lambda f_{ u_t ( \uptau ) } (0) \boldsymbol{V}_{xx}$, uniformly for $0 < \lambda < 1$, 

\item[(b)] $\displaystyle \underset{ n \to \infty }{ \mathsf{plim} } \ \frac{1}{ n^{ 1 + \upgamma_x } } \sum_{t=1}^{ \floor{ \lambda n } } \boldsymbol{x}_{t-1} \boldsymbol{x}_{t-1}^{\prime}  = \lambda \boldsymbol{V}_{xx}$, uniformly for some $0 < \lambda < 1$, where $\boldsymbol{V}_{xx}$ is a $p \times p$ non-random positive definite matrix and $\upgamma_x \in (0,1)$, 

\item[(c)] $\mathbb{E} \left(  \boldsymbol{x}_{t-1} \boldsymbol{x}_{t-1}^{\prime} \right)^{ 2 + s} < L$ with $s > 0$ and $L < \infty$ for all $1 \leq t \leq n$, 

\item[(d)] there exists a $\delta > 0$ and an $M < \infty$, such that $n^{-1} \sum_{t=1}^n \mathbb{E} \norm{ \boldsymbol{x}_{t-1} }^{ 3 (1 + \delta )} < M$ and $\mathbb{E} \left( n^{-1} \sum_{t=1}^n \norm{ \boldsymbol{x}_{t-1} }^3 \right)^{ ( 1 + \delta )} < M$ hold for any $n$.
\end{itemize}
\end{assumption}

\medskip

Assumption \ref{assumption2} (a) and (b) are standard convergence in probability limits for the nonstationary quantile predictive regression model. Assumption \ref{assumption2} (c) is employed for the convergence of the weighted empirical process $n^{- 1 / 2} \sum_{t =1 }^{ \floor{ \lambda n } } \boldsymbol{x}_{t-1} \big( \uptau_0 - \mathds{1} \left\{ F_{y | x } ( y_t ) \leq \uptau_0 \right\} \big)$. Furthermore, Assumption \ref{assumption2} (d) ensures stochastic equicontinuity (see, Chapter 2 in \cite{van1996WeakConergence}) of the sequential empirical process based on estimated quantile regression residuals, which is needed to establish weak convergence of the tests (see \cite{bai1996testing}). Moreover, in the case of possibly nonstationary regressors, since standard quantile regression estimators follow a locally uniform weak convergence (see \cite{De2006extreme}) then invariance principles hold uniformly for $\lambda \in (0,1)$.

\subsubsection{Fluctuation type tests}

Specifically, since we assume that the true break point is unknown, we need to search over all possible candidate subsets within the full sample. Furthermore, according to \cite{qu2008testing} recentering $\hat{\mathcal{J}}_n \big( \lambda, \uptau_0, \widehat{ \boldsymbol{ \beta} }_n( \uptau_0 ) \big)$ by the quantity $\lambda \hat{\mathcal{J}}_n \big( 1 , \uptau_0, \widehat{ \boldsymbol{ \beta} }_n( \uptau_0 ) \big)$ often yields better finite sample performance. Such considerations lead to the following test statistic: 
\begin{align}
\mathcal{SQ}_n ( \lambda ; \uptau_0 ) = \underset{ \lambda \in [0,1] }{ \mathsf{sup} } \bigg\| \frac{1}{ \sqrt{\uptau_0 ( 1- \uptau_0) } } \bigg[ \hat{\mathcal{J}}_n \big( \lambda, \uptau_0, \widehat{ \boldsymbol{ \beta} }_n( \uptau_0 ) \big) - \lambda \hat{\mathcal{J}}_n \big( 1 , \uptau_0, \widehat{ \boldsymbol{ \beta} }_n( \uptau_0 ) \big) \bigg] \bigg\|_{ \infty } 
\end{align}
where $\norm{ . }_{ \infty }$ is the $\mathsf{sup}-$norm such that for a generic vector $\mathsf{z} = \left( \mathsf{z}_1,..., \mathsf{z}_p \right)$ implies that $\norm{ . }_{ \infty } := \mathsf{max} \left( | \mathsf{z}_1 |, ..., | \mathsf{z}_p | \right)$ (see, \cite{Koenker2002inference}). 

We focus on the implementation of two different estimation methodologies. Therefore, to investigate the practical use of the proposed fluctuation type test for structural break detection in the nonstationary quantile predictive regression model, we consider the asymptotic distribution of the test statistics according to the estimator employed to construct the test function. Thus, Proposition \ref{proposition1A} summarizes the formulations of the test according to the estimation methodology employed for a fixed quantile level $\uptau_0 \in (0,1)$. 

\newpage

\begin{proposition}
\label{proposition1A}
Under the null hypothesis $\mathcal{H}_0^{(A)}$ and given that Assumptions \ref{assumption1}-\ref{assumption2} hold, then the fluctuation type statistics weakly converge to the limit distributions below 
\begin{small}
\begin{align*}
\textbf{(\textit{i})} \ \mathcal{SQ}^{ols}_n ( \lambda ; \uptau_0 )
&:= 
\underset{ \lambda \in [0,1] }{ \mathsf{sup} } \ 
\bigg\| \frac{1}{ \sqrt{\uptau_0 ( 1- \uptau_0) } } \bigg[ \hat{\mathcal{J}}_n \left( \lambda, \uptau_0, \widehat{ \boldsymbol{\theta} }^{ols}_n( \uptau_0 ) \right) - \lambda \hat{\mathcal{J}}_n \left( 1 , \uptau_0, \widehat{ \boldsymbol{\theta} }^{ols}_n( \uptau_0 ) \right) \bigg] \bigg\|_{ \infty } 
\\
&\Rightarrow 
\begin{cases}
\underset{ \lambda \in [0,1] }{ \mathsf{sup} } \ 
\big\| \boldsymbol{ \mathcal{BB} }_{p+1}( \lambda) \big\|_{\infty}, & \ \ \text{when} \ \upgamma_x \in (0,1)
\\
\underset{ \lambda \in [0,1] }{ \mathsf{sup} } \ 
\mathbb{S}_{xx}^{-1 /2}  \times
\left\{ 
\begin{bmatrix}
\mathbf{ \mathcal{BB} }_{\psi_{\uptau}} ( \lambda )_{ \textcolor{red}{ ( 1 \times n ) } }
\\
\displaystyle \mathbf{\mathcal{JB}}_{\psi_{\uptau}} (\lambda)_{ \textcolor{red}{ ( p \times n ) } }
\end{bmatrix}_{ \textcolor{red}{ (p+1) \times n } }
\right\},  & \ \ \text{when} \ \upgamma_x = 1
\end{cases}
\\
\\
\textbf{(\textit{ii})} \ \mathcal{SQ}^{ivx}_n ( \lambda ; \uptau_0 ) 
&:= 
\underset{ \lambda \in [0,1] }{ \mathsf{sup} } \
 \bigg\| \frac{1}{ \sqrt{\uptau_0 ( 1- \uptau_0) } } \bigg[ \hat{\mathcal{J}}_n \left( \lambda, \uptau_0, \widehat{ \boldsymbol{ \beta} }^{ivx}_n( \uptau_0 ) \right) - \lambda \hat{\mathcal{J}}_n \left( 1 , \uptau_0, \widehat{ \boldsymbol{ \beta} }^{ivx}_n( \uptau_0 ) \right) \bigg] \bigg\|_{ \infty } 
\\
&\Rightarrow
\underset{ \lambda \in [0,1] }{ \mathsf{sup} } \ 
\big\| \boldsymbol{ \mathcal{BB}}_p( \lambda) \big\|_{\infty}, \ \ \ \ \ \ \ \ \ \ \ \ \ \ \ \ \ \ \ \ \ \ \  \ \ \ \ \ \  \ \text{when} \ \upgamma_x = (0, \upgamma_z)
\\
\\
\textbf{(\textit{iii})} \ \mathcal{SQ}^{ivz}_n ( \lambda ; \uptau_0 ) 
&:= 
\underset{ \lambda \in [0,1] }{ \mathsf{sup} } \ 
\bigg\| \frac{1}{ \sqrt{\uptau_0 ( 1- \uptau_0) } } \bigg[ \hat{\mathcal{J}}_n \left( \lambda, \uptau_0, \widehat{ \boldsymbol{ \beta} }^{ivz}_n( \uptau_0 ) \right) - \lambda \hat{\mathcal{J}}_n \left( 1 , \uptau_0, \widehat{ \boldsymbol{ \beta} }^{ivz}_n( \uptau_0 ) \right) \bigg] \bigg\|_{ \infty } 
\\
&\Rightarrow \underset{ \lambda \in [0,1] }{ \mathsf{sup} } \ \big\| \boldsymbol{ \mathcal{BB} }_p( \lambda) \big\|_{\infty}, \ \ \ \ \ \ \ \ \ \ \ \ \ \ \ \ \ \ \ \ \ \ \ \ \ \ \ \ \  \ \text{when} \ \upgamma_x = (0,1]
\end{align*}
\end{small}
where $\boldsymbol{ \mathcal{BB} }_p( . )$ is a vector of $p$ independent Brownian bridge processes\footnote{Note that $\boldsymbol{ \mathcal{BB} }_p( . )$ is known as the square of a standardized tied-down Bessel process of order $p$.} on $\mathcal{D}_{ \mathbb{R}^p } \left( [0,1] \right)$,
\begin{align*}
\mathbb{S}_{xx} := \begin{bmatrix}
1  &  \displaystyle \int_0^1 \boldsymbol{J}_c(r)^{\prime}  dr
\\
\displaystyle \int_0^1 \boldsymbol{J}_c(r) dr & \displaystyle \int_0^1 \boldsymbol{J}_c(r) \boldsymbol{J}_c(r)^{\prime}
\end{bmatrix}_{ \textcolor{red}{ (p + 1 ) \times (p + 1 ) } }  \ \text{with} \ \ 0 < r < 1
\end{align*}
where $\mathbb{S}_{xx}$ is a positive definite stochastic matrix, $\mathbf{ \mathcal{BB} }_{\psi_{\uptau}} ( \lambda ) : = B_{\psi_{\uptau}}(\lambda) - \lambda B_{\psi_{\uptau}}(1)$.
\end{proposition}
A necessary condition to apply weak convergence arguments that yields invariance principles for partial sum processes in the Skorokhod space $\mathcal{D} \left( [0,1] \right)$ follows   
\begin{small}
\begin{align*}
\underset{ \lambda \in (0,1) }{ \mathsf{sup} } \boldsymbol{D}_n^{-1} \bigg| \left[ \hat{\mathcal{J}}_n \big( \lambda, \uptau_0, \widehat{ \boldsymbol{ \beta} }_n( \uptau_0 ) \big) - \lambda \hat{\mathcal{J}}_n \big( 1 , \uptau_0, \hat{\boldsymbol{\beta}}_n( \uptau_0 ) \big)  \right] - \left[ \mathcal{J}_n \big( \lambda, \uptau_0, \boldsymbol{ \beta}_n( \uptau_0 ) \big) - \lambda \mathcal{J}_n \big( 1 , \uptau_0, \boldsymbol{ \beta}_n( \uptau_0 ) \big) \right] \bigg| = o_{ \mathbb{P} } (1).
\end{align*}
\end{small}
The proposed test statistics extend the fluctuation type tests studied by \cite{qu2008testing}, to the nonstationary quantile predictive regression model of our setting. We compare the instrumental variable based method to the classical OLS approach for constructing the fluctuation test. Moreover, we employ the IVZ estimator given by Proposition \ref{proposition1A} \textbf{(\textit{iii})}, which replaces the original covariate vector with the constructed instruments as a post-estimation correction method that permits to obtain further simplifications of asymptotic terms. Our asymptotic theory analysis shows that the fluctuation type test weakly converges into a Brownian bridge limit when the IVZ estimator is employed and the same limit holds for both OLS and IVX based tests under mild integratedness.  

\newpage

On the other hand, under high persistence the fluctuation type test based on the OLS estimator is proved to have a weak convergence into a nonstandard and nonpivotal asymptotic distribution.\footnote{Similar results are proved by \cite{Katsouris2021breaks} who propose structural break tests for the linear predictive regression model. Within the particular framework, the author proves that the asymptotic distribution of sup-Wald type statistics when the regressors of the model exhibit high persistence are nonstandard (when the OLS estimator is employed), since the asymptotic distribution of the test depend on the unknown persistence coefficient.} A similar asymptotic result holds for the corresponding IVX based test statistic when $\upgamma_x = 1$ such that 
\begin{align}
\hat{\mathcal{J}}_n &\left( \lambda, \uptau_0, \widehat{ \boldsymbol{\beta} }^{ivx}_n( \uptau_0 ) \right) - \lambda \hat{\mathcal{J}}_n \left( 1 , \uptau_0, \widehat{ \boldsymbol{\beta} }^{ivx}_n( \uptau_0 ) \right) 
\nonumber
\\
&\Rightarrow
\sqrt{\uptau_0 (1 - \uptau_0)} \times \boldsymbol{\Gamma}_{cxz}^{-1 / 2} \times
\bigg\{ \mathcal{N} \displaystyle \bigg( \boldsymbol{0}, \lambda \boldsymbol{V}_{cxz} \bigg) 
- \lambda 
\mathcal{N} \bigg( \boldsymbol{0},  \boldsymbol{V}_{cxz} \bigg) \bigg\} 
\end{align}
unless we assume that the exponent rate of persistence is such that $\upgamma_x (0, \upgamma_z)$. 

\medskip

Fluctuation type statistics have been previously examined as a detector for parameter instability in studies such as \cite{kuan1994implementing}, \cite{chu1996monitoring} and \cite{leisch2000monitoring}. More precisely, the statistical advantage of these statistics lies in the fact that they utilize properties of the maximum of Wiener processes (see \cite{revesz1982increments}) and consequently weakly convergence arguments as defined by \cite{billingsley1968convergence} can be employed to derive their asymptotic behaviour. Furthermore, these class of tests belong to the same class as CUSUM type tests although in the latter case the test function is constructed based on regression residuals (see \cite{kulperger2005high}). 

Overall within the nonstationary quantile predictive regression model framework of our study, we observe some important conclusions for the implementation of fluctuation type tests as structural break detectors. Firstly, the asymptotic distributionof the test statistic $\mathcal{SQ}( \lambda ; \uptau_0 )$ depends on the chosen estimator when constructing the test function as seen from the limiting distributions when constructing the test based on the two estimators, under high persistent regressors. On the other, under the assumption of mildly integrated regressors these test statistics weakly converge into a Brownian bridge type limit regardless of the chosen estimator when constructing the test function. Therefore, for the particular persistence class fluctuatuon type tests depends only on the number of parameters subject to structural break since the nuisance coefficient of persistence that captures the nonstationary properties of predictors is filtered out. Secondly, these test statistics do not require to  estimate the sparsity coefficient $f_{y | x } \big( F^{-1}_{y | x } ( \uptau_0 ) \big)$. According to \cite{qu2008testing} this occurs since the subgradient, when evaluated at the true parameter value $\boldsymbol{\beta}_0( \uptau_0 )$, does not depend on the distribution of the errors. 

Thus, conducting statistical inference with some prior information regarding the presence of persistence regressors, using fluctuation type tests as detectors is preferable to construct the test function based on the IVZ estimator which can lead to conventional inference methods (e.g., using tabulated critical values).

\newpage

Next, we examine the self-normalized\footnote{Specifically a relevant application of self normalized statistics is the construction of confidence intervals for model parameters as in the study of \cite{shao2010self}. We leave this aspect for future research.} property of Wald type tests by deriving the related asymptotic theory. Due to the assumptions and conditions under which we construct the proposed test statistics, to examine their limit distributions stochastic equicontinuity arguments are necessary in proofs (see, \cite{newey1991uniform}). 

\subsubsection{Wald type tests}

We now introduce the Wald type tests based on the two estimation methodologies which we focus on (OLS versus IVX based tests). The formulation of the model under the null and under the alternative hypothesis can change the interpretation in the notation we employ for model parameters. One approach is to employ the formulations given by expressions \eqref{form.null}-\eqref{form.alter}. In that case, the Wald type test is constructed for testing the null hypothesis that the parameter vector $\boldsymbol{\upbeta}_{(2)}(\uptau_0)$ which implies that we are testing the null hypothesis that $\boldsymbol{\beta}_1(\uptau_0) - \boldsymbol{\beta}_1(\uptau_0) = \boldsymbol{0}$. However, one has to be careful when constructing the covariance matrix as the regressors need to be adjusted accordingly. Furthermore, a second approach is to construct the stacked regressors that correspond to the time series observations from each of the two subsamples.  

To simplify the notation, we employ the second approach and denote with $\widehat{ \boldsymbol{\beta} }_{1} ( \lambda ; \uptau_0 )$ the estimator of $\boldsymbol{\beta}_0 ( \uptau_0 )$, using observations up to $\kappa = \floor{ \lambda n }$ for some $0 < \lambda < 1$ and with $\widehat{ \boldsymbol{\beta} }_{2} ( \lambda ; \uptau_0 )$ the corresponding parameter estimator based on the remaining observations in the sample. Moreover, denote with $\tilde{\boldsymbol{X}} \equiv \big[ \boldsymbol{X}_1 \  \boldsymbol{X}_2 \big]$ and $\boldsymbol{R} \equiv \big[ \boldsymbol{I}_p \ - \boldsymbol{I}_p \big]$ the selection matrix, then $\Delta \widehat{ \boldsymbol{\beta} }_n ( \lambda ; \uptau_0 ) := \big( \widehat{ \boldsymbol{\beta} }_{2} ( \lambda ; \uptau_0 ) - \widehat{ \boldsymbol{\beta} }_{1} ( \lambda ; \uptau_0 ) \big)$. Then, the Wald test for testing the null hypothesis that the two regimes have equivalent parameter vectors, based on the OLS estimator and some unknown break-point $\kappa = \floor{ \lambda n }$ is formulated as below
\begin{align}
\mathcal{W}_n ( \lambda, \uptau_0 ) = n \Delta \widehat{ \boldsymbol{\beta} }_n ( \lambda ; \uptau_0 )^{\prime} \big[ \widehat{\boldsymbol{V} }_n( \lambda ; \uptau_0 ) \big]^{-1} \Delta \widehat{ \boldsymbol{\beta} }_n ( \lambda ; \uptau_0 )
\end{align}
where $\widehat{\boldsymbol{V}}_n( \lambda ; \uptau_0 )$ is a consistent estimate of the limiting variance of $\Delta \widehat{ \boldsymbol{\beta} }_n ( \lambda ; \uptau_0 )$ under the null hypothesis, $\mathcal{H}_0^{(A)}$, of no parameter instability for a fixed quantile level $\uptau_0 \in (0,1)$. The variance estimator is a key quantity which will affect the robustness of Wald type tests and takes different forms depending on the estimation method we employ when fitting the nonstationary quantile predictive regression. 

Consider the following limiting variance estimate
\begin{align}
\underset{ n \to \infty }{ \mathsf{plim} } \ \big\{ \widehat{\boldsymbol{V}}_n( \lambda ; \uptau_0 ) \big\}
\equiv 
\left[ \frac{ \uptau_0 ( 1 - \uptau_0 )  }{ \lambda ( 1 - \lambda ) } \right] \boldsymbol{\Omega}_0, \ \ \ \ (\lambda, \uptau_0) \in  (0,1) \times (0,1),
\end{align}
where $\boldsymbol{\Omega}_0 = 
\boldsymbol{H}_0^{-1} \boldsymbol{D}_0 \boldsymbol{H}_0^{-1}$ is the unknown variance of the OLS-Wald test. 

\newpage

Furthermore, define with 
\begin{align}
\boldsymbol{H}_0 
&=  
\underset{ n \to \infty }{ \mathsf{plim} } \ \frac{1}{n} \sum_{t=1}^n f_{y | x } \left( y_t | \boldsymbol{x}_{t-1} \right) \boldsymbol{x}_{t-1}^{\prime} \boldsymbol{x}_{t-1} 
\\
\boldsymbol{D}_0 
&= 
\underset{ n \to \infty }{ \mathsf{plim} } \ \frac{1}{n} \sum_{t=1}^n  \boldsymbol{x}_{t-1}^{\prime} \boldsymbol{x}_{t-1}   
\end{align}
where $f_{y | x } ( . | \boldsymbol{x}_{t-1} )$ and $F_{y | x } ( . | \boldsymbol{x}_{t-1} )$ are the conditional density and conditional cumulative distribution function of $y_t$ respectively (see \cite{goh2009nonstandard} and \cite{aue2017piecewise}). 

However, the particular form of the asymptotic variance only holds under the assumption of stationarity in which case the variance estimator simplifies since it does not depend on any nuisance parameters (such as the coefficients of persistence) and thus equivalent matrix moments to the expressions in \cite{qu2008testing} (see also \cite{andrews1993tests}) hold. In our setting, we consider alternative variance estimators based on both the chosen estimator and the persistence class of regressors. 

In any of the aforementioned cases, the supremum Wald test is defined as below 
\begin{align}
\mathcal{SW}_n ( \lambda ; \uptau_0 ) := \underset{ \lambda \in \Lambda_{\eta} }{ \mathsf{sup} } \ \bigg\{ n \Delta \widehat{ \boldsymbol{\beta} }_n ( \lambda ; \uptau_0 )^{\prime} \big[ \widehat{\boldsymbol{V}}_n( \lambda ; \uptau_0 ) \big]^{-1} \Delta \widehat{ \boldsymbol{\beta} }_n ( \lambda ; \uptau_0 ) \bigg\}
\end{align}  
In practise, a symmetric trimming coefficient is employed such that $0 < \eta < 1 / 2$ which lead to the admissible set $\Lambda_{\eta} := [ \eta, 1 - \eta ]$, in order to ensure that the test statistics converge in distribution under the null hypothesis. Therefore, we  investigate the asymptotic behaviour of the OLS based Wald test for the nonstationary quantile predictive regression model given by \eqref{model1}-\eqref{model2} which encompasses the case of stationary regressors. Under the assumption of stable regressors the asymptotic variance of the OLS-Wald test is equivalent to the case when regressors are stationary and ergodic. 

Under the assumption of nonstationarity, the formulation of the OLS-Wald test statistic requires to determine the asymptotic behaviour of the following quantities
\begin{small}
\begin{align}
\widehat{ \boldsymbol{\beta} }_{1}^{ols} ( \lambda; \uptau_0 )
&= \left( \frac{1}{\kappa} \sum_{t=1}^{\floor{\lambda n}} \boldsymbol{x}_{t-1} \boldsymbol{x}_{t-1}^{\prime} \right)^{-1} \left( \frac{1}{\kappa} \sum_{t=1}^{\floor{\lambda n}} \boldsymbol{x}_{t-1} y_{t}  \right) 
\\
\widehat{\boldsymbol{\beta}}_{2}^{ols} ( \lambda; \uptau_0 )
&= \left( \frac{1}{n-\kappa} \sum_{t= \floor{\lambda n} + 1}^n \boldsymbol{x}_{t-1} \boldsymbol{x}_{t-1}^{\prime} \right)^{-1} \left( \frac{1}{n-\kappa} \sum_{t= \floor{\lambda n} + 1}^n \boldsymbol{x}_{t-1} y_{t} \right)
\end{align}
\end{small}
and  $\Delta \widehat{ \boldsymbol{\beta} }^{ols}_n ( \lambda ; \uptau_0 ) =  \widehat{ \boldsymbol{\beta} }_{2}^{ols} ( \lambda ; \uptau_0 ) - \widehat{ \boldsymbol{\beta} }_{1}^{ols} ( \lambda ; \uptau_0 )$, for some $0 < \lambda < 1$ and $\uptau \in (0,1)$. Then, due to orthogonality of the two set of regressors the covariance matrix simplifies into the following expression:
\begin{small}
\begin{align}
\widehat{\boldsymbol{V}}_n^{ols}( \lambda ; \uptau_0 ) 
:= 
\bigg[ \boldsymbol{R} \big( \tilde{\boldsymbol{X}}^{\prime} \tilde{\boldsymbol{X}} \big)^{-1} \boldsymbol{R}^{\prime} \bigg]
\equiv 
\bigg[ 
\big( \boldsymbol{X}_1^{\prime} \boldsymbol{X}_1 \big)^{-1} + 
\big( \boldsymbol{X}_2^{\prime} \boldsymbol{X}_2 \big)^{-1}
\bigg]
\end{align} 
\end{small}

\newpage

\subsection{Asymptotic Theory}
\label{Section3.4}

As we discussed previously, the limit theory of the Wald type statistics for both the OLS and IVX estimators seems more difficult than the limit results for the fluctuation type tests, especially due to the dependence of regressors and parameter estimates to the nuisance parameter of persistence. Therefore, here we generalize the functionals introduced in Section \ref{Section3.2} and \ref{Section3.2} in order to study their asymptotic properties which can alleviate the difficulty in obtaining stochastic approximations under the presence of abstract degree of persistence; simplifying this way derivations for their limit distributions.   

\subsubsection{OLS-Wald test statistic}
\label{Section3.4.1}

We focus on the asymptotic theory for the OLS-Wald test statistic, which is employed as a structural break detection for the nonstationary quantile predictive regression model. In particular, we investigate the asymptotic behaviour of the partial sum processes for the OLS based functionals we introduced previously. For a general parameter vector $\boldsymbol{b} \in \mathbb{R}^p$ we denote with  
$S_n ( \lambda, \uptau_0 , \boldsymbol{b} )$ the partial sum given by the following expression 
\begin{align}
S_n \big( \lambda, \uptau_0, \boldsymbol{b} \big) = n^{- 1 / 2} \sum_{t=1}^{\floor{ \lambda n} } \boldsymbol{x}_{t-1} \psi_{ \uptau } \big( y_t - \boldsymbol{x}_{t-1}^{\prime} \boldsymbol{b} \big)
\end{align}
where $\psi_{ \uptau } ( \mathsf{u} )$ is such that $
\psi_{ \uptau } ( \mathsf{u} ) := \big[ \uptau - \mathds{1} \left\{ \mathsf{u} \leq 0 \right\} \big]$. Therefore, $S_n ( \lambda, \uptau_0, \boldsymbol{b} )$ is written as 
\begin{align}
\label{functional.Sn}
S_n \big( \lambda, \uptau_0, \boldsymbol{b} \big) = n^{- 1 / 2} \sum_{t=1}^{\floor{ \lambda n} } \boldsymbol{x}_{t-1} \big[ \uptau_0 - \mathds{1} \big\{ y_t - \boldsymbol{x}_{t-1}^{\prime} \boldsymbol{b} \leq 0 \big\} \big].  
\end{align}
Following conventional laws of invariance principles for $\textit{i.i.d}$ partial sums the induced sequence of increments are tight within a suitable topological space\footnote{Related theory to weak convergence arguments of partial sum processes can be found in various studies. For instance, \cite{WangPhillips2012specification} redefine the innovation sequence of their model, in the context of specification testing under nonstationarity, to a richer probability space which contains a standard Brownian motion. To do this, a triangular representation of the near unit process is employed in order to investigate the asymptotic behaviour of the transformed functional with respect to this triangular array. Although this would be an interesting way to represent our functionals we avoid the introduction of triangular arrays which could be more challenging to handle.} and in fact they converge weakly to Gaussian processes. Therefore, investigating the asymptotic behaviour and properties of these functionals is useful for the development of the asymptotic theory of the proposed test statistics as well as for other applications. In order to do this, we consider centering the quantity $\mathds{1} \big\{ y_t - \boldsymbol{x}_{t-1}^{\prime} \boldsymbol{b} \leq 0 \big\}$ at its expectation conditional on $\boldsymbol{x}_{t-1}$, instead around the quantile level $\uptau_0$. Furthermore, since we assume that the nonparametric functional given by expression \eqref{functional.Sn} can be employed as a stochastic process in $\mathcal{D} \left( [0,1] \right)$, which is the topological space of all right continuous functions with left limits then we can derive an invariance principle for this partial sum process. 

\newpage

To simplify derivations for the asymptotic theory, and following \cite{qu2008testing}, we define the quantity $\widetilde{S}_n \big( \lambda, \uptau_0, \boldsymbol{b} \big)$ with the expression below
\begin{align}
\widetilde{S}_n \big( \lambda, \uptau_0, \boldsymbol{b} \big) 
= 
n^{- 1 / 2} \sum_{t=1}^{\floor{ \lambda n} } \boldsymbol{x}_{t-1} \big[ F_{y | x } \big( \boldsymbol{x}_{t-1}^{\prime} \boldsymbol{b} \big) - \mathds{1} \big\{ y_t - \boldsymbol{x}_{t-1}^{\prime} \boldsymbol{b} \leq 0 \big\} \big].  
\end{align}  
where $F_{y | x } \big( \boldsymbol{x}_{t-1}^{\prime} \boldsymbol{b} \big)$ is assumed to be monotonic. A necessary and sufficient condition for the monotonicity property of the cumulative distribution function to hold is presented by Lemma \ref{lemma3A} (see, also Lemma A1 in \cite{qu2008testing}). Consequently, we obtain that 
\begin{align}
\label{functional.decomp}
S_n \big( \lambda, \uptau_0, \boldsymbol{b} \big) 
\equiv
\widetilde{S}_n \big( \lambda, \uptau_0, \boldsymbol{b} \big) + n^{- 1 / 2} \sum_{t=1}^{\floor{\lambda n }} \boldsymbol{x}_{t-1} \big[ \uptau_0 - F_{y | x } \big( \boldsymbol{x}_{t-1}^{\prime} \boldsymbol{b} \big) \big].
\end{align}
Our research objective here is to establish the weak convergence argument that holds for the random quantity $S_n \left( \lambda, \uptau_0, \boldsymbol{b} \right)$ on $\left( \mathcal{D} [0,1]  \right)^2$ by accommodating for the different convergence rates with appropriate matrix normalizations according to the estimator employed in each case. Moreover, the limit results of these functionals\footnote{Notice that the proposed functionals in this paper similar to the framework of \cite{qu2008testing}, clearly depend on the estimated parameter vector. Therefore, in our setting the assumption of a nonstationary quantile model contributes to some challenging asymptotic theory aspects, which we are motivated to tackle. Moreover, we shall note that a related large stream of literature considers functionals of estimated residuals with associated test statistics such as CUSUM and CUSUM-square commonly employed in the change-point literature. We avoid presenting the related literature here, as it beyond our scope.} can be utilized to show the following type of stochastic convergence 
\begin{align}
\sqrt{n} \left( \widehat{ \boldsymbol{\beta} }^{ols}_{1} ( \lambda ; \uptau_0 ) - \boldsymbol{\beta}_0 ( \uptau_0 ) \right) = \mathcal{O}_{ \mathbb{P} }(1),
\end{align}
where  $\widehat{\boldsymbol{\beta}}^{ols}_{1} ( \lambda ; \uptau_0 )$ is the quantile regression OLS based estimator\footnote{For instance, for the IVX estimator based on observations of the full sample the following order of convergence holds: $\textcolor{blue}{ n^{ \frac{1 + \upgamma_x }{2} } ( \widehat{ \boldsymbol{\beta} }^{ivx}_{n} ( \uptau ) - \boldsymbol{\beta}_0 ( \uptau ) ) = \mathcal{O}_{ \mathbb{P} }(1)}$, which is proved by Corollary \ref{corollary2} in the Appendix of the paper (see also Theorem 3.1 in \cite{lee2016predictive}).}  that corresponds to the subsample $1 \leq t \leq \floor{ \lambda n}$ for some $0 < \lambda < 1$ and $\uptau_0 \in (0,1)$, when the quantile regression has no model intercept. Intuitively, when the model structure incorporates both intercept and slopes then the different convergence rates of these coefficients due to the presence of nonstationarity is accommodated with the use of embedded normalization matrices. Therefore, for the remaining of this section, we suppose that the parameter vector is of the form $\boldsymbol{ \theta } ( \uptau ) = \big[ \alpha (\uptau), \boldsymbol{\beta} (\uptau)^{\prime} \big]^{\prime}$. 


\newpage

Consider the nonstationary quantile predictive regression \eqref{model1}-\eqref{model2} which includes a model intercept. Then, testing for a structural break via a Wald type formulation based on the OLS estimator implies to use the parameter vector $\boldsymbol{\theta}(\uptau)$ instead of $\boldsymbol{\beta}(\uptau)$. 

\medskip

\begin{proposition}
\label{proposition1}
Under the null hypothesis $\mathcal{H}_0^{(A)}$ and given that Assumptions \ref{assumption1}-\ref{assumption2} hold, then the Wald type statistics weakly converge to the limit distributions below 
\begin{align*}
\textbf{(\textit{i})} \ \ \mathcal{ SW }^{ols}_n \big(  \lambda ; \uptau_0 \big) 
&\Rightarrow 
\underset{ \lambda \in [0,1] }{ \mathsf{sup} } \ \frac{ \ \big\| \boldsymbol{ \mathcal{BB} }_{p+1}( \lambda) \ \big\|^2 }{ \lambda( 1 - \lambda)}, \ \text{for} \ \upgamma_x \in (0,1)  
\\
\textbf{(\textit{ii})} \ \ \mathcal{ SW }^{ols}_n \big(  \lambda ; \uptau_0 \big) 
&\Rightarrow 
\underset{ \lambda \in \Lambda_{\eta} }{ \mathsf{sup} } \  
\boldsymbol{\Delta}^{ols}_0 \big( \lambda; \uptau_0 \big) ^{\prime} \big[ \boldsymbol{\Sigma}^{-1} _0 \big( \lambda; \uptau_0 \big)  \big]\boldsymbol{\Delta}^{ols}_0 \big( \lambda; \uptau_0 \big), \text{for} \ \upgamma_x = 1
\end{align*}
\color{black}
where $\boldsymbol{ \mathcal{BB} }_{p+1}( . )$ is a vector of $(p+1)$ independent Brownian bridge processes on $\mathcal{D}_{ \mathbb{R}^{p+1} } \left( [0,1] \right)$. Denote with 
\begin{align*}
\boldsymbol{\Sigma}^{-1} _0 \big( \lambda; \uptau_0 \big)
&:=  
\textcolor{red}{ f_{ u_t(\uptau)}(0)^2 } \bigg[ \mathbb{S}_{xx}(\lambda) - \mathbb{S}_{xx}(\lambda) \mathbb{S}^{-1}_{xx}(1)  \mathbb{S}_{xx}(\lambda) \bigg]
\\
\boldsymbol{\Delta}^{ols}_0 \big( \lambda; \uptau_0 \big)  
&:=
\mathbb{S}^{-1}_{xx}(\lambda)
\begin{bmatrix}
\displaystyle B_{\psi_{\uptau}}(\lambda) 
\\
\displaystyle \int_0^{\lambda} \boldsymbol{J}_c(r) dB_{ \psi_{ \uptau} } 
\end{bmatrix}
-
\big[ \mathbb{S}_{xx}(1) - \mathbb{S}_{xx}(\lambda) \big]^{-1}
\begin{bmatrix}
\displaystyle B_{\psi_{\uptau}}(1) - B_{\psi_{\uptau}}(\lambda) 
\\
\displaystyle \int_0^{1} \boldsymbol{J}_c(r) dB_{ \psi_{ \uptau} } -  \int_0^{\lambda} \boldsymbol{J}_c(r) dB_{ \psi_{ \uptau} } 
\end{bmatrix}.
\end{align*}
where $\boldsymbol{\Sigma}^{-1} _0 \big( \lambda; \uptau_0 \big) \in \mathbb{R}^{ (p+1) \times (p+1)}$ and $\boldsymbol{\Delta}^{ols}_0 \big( \lambda; \uptau_0 \big) \in \mathbb{R}^{ (p+1) \times n}$ since the model included both an intercept and slopes.
\end{proposition}

The proof of Proposition \ref{proposition1} can be found in the Appendix of the paper. Notice that $\boldsymbol{\Sigma}^{-1} _0 \big( \lambda; \uptau_0 \big)$ represents the weakly convergence result of the inverse of the covariance matrix of the stochastic process $\sqrt{n} \big( \hat{\boldsymbol{\upbeta}}_{(2)}(\uptau_0)  - \boldsymbol{\upbeta}_{(2)}(\uptau_0) \big)$, in which case $\mathcal{W}_n(\uptau_0)$ is the Wald statistic for testing the null hypothesis $\mathcal{H}_0: \hat{\boldsymbol{\upbeta}}_{(2)}(\uptau_0) = \boldsymbol{0}$.

\newpage

\subsubsection{IVX-Wald test statistic}
\label{Section3.4.2}

The instrumentation methodology proposed by \cite{PhillipsMagdal2009econometric} has been proved to be robust in filtering abstract degree of persistence in predictive regression models (see also \cite{Phillips2013predictive, Phillips2016robust}). Our research objective in this section is to  study the asymptotic behaviour of the proposed structural break tests based on the endogenous instrumentation procedure in nonstationary quantile predictive regressions. In particular, the self-normalized sup IVX-Wald test function has been recently examined by \cite{Katsouris2021breaks} as break detector for coefficients of linear predictive regressions. Specifically, the supremum IVX-Wald test corresponds to the maximum\footnote{Further details regarding the formulation of Wald type tests and  asymptotic theory is presented in the seminal study of \cite{andrews1993tests}. The particular framework propose for structural change tests in linear regression models under the assumption of stationary and ergodic time series.} of a sequence of test statistics constructed based on sequential sample splitting locations such that $\kappa = \floor{ \lambda n }$ where $\lambda \in \Lambda_{\eta} := [ \eta, 1 - \eta]$ with $0 < \eta < 1/2$. Furthermore, we employ the dequantiled model structure and denote with $\widehat{\boldsymbol{\beta}}_{1}^{ivx} ( \lambda; \uptau_0 )$ and  $\widehat{\boldsymbol{\beta}}_{2}^{ivx} ( \lambda; \uptau_0 )$ the IVX based estimators for the two sub-samples occurred at each splitting step. Therefore, these estimators are computed via the following expressions
\begin{small}
\begin{align}
\widehat{ \boldsymbol{\beta} }_{1}^{ivx} ( \kappa; \uptau_0 )
&= \left( \frac{1}{ \kappa } \sum_{t=1}^{ \kappa + j } \tilde{\boldsymbol{z} }_{1,t-1} \boldsymbol{x}_{1,t-1}^{\prime} \right)^{-1} \left( \frac{1}{ \kappa } \sum_{t=1}^{ \kappa + j } \tilde{ \boldsymbol{z} }_{1,t-1} y_{t}  \right), 
\\
\widehat{\boldsymbol{\beta}}_{2}^{ivx} ( \kappa; \uptau_0 )
&= \left( \frac{1}{ n - \kappa } \sum_{t= \kappa + 1 + j}^n \tilde{ \boldsymbol{z} }_{2,t-1} \boldsymbol{x}_{2,t-1}^{\prime} \right)^{-1} \left( \frac{1}{n-\kappa} \sum_{t=\kappa + 1 + j}^n \tilde{ \boldsymbol{z} }_{2,t-1} y_{t} \right).
\end{align}
\end{small}
where $\kappa = \floor{ \lambda n }$ for some $0 < \lambda < 1$ and the indicator $j \in \left\{ 0,..., (n-\kappa) \right\}$ shows that a sequence of parameter estimates is obtained by moving along all proportions within the compact set $\Lambda_{\eta} = [ \eta, 1 - \eta]$, to compute the maximum Wald statistic. However, for notation convenience we drop the index notation $(\kappa + j)$ and $(\kappa + 1 + j)$ which can be confused with notation used for time-varying parameter estimates. Also, $\boldsymbol{x}_{1,t-1}:= \boldsymbol{x}_{t-1} \mathds{1} \left\{ t \leq \kappa \right\}$ and $\boldsymbol{x}_{2,t-1}:= \boldsymbol{x}_{t-1} \mathds{1} \left\{ t > \kappa \right\}$. Furthermore, $\tilde{\mathbf{Q}}_1 ( \lambda; \uptau_0 )$ and $\tilde{\mathbf{Q}}_2 ( \lambda; \uptau_0 )$ denotes the covariance matrices which correspond to the two subsample parameter estimates and permits to decompose the covariance matrix\footnote{The decomposition of the covariance matrix for the IVX-Wald statistic can be obtained using a formula for inverting
partitioned matrices. In particular, since $\boldsymbol{Z}_1^{\prime} \boldsymbol{X}_2 = \boldsymbol{Z}_2^{\prime} \boldsymbol{X}_1 = 0$ then the matrix inversion formula simplifies further, allowing us to obtain an expression for the variance of the test.} of the test with respect to each regime  
\begin{align*}
\tilde{\mathbf{Q}}_1 \big( \lambda; \uptau_0 \big)  = \left( \tilde{ \boldsymbol{Z} }_1^{\prime} \boldsymbol{X}_1 \right)^{-1} \left( \tilde{\boldsymbol{Z}}_1^{\prime} \tilde{\boldsymbol{Z}}_1 \right) \left(\boldsymbol{X}_1^{\prime} \tilde{\boldsymbol{Z}}_1 \right)^{-1} \ \  
\tilde{\mathbf{Q}}_2 \big( \lambda; \uptau_0 \big)  = \left(\tilde{\boldsymbol{Z}}_2^{\prime} \boldsymbol{X}_2 \right)^{-1} \left( \tilde{\boldsymbol{Z}}_2^{\prime} \tilde{\boldsymbol{Z}}_2 \right) \left(\boldsymbol{X}_2^{\prime} \tilde{\boldsymbol{Z}}_2 \right)^{-1}  
\end{align*}
Then, under the null hypothesis, $\mathcal{H}_0^{(A)}$, the sup IVX-Wald statistic is formulated as
\begin{small}
\begin{align}
\mathcal{SW}_{n}^{ivx} \big( \lambda; \uptau_0 \big) 
:= 
\underset{ \lambda \in \Lambda_{\eta} }{ \mathsf{sup} } \
\left\{
\Delta \widehat{\boldsymbol{\beta}}_n^{ivx} \big( \lambda; \uptau_0 \big) ^{\prime} \bigg[ \widehat{\boldsymbol{V}}_n^{ivx} \big( \lambda; \uptau_0 \big)  \bigg]^{-1} \Delta \widehat{\boldsymbol{\beta}}_n^{ivx} \big( \lambda; \uptau_0 \big)  
\right\}
\end{align} 
where $\Delta \widehat{\boldsymbol{\beta}}_n^{ivx} \big( \lambda; \uptau_0 \big)  := \left( \widehat{\boldsymbol{\beta}}_{1}^{ivx} ( \lambda ; \uptau_0 ) - \widehat{\boldsymbol{\beta}}_{2}^{ivx} ( \lambda ; \uptau_0 ) \right)$ and $\widehat{\boldsymbol{V}}_n^{ivx} \big( \lambda; \uptau_0 \big) := \tilde{\mathbf{Q}}_1 \big( \lambda; \uptau_0 \big)  + \tilde{\mathbf{Q}}_2 \big( \lambda; \uptau_0 \big)$. 
\end{small}

\newpage

\begin{theorem}
\label{Theorem1}
Under the null hypothesis and given that Assumptions \ref{assumption1}-\ref{assumption2} hold, then the sup IVX-Wald statistic weakly convergence to limit distribution below
\begin{align}
\mathcal{SW}_n^{ivx}\big( \lambda; \uptau_0 \big)  
\Rightarrow 
\underset{ \lambda \in \Lambda_{\eta} }{ \mathsf{sup} } \bigg\{ \boldsymbol{\Delta}^{ivx}_0 \big( \lambda; \uptau_0 \big)^{\prime} \big[ \boldsymbol{\Sigma}^{ivx}_0 \big( \lambda; \uptau_0 \big)  \big]^{-1} \boldsymbol{\Delta}^{ivx}_0 \big( \lambda; \uptau_0 \big)  \bigg\}
\end{align}
where $\Lambda_{\eta} := [ \eta,  1 - \eta ]$ with $0 < \eta < 1/2$ and
\begin{align}
\boldsymbol{\Delta}^{ivx}_0 \big( \lambda; \uptau_0 \big)   
&:= 
\boldsymbol{W}_p (\lambda) - \boldsymbol{\Psi}_c(\lambda) \boldsymbol{W}_p(1)
\\
\boldsymbol{\Sigma}^{ivx}_0 \big( \lambda; \uptau_0 \big)  
&:= 
\lambda \big( \boldsymbol{I}_p - \boldsymbol{\Psi}_c(\lambda)  \big) \big( \boldsymbol{I}_p - \boldsymbol{\Psi}_c(\lambda)  \big)^{\prime} + (1 - \lambda ) \boldsymbol{\Psi}_c(\lambda) \boldsymbol{\Psi}_c( \lambda)^{\prime} 
\end{align}
such that
\begin{equation*}
\boldsymbol{\Psi}_c (\lambda)
=
\begin{cases}
\displaystyle \left( \lambda  \boldsymbol{\Omega}_{xx} +  \int_0^{\lambda } \boldsymbol{J}_c^{\mu} (r) d\boldsymbol{J}_c^{\prime} \right) \left( \boldsymbol{\Omega}_{xx} + \int_0^{1} \boldsymbol{J}^{\mu}_c (r) d\boldsymbol{J}_c^{\prime} \right)^{-1}  & ,\text{for} \ \upgamma_x = 1 
\\
\\
\displaystyle \lambda \boldsymbol{I}_p  & , \text{for} \ \upgamma_x \in (0,1)
\end{cases}
\end{equation*}
where $\boldsymbol{W}_p(.)$ is a $p-$dimensional standard Brownian motion, $\boldsymbol{J}_c ( \lambda ) = \int_0^{\lambda} e^{(\lambda - s) \boldsymbol{C}_p} d \boldsymbol{B}(s)$ is an \textit{Ornstein-Uhkenbeck} process and we denote with $\boldsymbol{J}^{\mu}_c (\lambda) = \boldsymbol{J}_c (\lambda) - \int_0^1 \boldsymbol{J}_c(s) ds$ and $\boldsymbol{W}_p^{\mu} (\lambda ) = \boldsymbol{W}_p(\lambda) - \int_0^1 \boldsymbol{W}(s) ds$ the demeaned processes of $\boldsymbol{J}_c(\lambda)$ and $\boldsymbol{W}_p(\lambda)$ respectively.
\end{theorem}

\medskip

\begin{remark} Notice that inference on $\boldsymbol{\beta}_n^{ivx}(\uptau)$ critically depends on the estimator of the covariance matrix $\widehat{\boldsymbol{V}}_n^{ivx} ( \lambda ; \uptau_0 )$. Moreover, the estimation of the sparsity coefficient does not affect the estimation accuracy when constructing test statistics in quantile time series models due to the self-normalized property of Wald type tests. On the other hand, the robust estimation of the covariance matrix is ensured by employing fully modified type of transformations as in the linear model (see, \cite{kostakis2015Robust}). 
\end{remark}

Theorem \ref{Theorem1} presents the asymptotic distribution of the sup IVX-Wald test under the null hypothesis of a single unknown break-point. Furthermore, it covers some practical considerations that arise in empirical work especially with respect to the persistence properties of regressors. As we can observe from the asymptotic behaviour of the test for local unit root regressors (high persistence), it converges to a nonstandard and nonpivotal distribution. On the other hand, for mildly integrated regressors the test behaves in large samples similar to the sup OLS-Wald test which weakly converge into a Brownian bridge type of limit. In the former case, a comparison of the limit distributions of the two tests does not necessarily indicate which test statistic might have better performance in detecting structural breaks to the coefficients of nonstationary quantile predictive regressions. To investigate the particular aspect, we use simulation experiments where allow us to use a suitable experimental design that accommodates these conditions. The proof of Theorem \ref{Theorem1} can be found in the Appendix of the paper.

\newpage

Some important implications follow from Theorem \ref{Theorem1}. More precisely, our asymptotic theory analysis confirms some of the conclusions drawn in similar studies. For instance, \cite{hanson2002tests} and \cite{seo1998tests} (see also \cite{georgiev2018testing}) demonstrated that testing for structural breaks with integrated regressors based on the OLS estimation method converges to a nonstandard and nonpivotal limiting distribution. Furthermore, although the IVX-Wald statistic is proved to be robust to abstract degree of persistence when testing for parameter restrictions, within the structural break testing framework due to the presence of the unknown break-point location, the sup IVX-Wald test similar to the OLS counterpart is coverages to a nonpivotal limiting distribution, which is not Brownian bridge (as defined in the stationary case) even though it is still tied down.  

Nevertheless, some interesting further simplifications occur; for instance under the assumption of mildly integrated regressors, it can be easily proved that the limiting distribution of the sup IVX-Wald test converges to a normalized Brownian bridge limit. This occurs due to the asymptotic matrix moments such that, for $0 < \upgamma_x < 1$ it holds that $\sum_{t=1}^{ n } \boldsymbol{x}_{t-1} \tilde{\boldsymbol{z}}_{t-1}^{\prime} \Rightarrow  - \boldsymbol{\Omega}_{xx} \boldsymbol{C}_z^{-1}$ by expression (20) in \cite{PhillipsMagdal2009econometric}. Thus, it also holds that $\sum_{t=1}^{\floor{ \lambda n } } \boldsymbol{x}_{t-1} \tilde{\boldsymbol{z}}_{t-1}^{\prime}  \Rightarrow   - \lambda \boldsymbol{\Omega}_{xx} \boldsymbol{C}_z^{-1}$, which implies that $\boldsymbol{\Psi}_c (\lambda) = \lambda \boldsymbol{I}_p$.
Furthermore, we also consider the limiting distribution of the sup IVZ-Wald test.

\medskip

\begin{corollary}
\label{Corollary55AA}
Under the null hypothesis, $\mathcal{H}_0^{(A)}$ and suppose that Assumptions \ref{assumption1}-\ref{assumption2} hold, then the sup IVZ-Wald statistic weakly convergence to the asymptotic distribution below
\begin{align}
\mathcal{ SW }^{ivz}_n \big(  \lambda ; \uptau_0 \big) 
&\Rightarrow 
\underset{ \lambda \in \Lambda_{\eta} }{ \mathsf{sup} } \ \frac{ \ \big\| \boldsymbol{ \mathcal{BB} }_{p}( \lambda) \ \big\|^2 }{ \lambda( 1 - \lambda)}, \ \ \text{for} \ \ 0 < \upgamma_x \leq 1.
\end{align}
where $\boldsymbol{ \mathcal{BB} }_{p}( . )$ is a vector of $p$ independent Brownian bridge processes on $\mathcal{D}_{ \mathbb{R}^{p} } \left( [0,1] \right)$.
\end{corollary}

\medskip

In summary, in this section we show that the limiting distribution of the sup IVX-Wald test statistic under high persistence is nonstandard and nonpivotal. Furthermore, the particular limit result simplifies when regressors in the model are assumed to be mildly integrated resulting to weakly convergence into a Brownian bridge type limit. On the other hand, when we construct the test statistic based on the IVZ estimator then the sup IVZ-Wald test converges into a Brownian bridge type of limit regardless of the degree of persistence. Lastly, for a known break-point Wald type tests converge to a nuisance-parameter free limiting distribution, simplifying this way statistical inference. The proof of Corollary \ref{Corollary55AA} can be found in the Appendix.

Another important aspect we consider for the development of the asymptotic theory of the paper, is the classical result of Huber for models with nonstandard conditions such as quantile regression models. More specifically, the first order condition (FOC) defined as the right derivative of the objective function plays a key role in deriving the asymptotic theory of estimators for the quantile model. In particular, we can show that the parameter vector estimator solves these FOC and then apply a Bahadur representation for the estimator. All these results hold with almost surely convergence in large samples.

\newpage

\section{Monte Carlo Experiments}
\label{Section5}

Practically, it is unclear how well the asymptotic theory can provide reliable reference and guidance in finite samples when applied to time series data since usually they can exhibit abstract degree of persistence. However, under the assumption that regressors incorporated in the quantile predictive regression model are generated by near unit root processes for which their asymptotic behaviour is well-understood, then our test statistics can provide an indication regarding the ability of the testing procedures in detecting structural breaks in coefficients of nonstationary quantile predictive regression models. Thus, to investigate the finite sample performance of the proposed tests for their adequacy in detecting parameter instability we focus on the empirical size simulation results as well as on asymptotic power analysis under relevant sequence of local alternatives.

\subsection{Experimental Design}

We conduct a number of Monte Carlo experiments to evaluate the performance of the limit distribution of the Wald and fluctuation type statistics against the conventional $\chi^2$ asymptotic approximation. We simulate the following data generating process
\begin{align}
y_t &= \alpha + \sum_{j=1}^3 \beta_{j} x_{j,t-1} + u_t, \ \ \ \  1 \leq t \leq n
\\
\boldsymbol{x}_t &= 
\begin{bmatrix}
\varrho_n ( c_j, \upgamma_x  ) & 0 & 0
\\
0 & \varrho_n ( c_j, \upgamma_x  ) & 0
\\
0 & 0 & \varrho_n ( c_j, \upgamma_x  )
\end{bmatrix} \boldsymbol{x}_{t-1} + \boldsymbol{v}_{t}
\end{align}
where $\boldsymbol{x}_t = \big( x_{1t}, x_{1t}, x_{1t} \big)^{\prime}$, $\boldsymbol{v}_t = \big( v_{1t}, v_{1t}, v_{1t} \big)^{\prime}$ and $\varrho_n ( c_j, \upgamma_x  ) = \left( 1 + \displaystyle \frac{c_j}{ n^{ \upgamma_x }}  \right) $. Then, the innovation sequence of the model, denoted with $\boldsymbol{e}_t = \left( u_t, \boldsymbol{v}_t \right)^{\prime}$ is generated such that $\boldsymbol{e}_t \sim \mathcal{N} \left( \boldsymbol{0}_{(p+1) \times 1}, \boldsymbol{\Sigma}_{ee} \right)$, where $\boldsymbol{\Sigma}_{ee}$ is an $( p+1 ) \times (p+1)$ positive-definite covariance matrix with a pre-specified variance-covariance structure given as below
\begin{align}
\boldsymbol{\Sigma}_{ee} 
=
\begin{bmatrix}
\sigma^2_{uu} \ & \ \boldsymbol{\rho}^{\prime} 
\\
\boldsymbol{\rho} \ & \ \boldsymbol{\Sigma}_{vv}
\end{bmatrix}
\end{align}
where the matrix $\boldsymbol{\Sigma}_{vv}$ is of full rank $p$, resulting to a nonsingular matrix $\boldsymbol{\Sigma}_{ee}$. 

The coefficients of persistence are such that $c_j \in \left\{ -1, -2, -5 \right\}$ and $\upgamma_x$ is defined to be $\upgamma_x = 1$ to simulate near unit root predictors and $\upgamma_x = 0.75$ to simulate mildly integrated predictors. Moreover, we consider different sample size such that $n \in \left\{ 250, 500, 750, 1000 \right\}$. Under the null hypothesis of no parameter instability we use the following parameters $\alpha = 1$, $\beta_1 = 0.25, \beta_2 = 0.75, \beta_ 3 = -0.50$ and construct the proposed test statistics with significance level $\upalpha = 5 \%$. For the IVX instrumentation we use $c_z = 1$ and $\upgamma_z = 0.95$.

\newpage

In particular, to construct the test statistics the simulated pair $\left\{ y_t, \boldsymbol{x}_t \right\}_{t=1}^n$ is formulated: 
\begin{align*}
y_t =  \left( \alpha^{(1)} ( \uptau ) + \sum_{j=1}^3 \beta^{(1)}_{j} ( \uptau ) x_{j,t-1}   \right) \mathds{1} \big\{ t \leq \kappa \big\}  + \left( \alpha^{(2)}( \uptau ) + \sum_{j=1}^3 \beta^{(2)}_{j} ( \uptau ) x_{j,t-1}  \right) \mathds{1} \big\{ t > \kappa \big\} + u_t
\end{align*}
where $\kappa = \floor{ \lambda n }$ and the search over all possible subsets occurs for values of $\lambda \in \Lambda_{\eta}$. Denote with $\boldsymbol{\theta}^{(j)} ( \uptau) = \left(   \alpha^{(j)} ( \uptau ), \beta_1^{(j)} ( \uptau ), \beta_2^{(j)} ( \uptau ), \beta_3^{(j)} ( \uptau ) \right)^{\prime}$ with $j \in \left\{ 1, 2 \right\}$ to be the quantile dependent parameter vector of each of the two regimes, for a fixed quantile $\uptau_0$ that belongs in the compact set\footnote{Notice that the compact set $\mathcal{T}_{\eta}$  falls strictly within the unit interval to allow the conditional distribution to have an unbounded support.} $\mathcal{T}_{\eta}$ such that $0 < \eta < \uptau_0 < 1 - \eta < 1$.   

Then, the testing hypothesis of interest is formulated as below
\begin{align}
\mathcal{H}^{(A)}_0: \boldsymbol{\theta}_n^{(1)} ( \uptau) = \boldsymbol{\theta}_n^{(2)} ( \uptau) \ \ \ \text{versus} \ \ \ \mathcal{H}^{(A)}_1: \boldsymbol{\theta}_n^{(1)} ( \uptau) \neq \boldsymbol{\theta}_n^{(2)} ( \uptau)
\end{align}
with a fixed quantile $\uptau_0 \in \mathcal{T}_{\eta} := [ \eta, 1 - \eta ]$, for an unknown break-point location $\kappa= \floor{  \lambda n}$ where $0 < \lambda < 1$ and a significance level $\upalpha = 5 \%$. 

\medskip


\subsection{Implementation}

One important aspect for the correct implementation of the fluctuation based tests (which involves the estimation of a subgradient) is that no nuisance parameter is needed (in the case when the LUR process is not employed). More precisely, for the Wald type statistics, we need a consistent estimate of the variance-covariance matrix $\boldsymbol{\Omega}_0$. In particular, it requires estimating the following matrix\footnote{Notice that the estimation of the sparsity function will affect the finite-sample performance of the test statistics if not consistently estimated.} 
\begin{align}
\label{cov.robust.matrix}
\boldsymbol{H}_0 = \underset{ n \to \infty }{ \mathsf{plim} } \left(   n^{-1} \sum_{t=1}^n f_{y | x } \big( F^{-1}_{y | x } \left( \uptau \right)  \big) \boldsymbol{x}_{t-1} \boldsymbol{x}_{t-1}^{\prime} \right)^{-1}
\end{align}   
A discussion regarding methodologies for estimating the matrix given by expression \eqref{cov.robust.matrix} can be found in \cite{qu2008testing}. On the other hand, when we implement the quantile predictive regression model with the presence of persistence covariates, then we will need to examine the asymptotic properties of the above covariance estimator since it will be a function of the unknown coefficient of persistence. 

In summary, in this paper we take the position than when using Wald type statistics as structural break detectors in nonstationary quantile predictive regression models, the chosen estimator will affect the limit distribution of the tests and thus its finite-sample performance, especially under the presence of high persistence regressors. In particular, we have proved that when selecting the OLS estimator then the limit distribution is nonstandard and nonpivotal making inference challenging since critical values can be constructed only with the use of bootstrap-based methodologies. On the other hand, when the IVZ estimator is chosen then the limit distribution is proved to be nuisance-parameter free regardless of the persistence properties driving the behaviour of regressors employed when estimating the quantile predictive regression model.  

Another relevant aspect to investigate is the sensitivity of the  proposed test statistics to the break-point location within the full sample. Therefore, for instance we are interested to examine whether the proposed tests have roughly equal sensitivity to a break occurring early or late in the sample (see, also \cite{leisch2000monitoring}). We expect that the break-point location will not affect the power performance of the proposed test statistics especially due to the fact that we do not operate within a sequential monitoring scheme in which case parameter estimates and functionals are updated continuously.






\subsection{Simulation Results}

The simulation results of the Monte Carlo experiments focus on the performance of the tests by obtaining the empirical size and empirical power. In particular, we evaluate the performance of the tests at different quantiles by employing the test statistics that correspond to the fixed quantile level. More precisely, this allow us to check for structural breaks at the median, or at the upper and lower quantiles for example, thus observing the presence of parameter instability at different levels of the predictant with respect to persistent predictors. Our simulation experiments verify the empirical and theoretical results observed by  \cite{Katsouris2021breaks} in the case of the linear predictive regression model, that is, a trend of over-rejecting the null hypothesis when the OLS estimator is employed when constructing structural break tests\footnote{Notice that the concept of spurious break is discussed by \cite{hansen2000testing} who emphasize that instability in the exogenous variables can cause over-rejection in the standard OLS-based tests.}.

Overall, the finite-sample results reflect the main conjectures presented in the asymptotic theory of the paper and appear to be reasonable for practical use in testing for structural breaks in the coefficients of nonstationary quantile predictive regression models, especially with persistent regressors and endogeneity\footnote{Notice that exogeneity plays an important role in dealing with non-stationary variables. More specifically, in Chapter of \cite{banerjee1993co} it is mentioned that dynamic regression equations in which the conditioning is on weakly or strongly exogenous variables (for the parameter of interest) provide asymptotically unbiased estimates.}. In practise, in those cases in which the exact $\upalpha-$level critical values for $0 < \upalpha < 1$ depend on the unknown parameters of persistence, we employ bootstrap based resampling methods for inference purposes. Therefore, we can observe that these near unit root processes driving the regressors of the model, can affect the ability of the proposed structural break tests for detecting parameter instability in quantile predictive regression models. Specifically, in the simulation study of \cite{WangPhillips2012specification}, the authors mention that serial dependence can affect power. Furthermore, the lower long-run signal strength in the regressor tends to reduce  disciminatory power.


\newpage

\section{Empirical Application}
\label{Section6}

Our empirical application is concerned with the monitoring of the US housing price index returns (\textit{HPI}). Using macroeconomic variables with predictive regression models has been demonstrated in various studies. For instance, the empirical study of \cite{paye2012deja} verifies the episodic predictability conclusions documented in the literature such as in \cite{gonzalo2012regime, gonzalo2017inferring} (see also \cite{demetrescu2020testing}). The author finds statistical evidence of predictability in relation to countercyclical macroeconomic events when forecasting volatility using predictive regressions with macroeconomic covariates. Furthermore, \cite{atanasov2020consumption} investigate the impact of consumption fluctuations on predictability of expected returns using the IVX filter. Overall, our empirical study focuses in the implementation of the proposed test statistics\footnote{Notice, that a different stream of literature proposes testing procedures for detecting market exuberance and bubble effects. Our empirical application is concerned with the detection of structural breaks in the data based on the nonstationary quantile predictive regression.}; thus our research goal here is to test for structural breaks in the relation between the predictant and the predictors at various quantile levels of the underline data specific distribution.

\subsection{Data Description}

For the empirical application of our study, we utilize the dataset of \cite{yang2020testing} that includes the US housing price index returns along with ten common macroeconomic variables. Specifically, the HPI covers more transactions and longer time interval, and thus can well represent the trend of the national-wide housing price such as the housing bubble collapsed during the 2007 subprime mortgage crisis. Furthermore, based on the HPI the authors obtain the quarterly growth rate of the housing price and use this rate as the dependent variable. More precisely, the ten macroeconomic variables are collected from FRED, and all data are quarterly between 1975:Q1 and 2018:Q2.
\begin{small}
\begin{itemize}
\item \textcolor{blue}{CPI}: Consumer price index with all items less shelter for all urban consumers (Index 1982 to 1984 = 100).

\item \textcolor{blue}{DEF}: The implicit price deflator of the gross domestic product (Index 2012 = 100).

\item \textcolor{blue}{GDP}: $\%-$Change of the gross domestic product from the preceding period.

\item \textcolor{blue}{INC}: $\%-$Change of the real disposable personal income from the quarter one year ago. 

\item \textcolor{blue}{IND}: The industrial production index (Index 2012 = 100). An economic indicator that measures real output for all U.S. located facilities manufacturing, mining, and electric, and gas utilities (excluding those in U.S. territories).

\item \textcolor{blue}{INT}: The effective federal funds rate. The interest rate at which depository institutions trade federal funds (balances held at FRBs) with each other overnight.

\newpage

\item \textcolor{blue}{INV}: The shares of the residential fixed investment in the gross domestic product. Gross private domestic investment is a critical component of gross domestic product as it provides an indicator of the future productive capacity of the economy. Residential investment represents expenditures on residential structures and residential equipment that is owned by landlords and rented to tenants.

\item \textcolor{blue}{MOG}: 30-year mortgage rate. It represents contract interest rates on commitments for fixed-rate first mortgages.

\item \textcolor{blue}{RES}: The total reserve balances maintained with the Federal Reserve banks.

\item \textcolor{blue}{UNE}: The civilian unemployment rate. It represents the number of unemployed as a percentage of the labor force.

\end{itemize}
\end{small}

\subsection{Data Analysis}

We begin our analysis by applying standard unit root tests to the predictors\footnote{Notice that we have $t = 1,..., 174$ time series observations which correspond to quarterly economic indicators and macroeconomic variables.} employed for the quantile predictive regression model. Furthermore, we test each individual predictor separately for the presence of parameter instability using a toolkit of various structural break tests commonly employed in the literature. In particular, testing for breaks in housing price indices has been previously studied by \cite{canarella2012unit}. However, testing for quantile predictability as well as testing for breaks in nonstationary quantile regressions is a novel aspect not previously examined in the literature. 


Secondly, we implement the joint IVX-Wald statistic under the null hypothesis that all slope coefficients simultaneously equal to zero\footnote{We consider rejections of the null hypothesis at significance level $5\%$ to match the rejection probabilities employed in the simulation study of the paper.}. More precisely, the particular hypothesis correspond to the null hypothesis of no quantile predictability. Thus, formulating the model in this manner allow us to investigate whether there is a stable relation between regressand and regressors at a specific quantile level\footnote{In particular, the empirical study presented by  \cite{lee2016predictive} demonstrates statistical evidence of predictive ability using the nonstationary qunatile predictive regression model, at some specific quantiles of stock returns such as at lower or upper quantiles while on the other hand evidence of predictability disappear at the median of the conditional distribution of stock returns. } $\uptau_0 \in (0,1)$.


Thirdly, we implement the joint IVX-Wald statistic under the null hypothesis that at least two of the slope coefficients have no structural break throughout the sample.  






\newpage

\section{Conclusion}
\label{Section7}

In this paper we develop a framework for structural break detection for nonstationary quantile predictive regression models, under the null hypothesis of no structural break\footnote{Notice that we avoid to explicitly use the terminology of a null hypothesis of ``stationarity'' versus an alternative hypothesis of ``non-stationarity'' (see \cite{pitarakis2014joint} and \cite{Kwiatkowski1992testing}). Specifically, within our setting these two terms are interpreted in relation to the persistence properties of model predictors rather that with respect to the parameter constancy of coefficients in terms of temporal dependence. Related limit theory and conditions relevant to temporal dependence specifically for quantile and tail empirical processes can be found in Chapter 5 of \cite{De2006extreme}.}. A major challenge when deriving the asymptotic theory of these structural break tests is to obtain nuisance-parameter free limit distributions which is not a trivial task due to the stochastic approximation terms that depend on nuisance parameters, such as higher order covariance terms as functions of the coefficients of persistence that capture the time series properties of regressors. More precisely, we establish the asymptotic distributions for both Wald type (i.e., as in \cite{andrews1993tests}) and fluctuation type tests (i.e., as in \cite{qu2008testing}) with respect to two different estimation methods, that is, the OLS and IVX estimators\footnote{Notice in the study of \cite{Phillips1988asymptotic} the authors demonstrate the asymptotic equivalence of OLS and GLS based estimators in regression models with integrated regressors.}. Our test statistics show to have good finite-sample properties as shown by the Monte Carlo experiments in which we obtain the empirical size and power. 

Firstly, we verify that indeed the self-normalization of Wald type statistics when testing the null hypothesis of no predictability in quantile predictive regressions results to a nuisance-free distribution, that is, ensuring their pivotal property (as also proved by \cite{lee2016predictive} for abstract degree of persistence). Secondly, we demonstrate that the limit distribution of the proposed test statistics for structural break detection is not depending on the particular choice of the estimator of the quantile predictive regression model under mildly integrated; however under high persistence the choice of the estimator alters the limit theory due since different weakly convergence arguments apply. Furthermore, keeping the quantile level fixed versus testing for breaks across multiple quantile levels requires to consider extending the limit result into the two-parameter Gaussian process, for the latter case. Moreover, bootstrap-based methodologies can be applied when the limit distribution is nonstandard, allowing to infer regarding the presence of structural breaks under these conditions (e.g., high persistence).   

Further research worth mentioning includes to extend the current framework proposed in this paper, for quantile predictability tests robust to parameter instability. The specific application has important implications, from both the theoretical as well as the empirical perspective, especially under nonstationarity, since currently the common practise in the literature is the proposition of methods that investigate these two aspects separately. Additionally, our framework can be extended to the alternative hypothesis of multiple structural breaks as in \cite{qu2008testing} as well as within a multivariate setting such that the framework proposed by the study of \cite{qu2007estimating}.





\newpage

\begin{small}
\paragraph{Acknowledgements} This paper is part of my Ph.D thesis at the Department of Economics of the University of Southampton. I am deeply indebted to my academic advisors Jose Olmo and Tassos Magdalinos as well as to Jean-Yves Pitarakis for helpful discussions and constructive feedback. I am also grateful for given the chance to participate to various econometrics seminars and conferences during the Ph.D candidature. The author declares no conflicts of interests. 
\end{small}

\section{APPENDIX}
\label{Section8}

We provide detailed mathematical derivations for the proofs of main asymptotic theory. Relevant references are  \cite{PhillipsMagdal2009econometric} and \cite{qu2008testing}.

\subsection{Proofs of asymptotic theory results}
\label{Appendix8.1}

For $\boldsymbol{b} \in \mathbb{R}^p$ we define (see, \cite{lee2016predictive} and \cite{galvao2014testing}) the empirical process:  
\begin{align*}
\boldsymbol{G}_n \left( \uptau, \boldsymbol{b} \right) := n^{ - ( 1 + \upgamma_x ) / 2 } \sum_{t=1}^n \boldsymbol{z}_{t-1} \times \bigg\{ \psi_{\uptau} \big( u_t \left( \uptau \right) - \boldsymbol{x}_{t-1}^{\prime} \boldsymbol{b} \big) - \mathbb{E}_{ \mathcal{F}_{t - 1} } \left[ \psi_{\uptau} \big( u_t \left( \uptau \right) - \boldsymbol{x}_{t-1}^{\prime} \boldsymbol{b} \big) \right] \bigg\} 
\end{align*}
such that $\left( \uptau, \boldsymbol{b} \right) \in \mathcal{T}_{\iota} \times B \mapsto \boldsymbol{G}_n \left( \uptau, \boldsymbol{b} \right)$ is stochastically equicontinous for any $\epsilon > 0$. 

\subsubsection{Proof of Corollary \ref{corollary2}}

The consistency of the OLS quantile estimator is derived by \cite{koenker2006quantile}. Here, we prove the convergence rate for the IVX-QR estimator by employing standard methods from from the literature of extremum estimators and the regularity conditions imposed by \cite{bickel1975one}. Thus, we aim to show that 
\begin{align}
\label{convergence.rate}
\left( \widehat{ \boldsymbol{\beta} }_n^{ivx-qr} \left( \uptau \right) - \boldsymbol{\beta} \left( \uptau \right)  \right) = \mathcal{O}_{ \mathbb{P} } \left(  n^{ - (1 + \upgamma_z ) / 2 } \right).
\end{align}

\begin{proof}
For the remaining of the proof we denote with $\widehat{ \boldsymbol{\beta} }_n^{ivx-qr} ( \uptau ) := \widehat{ \boldsymbol{\beta} }_n^{\star} ( \uptau )$ to simplify notation. Then, consider the estimator distance such that $
\widehat{ \boldsymbol{\mathcal{E} } }_n^{\star} ( \uptau ) = \left( \widehat{ \boldsymbol{\beta} }_n^{\star} (\uptau) - \boldsymbol{\beta}^{\star} (\uptau) \right)$, which implies that by evaluating $\boldsymbol{G}_n ( \uptau, \boldsymbol{b} )$ at $\boldsymbol{b} = \widehat{ \boldsymbol{\mathcal{E} }}_n^{\star} ( \uptau )$ we obtain the following expression 
\begin{align*}
\boldsymbol{G}_n ( \uptau, \boldsymbol{b} ) \bigg|_{ \boldsymbol{b} = \widehat{ \boldsymbol{\mathcal{E} } }_n^{\star} ( \uptau ) }
&=
n^{ - \frac{(1 + \upgamma_z ) }{2} } \sum_{t=1}^n \tilde{ \boldsymbol{z} }_{t-1}  \bigg\{ \psi_{\uptau} \big( u_t \left( \uptau \right) - \boldsymbol{x}_{t-1}^{\prime} \widehat{ \boldsymbol{\mathcal{E} } }_n^{\star} ( \uptau ) \big) - \mathbb{E}_{ \mathcal{F}_{t - 1} } \left[ \psi_{\uptau} \big( u_t \left( \uptau \right) - \boldsymbol{x}_{t-1}^{\prime} \widehat{ \boldsymbol{\mathcal{E} } }_n^{\star} ( \uptau ) \big) \right] \bigg\}
\end{align*}
Next, we apply the result\footnote{The norm $\norm{.}$ represents the Euclidean norm, i.e., $\norm{x} = \left( \sum_{i=1}^p x_i^2 \right)^{1/2}$ for $x \in \mathbb{R}^p$.} given by Lemma \ref{lemma3}, which implies that for some constant $\mathcal{C}_1$ 
\begin{align}
\mathsf{sup} \big\{ \big\| \boldsymbol{G}_n \left( \uptau, \boldsymbol{b} \right) - \boldsymbol{G}_n \left( \uptau, \boldsymbol{0} \right) \big\| : \norm{ \boldsymbol{b} } \leq n^{(1+ \upgamma_x) / 2} \mathcal{C}_1 \big\} = o_{ \mathbb{P} } (1). 
\end{align}

\newpage 

Moreover, for $\boldsymbol{b} = \boldsymbol{0}$ the following empirical process holds
\begin{small}
\begin{align}
\boldsymbol{G}_n ( \uptau, \boldsymbol{0} )  
=
n^{ - \frac{(1 + \upgamma_z ) }{2} } \sum_{t=1}^n \tilde{ \boldsymbol{z} }_{t-1}  \bigg\{ \psi_{\uptau} \big( u_t \left( \uptau \right) \big) - \mathbb{E}_{ \mathcal{F}_{t - 1} } \left[ \psi_{\uptau} \big( u_t \left( \uptau \right) \big) \right] \bigg\} 
\end{align}
\end{small}
Therefore, the argument of Lemma \ref{lemma3} is expanded as below
\begin{align*}
\boldsymbol{G}_n ( \uptau, \boldsymbol{b} )  - \boldsymbol{G}_n ( \uptau, \boldsymbol{0} ) 
&=
n^{ - \frac{(1 + \upgamma_z ) }{2} } \sum_{t=1}^n \tilde{ \boldsymbol{z} }_{t-1}  \big\{ \psi_{\uptau} \big( u_t \left( \uptau \right) - \boldsymbol{x}_{t-1}^{\prime} \widehat{ \boldsymbol{\mathcal{E} } }_n^{\star} ( \uptau ) \big) - \mathbb{E}_{ \mathcal{F}_{t - 1} }  \left[ \psi_{\uptau} \big( u_t \left( \uptau \right) - \boldsymbol{x}_{t-1}^{\prime} \widehat{ \boldsymbol{\mathcal{E} } }_n^{\star} ( \uptau ) \big) \right] \big\} 
\\
&\ \ \ \ - 
n^{ - \frac{(1 + \upgamma_z ) }{2} } \sum_{t=1}^n \tilde{ \boldsymbol{z} }_{t-1}  \big\{ \psi_{\uptau} \big( u_t \left( \uptau \right) \big) - \mathbb{E}_{ \mathcal{F}_{t - 1} }  \left[ \psi_{\uptau} \big( u_t \left( \uptau \right) \big) \right] \big\} 
\\
&= 
n^{ - \frac{(1 + \upgamma_z ) }{2} } \sum_{t=1}^n \tilde{ \boldsymbol{z} }_{t-1} \psi_{\uptau} \big( u_t \left( \uptau \right) - \boldsymbol{x}_{t-1}^{\prime} \widehat{ \boldsymbol{\mathcal{E} } }_n^{\star} ( \uptau ) \big) \textcolor{red}{ \overset{ \mathbb{P} }{ \to } 0 }
\\
&\ \ \ \ - 
n^{ - \frac{(1 + \upgamma_z ) }{2} } \sum_{t=1}^n \tilde{ \boldsymbol{z} }_{t-1} \mathbb{E}_{ \mathcal{F}_{t - 1} }  \left[ \psi_{\uptau} \big( u_t \left( \uptau \right) - \boldsymbol{x}_{t-1}^{\prime} \widehat{ \boldsymbol{\mathcal{E} } }_n^{\star} ( \uptau ) \big) \right]
\\
&\ \ \ \ - 
n^{ - \frac{(1 + \upgamma_z ) }{2} } \sum_{t=1}^n \tilde{ \boldsymbol{z} }_{t-1} \psi_{\uptau} \big( u_t \left( \uptau \right) \big)
+ 
n^{ - \frac{(1 + \upgamma_z ) }{2} } \sum_{t=1}^n \mathbb{E}_{ \mathcal{F}_{t - 1} }  \left[ \psi_{\uptau} \big( u_t \left( \uptau \right) \big) \right] \textcolor{red}{ = 0 }
\end{align*}  
Taking the absolute value since the result holds for the Euclidean norm we obtain
\begin{small}
\begin{align*}
\bigg| \boldsymbol{G}_n ( \uptau, \boldsymbol{b} )  - \boldsymbol{G}_n ( \uptau, \boldsymbol{0} ) \bigg|
=
n^{ - \frac{(1 + \upgamma_z ) }{2} } \sum_{t=1}^n  \left\{      \tilde{ \boldsymbol{z} }_{t-1} \psi_{\uptau} \big( u_t \left( \uptau \right) \big) + \tilde{ \boldsymbol{z} }_{t-1} \mathbb{E}_{ \mathcal{F}_{t - 1} }  \left[ \psi_{\uptau} \big( u_t \left( \uptau \right) - \boldsymbol{x}_{t-1}^{\prime} \widehat{ \boldsymbol{\mathcal{E} } }_n^{\star} ( \uptau ) \big) \right]  \right\} + o_{ \mathbb{P} } (1).
\end{align*}
\end{small}
Similarly, with the embedded normalization matrices an equivalent expression holds
\begin{small}
\begin{align*}
\bigg| \boldsymbol{G}_n ( \uptau, \boldsymbol{b} )  - \boldsymbol{G}_n ( \uptau, \boldsymbol{0} ) \bigg|
=
\sum_{t=1}^n  \left\{ \tilde{ \boldsymbol{Z} }_{t-1,n} \psi_{\uptau} \big( u_t \left( \uptau \right) \big) + \tilde{ \boldsymbol{Z} }_{t-1,n} \mathbb{E}_{ \mathcal{F}_{t - 1} }  \left[ \psi_{\uptau} \big( u_t \left( \uptau \right) - \boldsymbol{X}_{t-1,n}^{\prime} \widehat{ \boldsymbol{\mathcal{E} } }_n^{\star} ( \uptau ) \big) \right]  \right\} + o_{ \mathbb{P} } (1).
\end{align*}
\end{small}
Then, the conditional expectation  
$\mathbb{E}_{ \mathcal{F}_{t - 1} }  \left[ \psi_{\uptau} \left( u_t (\uptau) -\boldsymbol{x}_{t-1}^{\prime} \widehat{ \boldsymbol{\mathcal{E} } }_n^{\star} ( \uptau ) \right) \right]$ can be expanded around the point $\boldsymbol{\mathcal{E} }( \uptau )  = \boldsymbol{0}$ using the first-order taylor expansion since the term $\widehat{ \boldsymbol{\beta} }^{\star}(\uptau) \boldsymbol{x}_{t-1}$ is strictly monotonic, uniformly on $\left\{ \norm{ \boldsymbol{x}_{t-1} } \leq \delta \right\}$ where $\iota \leq \uptau \leq 1 - \iota$ for some $\delta > 0$ (see, Theorem 1 of \cite{neocleous2008monotonicity}). Hence, we have that 
\begin{align*}
\mathbb{E}_{ \mathcal{F}_{t - 1} }  \left[ \psi_{\uptau} \left( u_t (\uptau) - \boldsymbol{x}_{t-1}^{\prime} \widehat{ \boldsymbol{\mathcal{E} } }_n^{\star} ( \uptau ) \right) \right] 
&\equiv 
\mathbb{E}_{ \mathcal{F}_{t - 1} }  \big[ \psi_{\uptau} \big( u_t (\uptau) - \boldsymbol{x}_{t-1}^{\prime}  \boldsymbol{\mathcal{E} }_n^{\star} ( \uptau ) \big) \big] \bigg|_{ \boldsymbol{\mathcal{E} } ( \uptau ) = 0 } 
\\
&\ \ \ \ \ \ 
+ \frac{ \partial \mathbb{E}_{ \mathcal{F}_{t - 1} }  \big[ \psi_{\uptau} \big( u_t (\uptau) - \boldsymbol{x}_{t-1} ^{\prime} \boldsymbol{\mathcal{E} } ( \uptau ) \big) \big] }{ \partial \boldsymbol{\mathcal{E} } ( \uptau ) } \bigg|_{ \boldsymbol{\mathcal{E} } ( \uptau ) = 0 } \times \widehat{ \boldsymbol{\mathcal{E} } }^{\star}_n ( \uptau ) 
+ o_{\mathbb{P}} \left( \widehat{ \boldsymbol{\mathcal{E} } }^{\star}_n ( \uptau ) \right).
\end{align*}
Note that $\boldsymbol{\mathcal{E} } ( \uptau ) = 0$ implies that $\boldsymbol{\beta} \left( \uptau \right) = \boldsymbol{\beta}^{\star} \left( \uptau \right)$. Moreover, by definition of $\psi_{\uptau}(.)$ and by applying standard results for the conditional expectation operator we obtain that 
\begin{align}
\label{cond.expect}
\mathbb{E}_{ \mathcal{F}_{t - 1} }  \big[ \psi_{\uptau} \big( u_t (\uptau) - \boldsymbol{\mathcal{E} } ( \uptau )^{\prime} \boldsymbol{x}_{t-1} \big) \big] 
&= 
\uptau - \mathbb{E}_{ \mathcal{F}_{t - 1} }  \big[ \mathds{1} \big\{ u_t ( \uptau ) < \boldsymbol{\mathcal{E} }( \uptau )^{\prime} \boldsymbol{x}_{t-1} \big\} \big] 
\nonumber
\\
&=
\uptau - \int_{ \infty }^{ \boldsymbol{x}_{t-1} ^{\prime} \boldsymbol{\mathcal{E} } ( \uptau ) } f_{ u_t ( \uptau ), t-1 } (s) ds
\end{align}

\newpage 

Hence, by differentiating expression \eqref{cond.expect} around the neighbourhood of $\boldsymbol{\mathcal{E} } ( \uptau )$ we get
\begin{align}
\frac{ \partial \mathbb{E}_{ \mathcal{F}_{t - 1} }  \big[ \psi_{\uptau} \big( u_t (\uptau) - \boldsymbol{x}_{t-1} ^{\prime} \boldsymbol{\mathcal{E} } ( \uptau ) \big) \big] }{ \partial \boldsymbol{\mathcal{E} } (\uptau) } \bigg|_{ \boldsymbol{\mathcal{E} } ( \uptau ) = 0 } = - \boldsymbol{x}_{t-1}^{\prime} f_{ u_t ( \uptau ), t-1 } (0) 
\end{align}
Therefore, it holds that 
\begin{align}
\label{limit.res}
\mathbb{E}_{ \mathcal{F}_{t - 1} }  \left[ \psi_{\uptau} \left( u_t (\uptau) - \boldsymbol{x}_{t-1} ^{\prime} \right) \right] = - \boldsymbol{x}_{t-1}^{\prime} f_{ u_t ( \uptau ), t-1 } (0) \widehat{ \boldsymbol{\mathcal{E} } }^{\star}_n ( \uptau )  + o_{\mathbb{P}}(1)
\end{align}
Next substituting the limit given by \eqref{limit.res} into the original expansion for the term $\big| \boldsymbol{G}_n ( \uptau, \boldsymbol{b} )  - \boldsymbol{G}_n ( \uptau, \boldsymbol{0} ) \big|$ as well as by replacing $\boldsymbol{x}_{t-1}$ with the corresponding embedded normalization matrix $\boldsymbol{X}_{t-1,n}$ we obtain the expression 
\begin{small}
\begin{align}
\label{expression.new}
\bigg| &\boldsymbol{G}_n ( \uptau, \boldsymbol{b} )  - \boldsymbol{G}_n ( \uptau, \boldsymbol{0} ) \bigg| 
\nonumber
\\
&\equiv  \boldsymbol{K}_{nz} \big( \uptau, \boldsymbol{\beta}^{\star} ( \uptau ) \big) 
- 
\sum_{t=1}^n f_{ u_t ( \uptau ), t-1 } (0) \tilde{ \boldsymbol{Z} }_{t-1,n} \boldsymbol{X}_{t-1}^{\prime} n^{ - \frac{1 + \upgamma_z }{2} } \left( \widehat{ \boldsymbol{\beta} }_n^{\star} (\uptau) - \boldsymbol{\beta}^{\star} (\uptau) \right) + o_{\mathbb{P}}(1)
\end{align}
\end{small}
where 
\begin{small}
\begin{align}
\boldsymbol{K}_{nz} \big( \uptau, \boldsymbol{\beta}^{\star} ( \uptau ) \big)  
:= 
\boldsymbol{D}_n^{-1} \sum_{t=1}^n \tilde{\boldsymbol{Z}}_{t-1} \psi_{\uptau} \big(  u_t ( \uptau ) \big)
\end{align}
\end{small}
Moreover, we define with 
\begin{align}
\boldsymbol{M}_{nz} \big( \uptau, \boldsymbol{\beta}^{\star} ( \uptau ) \big)  
:= 
\sum_{t=1}^n f_{ u_t ( \uptau ), t - 1 }(0) \tilde{\boldsymbol{Z}}_{t-1,n} \tilde{\boldsymbol{X}}_{t-1,n}^{\prime}  \
\end{align}
Next, by noting that from Lemma \ref{lemma3} it holds that $\mathsf{sup} \big\{ \norm{ \boldsymbol{G}_n ( \uptau, \boldsymbol{b} )  - \boldsymbol{G}_n ( \uptau, \boldsymbol{0} ) } \big\} = o_{ \mathbb{P} } (1)$, then by rearranging expression \eqref{expression.new} we obtain that 
\begin{align}
\label{limit.final}
n^{ \frac{1 + \upgamma_z }{2} } \big( \widehat{ \boldsymbol{\beta} }^{\star}_n ( \uptau ) - \boldsymbol{\beta}^{\star}( \uptau ) \big) 
=
\bigg[ \boldsymbol{M}_{nz} \big( \uptau, \boldsymbol{\beta}^{\star} ( \uptau ) \big) \bigg]^{-1} \boldsymbol{K}_{nz} \big( \uptau, \boldsymbol{\beta}^{\star} ( \uptau ) \big) + o_{ \mathbb{P} }(1)
\end{align}
which proves the result given by expression \eqref{convergence.rate}. In summary, we proved that $\widehat{\boldsymbol{\beta}}_n^{\star} \left( \uptau \right)$ is a consistent estimator of $\boldsymbol{\beta}^{\star} \left( \uptau \right)$ with convergence rate $\sqrt{n} \sqrt[\upgamma_z ]{n}$ where $\upgamma_z \in (0,1)$. Furthermore, using the result given by expression \eqref{limit.final}  we can prove the limit results summarized in Corollary \ref{corollary2} using the asymptotic results presented in Corollary \ref{corollary3A}. 
\end{proof}


\medskip

\begin{remark}
Notice that for the proof of Corollary \ref{corollary3A}, we use the fact that 
\begin{align}
\frac{1}{ \sqrt{n} } \sum_{t=1}^n \psi_{\uptau} \big( u_t (\uptau) \big) \overset{ \mathbb{P} }{ \to } \mathcal{N} \big( 0, \uptau (1 - \uptau) \big) \ \ \text{for some} \ \uptau \in (0,1), 
\end{align}
since due to the structure of the model the quantile regression induced innovation term $\psi_{\uptau} \big( u_t (\uptau) \big) \sim \mathsf{mds} \big( 0, \uptau (1 - \uptau) \big)$, i.e., it has a covariance which depends on the quantile $\uptau$.  
\end{remark}

\newpage

\subsubsection{Proof of Corollary \ref{corollary3A}}

\underline{\textbf{Part \textit{(i)}}}

We begin by considering the limiting distribution of the functional,
\begin{small}  
\begin{align}
\boldsymbol{K}_{nx}  \left( \uptau, \boldsymbol{\theta}^{ols}_n(\uptau) \right) 
:= 
\boldsymbol{D}_n^{-1} \sum_{t=1}^n \boldsymbol{X}_{t-1} \psi_{\uptau} \big(  u_t ( \uptau ) \big) \Rightarrow \boldsymbol{K}_{x}  \left( \uptau, \boldsymbol{\theta}^{ols}(\uptau) \right)
\end{align}
\begin{align*}
\boldsymbol{K}_{x}  \left( \uptau, \boldsymbol{\theta}^{ols}(\uptau) \right)
\equiv 
\begin{cases}
\begin{bmatrix}
B_{\psi_{\uptau}}(1) \\
\displaystyle \int_0^1 \boldsymbol{J}_c(r) dB_{ \psi_{\uptau} } 
\end{bmatrix}  & \ \ \ \textit{LUR},   
\\
\\
\mathcal{N} \displaystyle  \left( \boldsymbol{0}, \uptau (1 - \uptau) \times   
\begin{bmatrix}
   1  &  \boldsymbol{0}^{\prime} \\
   \boldsymbol{0}  & \boldsymbol{V}_{xx}
\end{bmatrix} 
\right)         
       & \ \ \ \textit{MI}.  
\end{cases}
\end{align*}
\end{small}
\begin{proof}

\textbf{\underline{Mildly Integrated:}} (MI) 

By expanding the expression for $\boldsymbol{K}_{nx}  \big( \uptau, \boldsymbol{\theta}^{ols}_n(\uptau) \big)$, for the mildly integrated regressors case we obtain the following expression  
\begin{small}
\begin{align}
\boldsymbol{K}_{nx}  \left( \uptau, \boldsymbol{\theta}^{ols}_n(\uptau) \right)
&= 
\boldsymbol{D}_n^{-1} \sum_{t=1}^n \boldsymbol{X}_{t-1} \psi_{\uptau} \big(  u_t ( \uptau ) \big)
\nonumber
\\
&= 
\begin{bmatrix}
\displaystyle  \frac{1}{ \sqrt{n} } & \boldsymbol{0}^{\prime} 
\nonumber
\\
\boldsymbol{0} & \displaystyle n^{ - \frac{ 1 + \upgamma_x }{2} } \boldsymbol{I}_p
\end{bmatrix}_{ \textcolor{red}{ ( p + 1 ) \times ( p + 1 ) }  } \times
\sum_{t=1}^n 
\begin{bmatrix}
\boldsymbol{1} 
\nonumber
\\
\boldsymbol{x}^{\prime}_{t-1}
\end{bmatrix}_{ \textcolor{red}{ ( p + 1 ) \times n }  } 
\times \psi_{\uptau} \big(  u_t ( \uptau ) \big)
\\
&=
\begin{bmatrix}
\displaystyle  \frac{1}{ \sqrt{n} } \sum_{t=1}^n \psi_{\uptau} \big(  u_t ( \uptau ) \big) \otimes \boldsymbol{1}^{\prime}_{ \textcolor{red}{ ( 1 \times n ) }  } 
\\
\nonumber
\\
\displaystyle  n^{ - \frac{ 1 + \upgamma_x }{2} } \sum_{t=1}^n \boldsymbol{x}_{t-1} \psi_{\uptau} \big(  u_t ( \uptau ) \big) \otimes \boldsymbol{I}_p
\end{bmatrix}_{ \textcolor{red}{ ( p + 1 ) \times n }  } 
\Rightarrow
\mathcal{N}_{p+1} \displaystyle  \left( \boldsymbol{0}, \uptau (1 - \uptau) \times   
 \begin{bmatrix}
 1  &  \boldsymbol{0}^{\prime} \\
 \boldsymbol{0}  & \boldsymbol{V}_{xx}
 \end{bmatrix} 
\right).
\end{align}
\end{small}
Since the two scalar processes $\displaystyle \bigg\{ \frac{1}{ \sqrt{n} } \sum_{t=1}^n \psi_{\uptau} \big(  u_t ( \uptau ) \big) \bigg\}$ and  $\displaystyle \bigg\{ n^{ - \frac{ 1 + \upgamma_x }{2} } \sum_{t=1}^n \boldsymbol{x}_{t-1} \psi_{\uptau} \big(  u_t ( \uptau ) \big) \bigg\}$ are uncorrelated with negligible higher-order moment terms, thus mutually independent Gaussian with zero mean and variance 1 and and $\boldsymbol{V}_{xx}$ respectively. Therefore, a joint convergence to a Gaussian random variate holds due to their conditional independence. To see this, consider the following limit 
\begin{align}
\label{limit.1}
n^{ - \frac{1 + \upgamma_x }{2} } \sum_{t=1}^n \boldsymbol{x}_{t-1} \psi_{\uptau} \big(  u_t ( \uptau ) \big) \otimes \boldsymbol{I}_p \Rightarrow \mathcal{N} \big( \boldsymbol{0}, \uptau (1 - \uptau) \boldsymbol{V}_{xx} \big)
\end{align}
when $0 < \upgamma_x < 1$.

\newpage 

\begin{remark}
Further details regarding these derivations can be found in the proof of Theorem 1 in \cite{xiao2009quantile}. Notice that although the proof of Theorem 1 in \cite{xiao2009quantile} corresponds to the framework of quantile cointegrating regression model, after appropriate modifications we can obtain the limit given by expression \eqref{limit.1}. Furthermore, joint convergence of these two terms holds to a Gaussian random variable with mean vector zero and covariance matrix determined by the covariance of each individual term. 
\end{remark}

Additionally, the following invariance principle for the corresponding partial sum process holds for the exponent rate $0 < \upgamma_x < 1$ such that
\begin{align}
n^{ - \frac{1 + \upgamma_x}{2} } \sum_{t=1}^{ \floor{\lambda n} } \boldsymbol{x}_{t-1} \psi_{\uptau} \big(  u_t ( \uptau ) \big) \otimes \boldsymbol{I}_p \Rightarrow \mathcal{N} \big( \boldsymbol{0}, \uptau (1 - \uptau) \lambda \boldsymbol{V}_{xx} \big).
\end{align}
which is useful when deriving the convergence limit for the OLS based functionals.   

\medskip

\textbf{\underline{Local Unit Root:}} (LUR) 
\begin{small}
\begin{align}
\boldsymbol{K}_{nx}  \left( \uptau, \boldsymbol{\theta}^{ols}_n(\uptau) \right)
&= 
\boldsymbol{D}_n^{-1} \sum_{t=1}^n \boldsymbol{X}_{t-1} \psi_{\uptau} \big(  u_t ( \uptau ) \big)
\nonumber
\\
&= 
\begin{bmatrix}
\displaystyle \frac{1}{ \sqrt{n} } & \boldsymbol{0}^{\prime} 
\\
\boldsymbol{0} & \displaystyle \frac{1}{n} \boldsymbol{I}_p
\end{bmatrix}_{ \textcolor{red}{ ( p + 1 ) \times ( p + 1 ) }  } \times
\sum_{t=1}^n 
\begin{bmatrix}
\boldsymbol{1}^{\prime}_{ \textcolor{red}{ ( 1 \times n ) }  } 
\\
\boldsymbol{x}_{t-1}
\end{bmatrix}_{ \textcolor{red}{ ( p + 1 ) \times n }  } 
\times \psi_{\uptau} \big(  u_t ( \uptau ) \big)
\nonumber
\\
&=
\begin{bmatrix}
\displaystyle  \frac{1}{ \sqrt{n} } \sum_{t=1}^n \psi_{\uptau} \big(  u_t ( \uptau ) \big) \otimes \boldsymbol{1}^{\prime}  
\\
\nonumber
\\
\displaystyle  \frac{1}{n} \sum_{t=1}^n \boldsymbol{x}_{t-1} \psi_{\uptau} \big(  u_t ( \uptau ) \big) \otimes \boldsymbol{I}_p
\end{bmatrix}_{ \textcolor{red}{ ( p + 1 ) \times n }  } 
\Rightarrow
\begin{bmatrix}
B_{\psi_{\uptau}}(1) 
\\
\displaystyle \int_0^1 \boldsymbol{J}_c(r) dB_{ \psi_{ \uptau} } 
\end{bmatrix}_{ \textcolor{red}{ ( p + 1 ) \times n }  } 
\end{align}
\end{small}
where $\boldsymbol{x}_t  = \left( \boldsymbol{I}_p + \frac{ \boldsymbol{C}_p }{n} \right) \boldsymbol{x}_{t-1} + \boldsymbol{v}_t$, for $1 \leq t \leq n$, since $\boldsymbol{X}_{t-1} = \big[ \boldsymbol{1} \ \boldsymbol{x}_{t-1}^{\prime} \big]^{\prime}$. 

Therefore, for deriving the limit result above the following weakly convergence arguments can be applied 
\begin{align}
\frac{1}{ \sqrt{n} } \sum_{t=1}^{ n } \psi_{\uptau} \big( u_t ( \uptau ) \big) \Rightarrow B_{\psi_{\uptau}}(1) \ \ \ \text{and} \ \ \ \frac{1}{ \sqrt{n} } \sum_{t=1}^{ \floor{\lambda n} } \psi_{\uptau} \big( u_t ( \uptau ) \big) \Rightarrow B_{\psi_{\uptau}}( \lambda ) 
\end{align}
for some $0 < \lambda < 1$ and $\uptau \in (0,1)$, where $B_{\psi_{\uptau}}(1)$ is a standard Brownian motion that corresponds to the error function $\psi_{\uptau} \big( u_t ( \uptau ) \big)$. 

\medskip

\begin{remark}
Overall, in the case of LUR regressors (i.e., under high persistence) the convergence to Brownian motion functionals occurs due to the different normalization rates employed for these terms. Furthermore, since the local unit root coefficient is a general case that covers abstract degrees of persistence, then in practise we have convergence to correlated Brownian motions. 
\end{remark}


\newpage 

\underline{\textbf{Part \textit{(ii)}}}

Next, we consider the limiting distribution of the functional, $\boldsymbol{L}_{nx} \big( \uptau, \boldsymbol{\theta}^{ols}_n(\uptau) \big)$ for the two persistence classes separately as explained below. 

\textbf{\underline{Local Unit Root:}} (LUR) 

We obtain the following expression  
\begin{small}
\begin{align*}
\boldsymbol{L}_{nx} \big( \uptau, \boldsymbol{\theta}^{ols}_n(\uptau) \big)
&\overset{ \mathsf{def} }{ = }
\boldsymbol{D}_n^{-1} \left[ \sum_{t=1}^n f_{ u_t ( \uptau ), t-1 } (0) \boldsymbol{X}_{t-1} \boldsymbol{X}_{t-1}^{\prime} \right] \boldsymbol{D}_n^{-1} 
\\
&=
\begin{bmatrix}
\displaystyle \frac{1}{ \sqrt{n} } & 0 
\\
0 & \displaystyle \frac{1}{n} \boldsymbol{I}_p 
\end{bmatrix}
\begin{bmatrix}
\displaystyle \sum_{t=1}^n  \textcolor{red}{f_{ u_t ( \uptau ), t-1 }(0)}   &  \displaystyle  \sum_{t=1}^n \textcolor{red}{f_{ u_t ( \uptau ), t-1 }(0)} \boldsymbol{x}_{t-1}^{\prime}  
\\
\displaystyle \sum_{t=1}^n \textcolor{red}{f_{ u_t ( \uptau ), t-1 }(0)} \boldsymbol{x}_{t-1} & \displaystyle \sum_{t=1}^n \textcolor{red}{f_{ u_t ( \uptau ), t-1 }(0)} \boldsymbol{x}_{t-1} \boldsymbol{x}_{t-1}^{\prime} 
\end{bmatrix}
\begin{bmatrix}
\displaystyle \frac{1}{ \sqrt{n} } & 0 
\\
0 & \displaystyle \frac{1}{n} \boldsymbol{I}_p 
\end{bmatrix}
\\
&= 
\begin{bmatrix}
\displaystyle \frac{1}{ \sqrt{n} } \sum_{t=1}^n  \textcolor{red}{f_{ u_t ( \uptau ), t-1 }(0)}   &  \displaystyle   \frac{1}{ \sqrt{n} } \sum_{t=1}^n \textcolor{red}{f_{ u_t ( \uptau ), t-1 }(0)} \boldsymbol{x}_{t-1}^{\prime}  
\\
\displaystyle \frac{1}{n} \sum_{t=1}^n \textcolor{red}{f_{ u_t ( \uptau ), t-1 }(0)} \boldsymbol{x}_{t-1} \otimes \boldsymbol{I}_p & \displaystyle  \frac{1}{n} \sum_{t=1}^n \textcolor{red}{f_{ u_t ( \uptau ), t-1 }(0)} \boldsymbol{x}_{t-1} \boldsymbol{x}_{t-1}^{\prime} \otimes \boldsymbol{I}_p
\end{bmatrix}
\begin{bmatrix}
\displaystyle \frac{1}{ \sqrt{n} } & 0 
\\
0 & \displaystyle \frac{1}{n} \boldsymbol{I}_p 
\end{bmatrix}
\\
&=
\begin{bmatrix}
\displaystyle \frac{1}{ n } \sum_{t=1}^n \textcolor{red}{f_{ u_t ( \uptau ), t-1 }(0)}   &  \displaystyle    \frac{1}{n} \frac{1}{ \sqrt{n} } \sum_{t=1}^n \textcolor{red}{f_{ u_t ( \uptau ), t-1 }(0)} \boldsymbol{x}_{t-1}^{\prime} \otimes \boldsymbol{I}_p 
\\
\displaystyle \frac{1}{n} \frac{1}{ \sqrt{n} }  \sum_{t=1}^n \textcolor{red}{f_{ u_t ( \uptau ), t-1 }(0)} \boldsymbol{x}_{t-1} \otimes \boldsymbol{I}_p & \displaystyle \frac{1}{n^2} \sum_{t=1}^n \textcolor{red}{f_{ u_t ( \uptau ), t-1 }(0)} \boldsymbol{x}_{t-1} \boldsymbol{x}_{t-1}^{\prime} \otimes \boldsymbol{I}_p
\end{bmatrix}
\\
&\Rightarrow
f_{ u_t ( \uptau )} (0) \times 
\begin{bmatrix}
1  &  \displaystyle \int_0^1 \boldsymbol{J}_c(r)^{\prime} \\
\displaystyle \int_0^1 \boldsymbol{J}_c(r)  &  \displaystyle \int_0^1 \boldsymbol{J}_c(r) \boldsymbol{J}_c(r)^{\prime}
\end{bmatrix} 
\end{align*}
\end{small}
Since, the following convergence in probability holds
\begin{small}
\begin{align}
\frac{1}{ n } \sum_{t=1}^n \textcolor{red}{f_{ u_t ( \uptau ), t-1 }(0)} \overset{ \mathbb{P} }{ \to } \mathbb{E} \big[ \textcolor{red}{f_{ u_t ( \uptau ), t-1 }(0)} \big] =:  f_{ u_t ( \uptau )} (0)
\end{align}
\end{small}
Moreover, recall that for the case of LUR regressors it holds that
\begin{small} 
\begin{align*}
\boldsymbol{D}_n \big( \widehat{ \boldsymbol{\theta}}_n^{qr}(\uptau) - \boldsymbol{\theta}^{qr}( \uptau ) \big) 
&=
\begin{bmatrix}
\displaystyle \sqrt{n} & 0 
\\
0 & \displaystyle n \boldsymbol{I}_p 
\end{bmatrix}
\begin{pmatrix}
\widehat{\alpha}_n^{qr}( \uptau ) - \alpha( \uptau )
\nonumber
\\
\widehat{ \boldsymbol{\beta} }_n^{qr}( \uptau ) - \boldsymbol{\beta}( \uptau )
\end{pmatrix}
\\
&\equiv 
\left\{ \boldsymbol{D}_n^{-1} \left[ \sum_{t=1}^n f_{ u_t ( \uptau ), t-1 } (0) \boldsymbol{X}_{t-1} \boldsymbol{X}_{t-1}^{\prime} \right] \boldsymbol{D}_n^{-1} \right\}^{-1} 
\times 
\left\{ \boldsymbol{D}_n^{-1} \sum_{t=1}^n \boldsymbol{X}_{t-1}  \psi_{ \uptau } \big( u_t ( \uptau ) \big) \right\}
\nonumber
\\
&\Rightarrow 
\left\{ 
f_{ u_t ( \uptau )} (0) \times 
\begin{bmatrix}
1  &  \displaystyle \int_0^1 \boldsymbol{J}_c(r)^{\prime} 
\\
\displaystyle \int_0^1 \boldsymbol{J}_c(r)  &  \displaystyle \int_0^1 \boldsymbol{J}_c(r) \boldsymbol{J}_c(r)^{\prime}
\end{bmatrix} 
\right\}^{-1}
\times
\begin{bmatrix}
B_{\psi_{\uptau}}(1) 
\\
\displaystyle \int_0^1 \boldsymbol{J}_c(r) dB_{ \psi_{ \uptau} } 
\end{bmatrix}
\end{align*}
\end{small}
which follows by an application of the continuous mapping theorem to the first term of the expression above. Notice that in the case of persistent regressors, the limiting distribution of the normalized quantile OLS based estimator is nonstandard and nonpivotal.

\newpage 

\textbf{\underline{Mildly Integrated:}} (MI) 

We obtain the following expression  
\begin{align}
\boldsymbol{L}_{nx} \big( \uptau, \boldsymbol{\theta}^{ols}_n ( \uptau ) \big) 
&\overset{ \mathsf{def} }{ = }
\boldsymbol{D}_n^{-1} \left[ \sum_{t=1}^n f_{ u_t ( \uptau ), t-1 } (0) \boldsymbol{X}_{t-1} \boldsymbol{X}_{t-1}^{\prime} \right] \boldsymbol{D}_n^{-1} 
\nonumber
\\
&=
\begin{bmatrix}
\displaystyle \frac{1}{n} \sum_{t=1}^n \textcolor{red}{f_{ u_t ( \uptau ), t-1 }(0)}   &  \displaystyle  \frac{1}{ n^{ 1 + \frac{\upgamma_x}{2} } } \sum_{t=1}^n \textcolor{red}{f_{ u_t ( \uptau ), t-1 }(0)} \boldsymbol{x}_{t-1}^{\prime} \otimes \boldsymbol{I}_p 
\nonumber
\\
\displaystyle \frac{1}{ n^{ 1 + \frac{\upgamma_x}{2} } }  \sum_{t=1}^n \textcolor{red}{f_{ u_t ( \uptau ), t-1 }(0)} \boldsymbol{x}_{t-1} \otimes \boldsymbol{I}_p & \displaystyle \frac{1}{ n^{ 1 + \upgamma_x} } \sum_{t=1}^n \textcolor{red}{f_{ u_t ( \uptau ), t-1 }(0)} \boldsymbol{x}_{t-1} \boldsymbol{x}_{t-1}^{\prime} \otimes \boldsymbol{I}_p
\end{bmatrix}
\\
&\Rightarrow
f_{ u_t ( \uptau )} (0) \times 
\begin{bmatrix}
1  &  \boldsymbol{0}^{\prime} 
\\
\boldsymbol{0}  & \boldsymbol{V}_{xx}
\end{bmatrix} 
\end{align}
Since it holds that, 
\begin{align*}
\frac{1}{ n^{ 1 + \frac{\upgamma_x}{2} } } \sum_{t=1}^n \boldsymbol{x}_{t-1} \overset{ \mathbb{P} }{ \to} \boldsymbol{0} \ \ \ \ \text{and} \ \ \ \ \frac{1}{ n^{1 + \upgamma_x} } \sum_{t=1}^n \boldsymbol{x}_{t-1} \boldsymbol{x}_{t-1}^{\prime} \overset{ \mathbb{P} }{ \to } \boldsymbol{V}_{xx} \ \ \ \text{when} \ \upgamma_x \in (0,1),
\end{align*}
where the second limit follows by Lemma B3 in the Appendix of \cite{kostakis2015Robust}. Therefore, for the case of mildly integrated regressors in the model we obtain 
\begin{align}
\boldsymbol{D}_n \big( \widehat{ \boldsymbol{\theta}}_n^{qr}(\uptau) - \boldsymbol{\theta}^{qr}( \uptau ) \big) 
&=
\begin{bmatrix}
\displaystyle \sqrt{n} & 0 
\\
0 & \displaystyle n^{ \frac{1 + \upgamma_x }{2} } \boldsymbol{I}_p 
\end{bmatrix}
\begin{pmatrix}
\widehat{\alpha}_n^{qr}( \uptau ) - \alpha( \uptau )
\\
\widehat{ \boldsymbol{\beta} }_n^{qr}( \uptau ) - \boldsymbol{\beta}( \uptau )
\end{pmatrix}
\nonumber
\\
&\equiv 
\left\{ \boldsymbol{D}_n^{-1} \left[ \sum_{t=1}^n f_{ u_t ( \uptau ), t-1 } (0) \boldsymbol{X}_{t-1} \boldsymbol{X}_{t-1}^{\prime} \right] \boldsymbol{D}_n^{-1} \right\}^{-1} \times \left\{ \boldsymbol{D}_n^{-1} \sum_{t=1}^n \boldsymbol{X}_{t-1}  \psi_{ \uptau } \big( u_t ( \uptau ) \big) \right\}
\nonumber
\\
&\Rightarrow 
\left\{ 
f_{ u_t ( \uptau )} (0) \times 
\begin{bmatrix}
1  &  \boldsymbol{0}^{\prime} 
\\
\boldsymbol{0}  & \boldsymbol{V}_{xx}
\end{bmatrix}
\right\}^{-1} 
\times
\mathcal{N}
\left( 
0, \uptau ( 1 - \uptau ) 
\times
\begin{bmatrix}
1  &  \boldsymbol{0}^{\prime} 
\\
\boldsymbol{0}  & \boldsymbol{V}_{xx}
\end{bmatrix}
\right)
\nonumber
\\
&=
\frac{1}{ f_{ u_t ( \uptau )} (0)  } \times
\begin{bmatrix}
1  &  \boldsymbol{0}^{\prime} 
\\
\boldsymbol{0}  & \boldsymbol{V}_{xx}^{-1}
\end{bmatrix}
\times
\mathcal{N}
\left( 
0, \uptau ( 1 - \uptau ) 
\times
\begin{bmatrix}
1  &  \boldsymbol{0}^{\prime} 
\\
\boldsymbol{0}  & \boldsymbol{V}_{xx}
\end{bmatrix}
\right)
\nonumber
\\
&\equiv
\mathcal{N}
\left( 
0, \frac{ \uptau ( 1 - \uptau ) }{ f_{ u_t ( \uptau )} (0)^2   }   
\times
\begin{bmatrix}
1  &  \boldsymbol{0}^{\prime} 
\\
\boldsymbol{0}  & \boldsymbol{V}^{-1}_{xx}
\end{bmatrix}
\right)
\end{align}  
by an application of the continuous mapping theorem to the first term. 
\end{proof}

\medskip

\begin{remark}
Notice also that the limit results given by Corollary \ref{corollary3A} can be used to prove the limiting distribution provided by Corollary \ref{corollary1} of the paper. This can be done, by employing the trick presented in the proof of Theorem 1 in \cite{xiao2009quantile}. In particular, by linearizing  the optimization function in terms of an arbitrary centered quantity $\boldsymbol{D}_n \left( \widehat{\boldsymbol{\theta}}_n (\uptau)  - \boldsymbol{\theta}(\uptau) \right)$. Thus, using the convexity lemma we take the distributional limit of the linearized part and then minimize to get the desired expression as in \eqref{limit.final}. 
\end{remark}

\newpage 

\subsubsection{Proof of Corollary \ref{corollary4A}}

\begin{proof}
For the IVX based estimation of the quantile regression model, we use the dequantile procedure proposed by \cite{lee2016predictive}. Thus, $y_t (\uptau) = y_t - \alpha (\uptau) + \mathcal{O}_{\mathbb{P}} ( n^{-1/2} )$. 
Furthermore, we employ the following embedded normalization matrices 
\begin{align}
\tilde{\boldsymbol{Z}}_{t-1,n} := \tilde{\boldsymbol{D}}_n \tilde{\boldsymbol{z} }_{t-1} \ \ \ \ \text{and} \ \ \ \ \tilde{ \boldsymbol{X} }_{t-1,n} := \tilde{\boldsymbol{D}}_n \boldsymbol{x}_{t-1}, \ \ \ \text{where} \ \ \tilde{\boldsymbol{D}}_n := n^{ \frac{1 + ( \upgamma_x \wedge \upgamma_z ) }{2} } \boldsymbol{I}_p
\end{align} 

\underline{\textbf{Part \textit{(i)}}}

The limit holds regardless of the stochastic dominance of the exponent rates $\upgamma_x$ and $\upgamma_z$
\begin{align}
\boldsymbol{K}_{nz}  \big( \uptau, \boldsymbol{\beta}^{ivx}_n(\uptau) \big) 
&:= 
\sum_{t=1}^n \tilde{\boldsymbol{Z}}_{t-1} \psi_{\uptau} \big(  u_t ( \uptau ) \big)
\equiv 
n^{ \frac{1 + ( \gamma_x \wedge \gamma_z ) }{2} }  \sum_{t=1}^n \tilde{\boldsymbol{z}}_{t-1} \psi_{\uptau} \big(  u_t ( \uptau ) \big) \otimes \boldsymbol{I}_p
\nonumber
\\
&\Rightarrow
\mathcal{N} \big( \boldsymbol{0}, \uptau (1 - \uptau) \boldsymbol{V}_{cxz} \big).
\end{align}

\underline{\textbf{Part \textit{(ii)}}}
\begin{align}
\boldsymbol{M}_{nz} \big( \uptau, \boldsymbol{\beta}^{ivx}_n ( \uptau ) \big) 
:= 
\sum_{t=1}^n f_{ u_t ( \uptau ), t - 1 }(0) \tilde{\boldsymbol{Z}}_{t-1,n} \tilde{\boldsymbol{Z}}_{t-1,n}^{\prime} \Rightarrow f_{ u_t ( \uptau )}(0) \times \boldsymbol{V}_{cxz}
\end{align}
which can be easily shown, since $\sum_{t=1}^n  \tilde{\boldsymbol{z}}_{t-1} \tilde{\boldsymbol{z}}_{t-1}^{\prime} \overset{ \mathbb{P} }{ \to } \boldsymbol{V}_{cxz}$.

\underline{\textbf{Part \textit{(iii)}}}

Then, $\tilde{\boldsymbol{Z}}_{t-1,n} \tilde{\boldsymbol{X}}_{t-1,n}^{\prime} \equiv \tilde{\boldsymbol{D}}_n \tilde{\boldsymbol{z} }_{t-1} \tilde{\boldsymbol{x} }_{t-1}^{\prime} \tilde{\boldsymbol{D}}_n^{\prime}$. 
Therefore, the limit result follows as below
\begin{align*}
\boldsymbol{M}_{nz} \big( \uptau, \boldsymbol{\beta}^{ivx}_n ( \uptau ) \big) 
&:=  
\left[ \sum_{t=1}^n f_{ u_t ( \uptau ), t-1 } (0) \tilde{\boldsymbol{Z}}_{t-1,n} \tilde{ \boldsymbol{X} }_{t-1,n}^{\prime} \right] 
\equiv
\textcolor{magenta}{ \tilde{\boldsymbol{D}}_n } \left[ \sum_{t=1}^n f_{ u_t ( \uptau ), t-1 } (0) \tilde{\boldsymbol{z}}_{t-1} \tilde{ \boldsymbol{x} }_{t-1}^{\prime} \right] \textcolor{magenta}{ \tilde{\boldsymbol{D}}_n^{\prime} } 
\\
&\Rightarrow
f_{ u_t ( \uptau )} (0) \times \boldsymbol{\Gamma}_{cxz}
\end{align*}
Since, a convergence in probability holds $
\frac{1}{ n } \sum_{t=1}^n \textcolor{red}{f_{ u_t ( \uptau ), t-1 }(0)} \overset{ \mathbb{P} }{ \to } \mathbb{E} \big[ \textcolor{red}{f_{ u_t ( \uptau ), t-1 }(0)} \big] =:  f_{ u_t ( \uptau )} (0)$. Moreover, we have that 
\begin{align*}
\tilde{\boldsymbol{D}}_n &\big( \widehat{ \boldsymbol{\beta}}_n^{ivx-qr}(\uptau) - \boldsymbol{\beta}( \uptau ) \big) 
\\
&\equiv 
\left\{ \textcolor{magenta}{ \tilde{\boldsymbol{D}}_n } \left[ \sum_{t=1}^n f_{ u_t ( \uptau ), t-1 } (0) \tilde{\boldsymbol{z}}_{t-1} \boldsymbol{x}_{t-1}^{\prime} \right] \textcolor{magenta}{ \tilde{\boldsymbol{D}}_n^{\prime} }  \right\}^{-1} 
\times 
\left\{ \textcolor{magenta}{ \tilde{\boldsymbol{D}}_n } \sum_{t=1}^n \tilde{\boldsymbol{z}}_{t-1}  \psi_{ \uptau } \big( u_t ( \uptau ) \big) \right\}
\\
&\Rightarrow 
\bigg\{ 
f_{ u_t ( \uptau )} (0) \times \boldsymbol{\Gamma}_{cxz} 
\bigg\}^{-1}
\times
\mathcal{N} \big( \boldsymbol{0}, \uptau (1 - \uptau) \boldsymbol{V}_{cxz} \big)
\\
&\equiv 
\frac{1}{ f_{ u_t ( \uptau )} (0) } \times \boldsymbol{\Gamma}_{cxz}^{-1} \times
\mathcal{N} \big( \boldsymbol{0}, \uptau (1 - \uptau) \boldsymbol{V}_{cxz} \big)
\\
&= 
\mathcal{N} \left( \boldsymbol{0}, \frac{\uptau (1 - \uptau)}{ f_{ u_t ( \uptau )} (0)^2 }  \boldsymbol{\Gamma}_{cxz}^{-1} \boldsymbol{V}_{cxz} \left( \boldsymbol{\Gamma}_{cxz}^{-1} \right)^{\prime} \right)
\equiv 
\mathcal{N} \left( \boldsymbol{0}, \frac{\uptau (1 - \uptau)}{ f_{ u_t ( \uptau )} (0)^2 } \left( \boldsymbol{\Gamma}_{cxz} \boldsymbol{V}_{cxz}^{-1} \boldsymbol{\Gamma}_{cxz}^{\prime} \right)^{-1} \right).
\end{align*}

\newpage

The above result proves the Gaussian random variable limit given by Corollary \ref{corollary2} which holds for both the cases of local unit root and mildly integrated regressors in the quantile predictive regression model. Furthermore, in the case we employ the alternative IVX-QR estimator proposed by \cite{lee2016predictive} (IVZ estimator); in which case the set of nonstationary regressors, $\boldsymbol{x}_{t-1}$, is replaced by the mildly integrated instruments, $\tilde{\boldsymbol{z}}_{t-1}$, we obtain the following limit result
\begin{align}
\tilde{\boldsymbol{D}}_n \big( \widehat{ \boldsymbol{\beta}}_n^{ivz-qr}(\uptau) - \boldsymbol{\beta}( \uptau ) \big) 
&\Rightarrow 
\bigg\{ 
f_{ u_t ( \uptau )} (0) \times \boldsymbol{V}_{cxz} 
\bigg\}^{-1}
\times
\mathcal{N} \big( \boldsymbol{0}, \uptau (1 - \uptau) \boldsymbol{V}_{cxz} \big)
\nonumber
\\
&= 
\frac{1}{ f_{ u_t ( \uptau )} (0) } \times \boldsymbol{V}_{cxz}^{-1} \times
\mathcal{N} \big( \boldsymbol{0}, \uptau (1 - \uptau) \boldsymbol{V}_{cxz} \big)
\nonumber
\\
&=
\mathcal{N} \left( \boldsymbol{0}, \frac{\uptau (1 - \uptau)}{ f_{ u_t ( \uptau )} (0) } \boldsymbol{V}_{cxz}^{-1} \right)
\end{align}
since, $\sum_{t=1}^n f_{ u_t ( \uptau ), t - 1 }(0) \tilde{\boldsymbol{Z}}_{t-1,n} \tilde{\boldsymbol{Z}}_{t-1,n}^{\prime} \Rightarrow f_{ u_t ( \uptau )}(0) \times \boldsymbol{V}_{cxz}$, which is nuisance-parameter free for both the case of local unit root or mildly integrated regressors in the model. 
\end{proof}




\medskip

\begin{remark}
Notice that in a standard time series quantile regression with stationary covariates it holds that the regression $\uptau-$quantile is asymptotically normal, with
\begin{align}
\sqrt{n} \left( \widehat{\boldsymbol{\beta}}_n - \boldsymbol{\beta}(\uptau) \right) \overset{ d }{ \to } \mathcal{N} \big(  \boldsymbol{0}, \uptau(1 - \uptau) \boldsymbol{D}_1^{-1}( \uptau)    \boldsymbol{D}_0 \boldsymbol{D}_1^{-1}( \uptau) \big).
\end{align}
with an appropriate defined covariance matrix which is a function of the moments of the underline error distribution (see, \cite{goh2009nonstandard}). Therefore, we can clearly see that under nonstationarity the covariance matrix of the Gaussian random variate is stochastic due to the presence of the nuisance parameter of persistence. 
\end{remark}

\medskip

\begin{remark}
Notice that the objective function for the setting of the nonstationary quantile predictive regression model, becomes globally convex in the parameter. Hence, the method based on the convexity lemma by Pollard (1991) is applicable. Therefore, our proofs for the asymptotic theory of test of parameter restrictions  is based on the asymptotic theory framework proposed by \cite{xiao2009quantile}. Consequently, the limit results for the linear parameter restrictions can be employed when constructing the parameter specific restrictions that correspond to structural break tests. 
\end{remark}

\newpage

\subsubsection{Alternative IVZ-QR estimator}

Following the framework proposed by \cite{lee2016predictive} we also consider the limiting distribution of the IVX-QR estimator when the original persistent regressors are replaced by the instrumental variables in the optimization function. The particular approach is convenient as it significantly reduces the computational time by avoiding the nonconvex optimization procedure given by expression \eqref{min.function} which requires to use a grid search with several local optima. More specifically, we consider 
\begin{align}
\widehat{ \boldsymbol{ \gamma } }_n^{ivx-qr} \left( \uptau \right) := \underset{ \boldsymbol{ \gamma } \in \mathbb{R}^{p} }{ \mathsf{arg \ min} } \ \sum_{t=1}^n \uprho_{\tau} \big( y_{t} ( \uptau ) - \tilde{ \boldsymbol{z} }_{t-1}^{\prime} \boldsymbol{ \gamma } \big). 
\end{align}

\begin{corollary}
Under the null hypothesis $\mathcal{H}_0 : \boldsymbol{ \beta } ( \uptau ) = 0$, it holds that  
\begin{align}
\tilde{\boldsymbol{D} }_n \big( \widehat{ \boldsymbol{ \gamma } }_n^{ivx-qr} ( \uptau ) - \boldsymbol{ \beta } ( \uptau ) \big) \Rightarrow 
\mathcal{N} \left( \boldsymbol{0}, \frac{ \displaystyle \uptau ( 1 - \uptau) }{ \displaystyle f_{ u_t ( \uptau )}(0)^2 } \boldsymbol{V}_{cxz}^{-1} \right)
\end{align} 
both for near unit root and mildly integrated predictors, where $\tilde{\boldsymbol{D} }_n = n^{\frac{ 1 + \upgamma_x \wedge \upgamma_z }{2}} \boldsymbol{I}_p$. 
\end{corollary}

\medskip

\begin{lemma}
\label{lemma6A}
(Self-normalized IVX-QR) Under Assumption \ref{assumption1} it holds that, 
\begin{align}
\frac{ \widehat{f_{ u_t ( \uptau )} }(0)^2   }{  \uptau ( 1- \uptau)  } \left( \widehat{\boldsymbol{\gamma}}^{ivx-qr}_n ( \uptau ) - \boldsymbol{\beta} ( \uptau ) \right)^{\prime}  \big( \tilde{ \boldsymbol{Z}}^{\prime} \tilde{ \boldsymbol{Z} } \big)^{-1} \left( \widehat{\boldsymbol{\gamma}}^{ivx-qr}_n ( \uptau ) - \boldsymbol{\beta} ( \uptau ) \right) \Rightarrow \chi^2_p 
\end{align}
such that $\widehat{f_{ u_t ( \uptau )} }(0)^2$ is a consistent estimator of $f_{ u_t ( \uptau )}(0)^2$ and $p$ degrees of freedom. 
\end{lemma}
Therefore, Lemma \ref{lemma6A} provides a uniform inference limit result which allows to easily obtain critical values since is nuisance-parameter free. Furthermore, if we are interested to test for example the predictability of a specific subgroup among the predictors, say $\mathcal{H}_0 : \beta_{1} (\uptau) = \beta_{2}(\uptau) = 0$, then the formulation of the Wald statistic with the linear restrictions matrix can be employed. In the particular example, the restrictions matrix takes the form $\boldsymbol{R} = \big[ \boldsymbol{I}_2, \boldsymbol{0}_{ 2 \times (p-2) } \big]$. Then, generalizing the specific example for testing a set of linear restrictions, implies that the null hypothesis is formulated as $\mathcal{H}_0 : \boldsymbol{R} \boldsymbol{\beta} ( \uptau ) = 0$ where $\boldsymbol{R}$ is a $r \times p$ known restriction matrix. 

Then, the limiting distribution for the IVX-Wald statistic for the quantile predictive regression is given by the following expression  
\begin{align*}
\frac{ \widehat{f_{ u_t ( \uptau )} }(0)^2 }{ \uptau ( 1- \uptau)  } \left( \boldsymbol{R} \widehat{ \boldsymbol{\gamma} }^{ivx-qr}_n ( \uptau ) \right)^{\prime}  \bigg[ \boldsymbol{R} \big( \tilde{ \boldsymbol{Z} }^{\prime} \tilde{ \boldsymbol{Z} } \big)^{-1} \boldsymbol{R}^{\prime} \bigg]^{-1} \left( \boldsymbol{R} \widehat{ \boldsymbol{\gamma} }^{ivx-qr}_n ( \uptau ) \right) \Rightarrow \chi^2_{p-2} 
\end{align*}
where $\chi^2_{p-2}$ denotes the chi-square random variate with $(p-2)$ degrees of freedom such that $\mathbb{P} \left( \chi^2 \geq \chi^2_{ p - 2 ; \upalpha  } \right) = \upalpha$, where $0 < \upalpha < 1$ denotes the significance level. 

\newpage 

\subsubsection{Proof of Lemma \ref{lemma1}}

We have that
\begin{small}
\begin{align}
\mathcal{W}^{ivx-qr}_n ( \uptau ) 
=
\frac{ \widehat{f_{ u_t ( \uptau )} }(0)^2   }{  \uptau ( 1- \uptau)  } \left( \widehat{\boldsymbol{\beta}}^{ivx-qr}_n ( \uptau ) - \boldsymbol{\beta} ( \uptau ) \right)^{\prime}  \bigg( \boldsymbol{X}^{\prime} \boldsymbol{P}_{ \tilde{ \boldsymbol{Z} }  } \boldsymbol{X} \bigg) \left( \widehat{\boldsymbol{\beta}}^{ivx-qr}_n ( \uptau ) - \boldsymbol{\beta} ( \uptau ) \right) \Rightarrow \chi^2_p 
\end{align}
where
\begin{align*}
\bigg( \boldsymbol{X}^{\prime} \boldsymbol{P}_{ \tilde{ \boldsymbol{Z} } } \boldsymbol{X} \bigg) 
:= \left( \boldsymbol{X}^{\prime} \tilde{\boldsymbol{Z}} \right) \left( \tilde{\boldsymbol{Z}}^{\prime} \tilde{\boldsymbol{Z}} \right)^{-1} \left( \tilde{\boldsymbol{Z}}^{\prime}\boldsymbol{X}  \right) 
\equiv 
\left( \sum_{t=1}^n \boldsymbol{x}_{t-1} \tilde{\boldsymbol{z} }^{\prime}_{t-1} \right) \left( \sum_{t=1}^n \tilde{\boldsymbol{z}}_{t-1} \tilde{\boldsymbol{z} }^{\prime}_{t-1} \right)^{-1}  \left( \sum_{t=1}^n \boldsymbol{x}_{t-1} \tilde{\boldsymbol{z} }^{\prime}_{t-1} \right)^{\prime}
\end{align*}
Moreover, we use the embedded normalization matrices such that 
\begin{align*}
\left( \sum_{t=1}^n f_{ u_t ( \uptau ), t-1 } (0) \right) \times  \left( \sum_{t=1}^n \tilde{\boldsymbol{Z}}_{t-1,n} \boldsymbol{X}^{\prime}_{t-1,n} \right) 
\equiv 
\left( \sum_{t=1}^n f_{ u_t ( \uptau ), t-1 } (0) \tilde{\boldsymbol{Z}}_{t-1,n} \boldsymbol{X}^{\prime}_{t-1,n}   \right) 
\overset{ \mathbb{P} }{\to} 
f_{ u_t ( \uptau )}(0) \boldsymbol{\Gamma}_{cxz}
\end{align*}
and the fact that $\displaystyle \left( \sum_{t=1}^n \boldsymbol{z}_{t-1} \tilde{\boldsymbol{z} }^{\prime}_{t-1} \right) \overset{ \mathbb{P}  }{\to} \boldsymbol{V}_{cxz}$. 
\end{small}
\begin{proof}
From Corollary \ref{corollary2} we have that 
\begin{align*}
\tilde{\boldsymbol{D}}_n \left( \widehat{\boldsymbol{\beta}}_n^{ivx-qr} \left( \uptau \right) - \boldsymbol{\beta} \left( \uptau  \right) \right) 
&=
\left( \sum_{t=1}^n f_{ u_t ( \uptau ), t-1 } (0) \tilde{\boldsymbol{Z}}_{t-1,n} \boldsymbol{X}^{\prime}_{t-1,n}   \right)^{-1} \left( \sum_{t=1}^n f_{ u_t ( \uptau ), t-1 }(0) \tilde{\boldsymbol{Z}}_{t-1,n} \psi_{\uptau} \big( u_t ( \uptau) \big) \right) 
\\
&\Rightarrow
\mathcal{N} \displaystyle  \left( 0, \frac{  \uptau  (1 -  \uptau ) }{ f_{u_t ( \uptau) }(0)^2  } \times \big( \boldsymbol{\Gamma}_{cxz}\boldsymbol{V}_{cxz}^{-1} \boldsymbol{\Gamma}_{cxz}^{\prime} \big)^{-1} \right) 
\end{align*}
Denote the consistent sample estimator of $\widehat{f_{ u_t ( \uptau )} }(0)$, with $f_{ u_t ( \uptau )}(0)$ where 
\begin{align}
\widehat{f_{ u_t ( \uptau )} }(0) = \sum_{t=1}^n f_{ u_t ( \uptau ), t-1 } (0)  
\end{align} 
Therefore, the following asymptotic convergence result follows
\begin{align*}
\bigg( \boldsymbol{X}^{\prime} \boldsymbol{P}_{ \tilde{ \boldsymbol{Z} } } \boldsymbol{X} \bigg)  
\Rightarrow 
\big[ f_{ u_t ( \uptau )}(0) \boldsymbol{\Gamma}_{cxz} \big] \times \boldsymbol{V}_{cxz}^{-1} \times \big[ f_{ u_t ( \uptau )}(0) \boldsymbol{\Gamma}_{cxz} \big]^{\prime}
\equiv 
f_{ u_t ( \uptau )}(0)^2 \big( \boldsymbol{\Gamma}_{cxz} \boldsymbol{V}_{cxz}^{-1} \boldsymbol{\Gamma}_{cxz}^{\prime}  \big)
\end{align*}
Thus, we obtain 
\begin{align*}
\mathcal{W}^{ivx-qr}_n (\uptau)
&\Rightarrow 
\frac{ 1  }{  \uptau ( 1- \uptau)  } \times \mathcal{N} \displaystyle  \left( 0, \frac{  \uptau  (1 -  \uptau ) }{ f_{u_t ( \uptau) }(0)^2  } \times \big( \boldsymbol{\Gamma}_{cxz} \boldsymbol{V}^{-1}_{cxz} \boldsymbol{\Gamma}_{cxz}^{\prime} \big)^{-1} \right) 
\\
&\ \ \ \ \ \ \ \ \ \ \ \ \times
\bigg\{ f_{ u_t ( \uptau )}(0)^2 \times \big( \boldsymbol{\Gamma}_{cxz} \boldsymbol{V}_{cxz}^{-1} \boldsymbol{\Gamma}_{cxz}^{\prime}  \big) \bigg\}
\\
&\ \ \ \ \ \ \ \ \ \ \ \ \ \ \ \ \ \times 
\mathcal{N} \displaystyle  \left( 0, \frac{  \uptau  (1 -  \uptau ) }{ f_{u_t ( \uptau) }(0)^2  } \times \big( \boldsymbol{\Gamma}_{cxz} \boldsymbol{V}^{-1}_{cxz} \boldsymbol{\Gamma}_{cxz}^{\prime} \big)^{-1}  \right)
\end{align*}

\newpage 

\begin{align*}
\mathcal{W}^{ivx-qr}_n (\uptau)
&\Rightarrow 
\frac{ f_{ u_t ( \uptau ) }(0)^2   }{  \uptau ( 1- \uptau)  } \frac{\uptau ( 1- \uptau)}{ f_{ u_t ( \uptau ) }(0)^2 } \times \bigg[ \mathcal{N} \left( 0, \big( \boldsymbol{\Gamma}_{cxz} \boldsymbol{V}^{-1}_{cxz} \boldsymbol{\Gamma}_{cxz}^{\prime} \big)^{-1}  \right)   \bigg]^{\prime}
\\
&\ \ \times
\big( \boldsymbol{\Gamma}_{cxz} \boldsymbol{V}_{cxz}^{-1} \boldsymbol{\Gamma}_{cxz}^{\prime}  \big)^{-1} \times \bigg[ \mathcal{N} \left( 0, \big( \boldsymbol{\Gamma}_{cxz} \boldsymbol{V}^{-1}_{cxz} \boldsymbol{\Gamma}_{cxz}^{\prime} \big)^{-1}  \right) \bigg] 
\\
&= 
\mathcal{N} (0,1) 
\left[ \big( \boldsymbol{\Gamma}_{cxz} \boldsymbol{V}^{-1}_{cxz} \boldsymbol{\Gamma}_{cxz}^{\prime} \big)^{-1/2}  \right]^{\prime} 
\times \big( \boldsymbol{\Gamma}_{cxz} \boldsymbol{V}_{cxz}^{-1} \boldsymbol{\Gamma}_{cxz}^{\prime}  \big)
\times
\left[ \big( \boldsymbol{\Gamma}_{cxz} \boldsymbol{V}^{-1}_{cxz} \boldsymbol{\Gamma}_{cxz}^{\prime} \big)^{-1/2}  \right]
\mathcal{N} (0,1)
\\
&= \big[ \mathcal{N} (0,1) \big]^2 \equiv \chi^2_2.
\end{align*}
which is a standard $\chi^2-$distribution with $2$ degrees of freedom. 
\end{proof}

\subsubsection{Proof of Proposition \ref{proposition1A}}

\begin{proof}
\

\underline{ \textbf{Part (\textit{i})} }
 
First we consider the limit expression for the case of mildly integrated regressors
\begin{small}
\begin{align*}
\mathcal{L} 
&:=
\hat{\mathcal{J}}_n \left( \lambda, \uptau_0, \widehat{ \boldsymbol{\theta} }^{ols}_n( \uptau_0 ) \right) - \lambda \hat{\mathcal{J}}_n \left( 1 , \uptau_0, \widehat{ \boldsymbol{\theta} }^{ols}_n( \uptau_0 ) \right)
\\
&= 
\left( \boldsymbol{D}_n^{-1} \left[ \sum_{t=1}^n \boldsymbol{X}^{\prime}_{t-1} \boldsymbol{X}_{t-1} \right] \boldsymbol{D}_n^{-1} \right)^{- 1 / 2} \left\{ \boldsymbol{D}_n^{-1} \sum_{t=1}^{ \floor{\lambda n} } \boldsymbol{X}_{t-1} \psi_{\uptau} \big( u_t (\uptau_0) \big) - \lambda \boldsymbol{D}_n^{-1} \sum_{t=1}^{n} \boldsymbol{X}_{t-1} \psi_{\uptau} \big( u_t (\uptau_0) \big) \right\}
\\
&\Rightarrow
\left\{ 
\begin{bmatrix}
  1  &  \boldsymbol{0}^{\prime} \\
  \boldsymbol{0}  & \boldsymbol{V}_{xx}
  \end{bmatrix}  \right\}^{-1 / 2}
\left\{ \mathcal{N} \displaystyle  \left( \boldsymbol{0}, \uptau_0 (1 - \uptau_0)   
\lambda 
\begin{bmatrix}
  1  &  \boldsymbol{0}^{\prime} \\
  \boldsymbol{0}  & \boldsymbol{V}_{xx}
  \end{bmatrix} 
\right) - \lambda \mathcal{N} \left( \boldsymbol{0}, \uptau_0 (1 - \uptau_0) 
\begin{bmatrix}
1 & \boldsymbol{0}^{\prime}
\\
\boldsymbol{0} & \boldsymbol{V}_{xx}
\end{bmatrix}  \right) \right\} 
\\
&=
\sqrt{ \uptau_0 (1 - \uptau_0) }
\left\{ 
\begin{bmatrix}
  1  &  \boldsymbol{0}^{\prime} \\
  \boldsymbol{0}  & \boldsymbol{V}_{xx}
  \end{bmatrix}  \right\}^{-1 / 2}
\left\{ 
\mathcal{N} \displaystyle  \left( \boldsymbol{0},    
\lambda   
  \begin{bmatrix}
  1  &  \boldsymbol{0}^{\prime} \\
  \boldsymbol{0}  & \boldsymbol{V}_{xx}
  \end{bmatrix} 
\right) - \lambda \mathcal{N} \left( \boldsymbol{0}, 
\begin{bmatrix}
1 & \boldsymbol{0}^{\prime}
\\
\boldsymbol{0} & \boldsymbol{V}_{xx}
\end{bmatrix}  \right) \right\}
\\
&=
\sqrt{ \uptau_0 (1 - \uptau_0) }
\left\{ 
\begin{bmatrix}
  1  &  \boldsymbol{0}^{\prime} \\
  \boldsymbol{0}  & \boldsymbol{V}_{xx}
  \end{bmatrix}  \right\}^{-1 / 2} \left\{ 
\begin{bmatrix}
  1  &  \boldsymbol{0}^{\prime} \\
  \boldsymbol{0}  & \boldsymbol{V}_{xx}
  \end{bmatrix} \right\}^{1 / 2} \times 
\bigg\{ 
\mathcal{N} \displaystyle  \left( \boldsymbol{0},    
\lambda \boldsymbol{I}_p \right) - \lambda \mathcal{N} \left( \boldsymbol{0}, \boldsymbol{I}_p \right) \bigg\}  
\\
&\equiv 
\sqrt{ \uptau_0 (1 - \uptau_0) } \bigg[ \boldsymbol{W}_p ( \lambda ) - \lambda \boldsymbol{W}_p ( 1 ) \bigg].
\end{align*}
\end{small}
Therefore, for some $0 < \lambda < 1$ and $\uptau_0 \in (0,1)$ 
\begin{align}
\label{BB.limit}
\mathcal{SQ}^{ols}_n ( \lambda ; \uptau_0 ) \Rightarrow 
\underset{ \lambda \in [0,1] }{ \mathsf{sup} } \ \big\| \boldsymbol{ \mathcal{BB} }_p( \lambda) \big\|_{\infty}.
\end{align}
which is a nuisance-parameter free distribution that holds under the null hypothesis. In summary, suppose that the data are generated by the quantile predictive regression model and Assumptions \ref{assumption1}-\ref{assumption2} are satisfied. Then, under the null null hypothesis $\mathcal{H}_0^{(A)}$ the fluctuation type statistics weakly converge to the limiting distribution given by expression \eqref{BB.limit} for mildly integrated regressors for some unknown break-point location $0 < \lambda < 1$.

\newpage 

Second, for local unit root regressors (high persistence) then, following limit holds
\begin{small}
\begin{align}
\mathcal{L} 
&:=
\hat{\mathcal{J}}_n \left( \lambda, \uptau_0, \widehat{ \boldsymbol{\theta} }^{ols}_n( \uptau_0 ) \right) - \lambda \hat{\mathcal{J}}_n \left( 1 , \uptau_0, \widehat{ \boldsymbol{\theta} }^{ols}_n( \uptau_0 ) \right)
\nonumber
\\
&=
\left( \boldsymbol{D}_n^{-1} \left[ \sum_{t=1}^n \boldsymbol{X}^{\prime}_{t-1} \boldsymbol{X}_{t-1} \right] \boldsymbol{D}_n^{-1} \right)^{- 1 / 2} \left\{ \boldsymbol{D}_n^{-1} \sum_{t=1}^{ \floor{\lambda n} } \boldsymbol{X}_{t-1} \psi_{\uptau} \big( u_t (\uptau_0) \big) - \lambda \boldsymbol{D}_n^{-1} \sum_{t=1}^{n} \boldsymbol{X}_{t-1} \psi_{\uptau} \big( u_t (\uptau_0) \big) \right\}
\nonumber
\\
&\Rightarrow
\begin{bmatrix}
  1  &  \displaystyle \int_0^1 \boldsymbol{J}_c(r)^{\prime} \\
  \displaystyle \int_0^1 \boldsymbol{J}_c(r)  &      \displaystyle \int_0^1 \boldsymbol{J}_c(r) \boldsymbol{J}_c(r)^{\prime}
\end{bmatrix}^{-1 / 2}
\times
\left\{ 
\begin{bmatrix}
B_{\psi_{\uptau}}(\lambda) 
\\
\displaystyle \int_0^{\lambda} \boldsymbol{J}_c(r) dB_{ \psi_{ \uptau} } 
\end{bmatrix} 
- \lambda 
\begin{bmatrix}
B_{\psi_{\uptau}}(1) 
\\
\displaystyle \int_0^1 \boldsymbol{J}_c(r) dB_{ \psi_{ \uptau} } 
\end{bmatrix} \right\} 
\nonumber
\\
&\equiv
\begin{bmatrix}
  1  &  \displaystyle \int_0^1 \boldsymbol{J}_c(r)^{\prime} \\
  \displaystyle \int_0^1 \boldsymbol{J}_c(r)  &      \displaystyle \int_0^1 \boldsymbol{J}_c(r) \boldsymbol{J}_c(r)^{\prime}
\end{bmatrix}^{-1 / 2}
\times
\begin{bmatrix}
B_{\psi_{\uptau}}(\lambda) - \lambda B_{\psi_{\uptau}}(1) 
\\
\displaystyle \int_0^{\lambda} \boldsymbol{J}_c(r) dB_{ \psi_{ \uptau} } - \lambda \int_0^{1} \boldsymbol{J}_c(r) dB_{ \psi_{ \uptau} }
\end{bmatrix} 
\nonumber
\\
&\equiv 
\begin{bmatrix}
1  &  \displaystyle \int_0^1 \boldsymbol{J}_c(r)^{\prime}_{ \textcolor{red}{ ( 1 \times p ) } } 
\\
\displaystyle \int_0^1 \boldsymbol{J}_c(r)_{ \textcolor{red}{ ( p \times 1 ) } } & \displaystyle \int_0^1 \boldsymbol{J}_c(r) \boldsymbol{J}_c(r)^{\prime}_{ \textcolor{red}{ ( p \times p ) } }
\end{bmatrix}^{-1 / 2}_{ \textcolor{red}{ (p + 1 ) \times (p + 1 ) } }
\times
\begin{bmatrix}
\mathbf{ \mathcal{BB} }_{\psi_{\uptau}} ( \lambda )_{ \textcolor{red}{ ( 1 \times n ) } }
\\
\displaystyle \mathbf{\mathcal{JB}}_{\psi_{\uptau}} (\lambda)_{ \textcolor{red}{ ( p \times n ) } }
\end{bmatrix}_{ \textcolor{red}{ (p+1) \times n } } 
\end{align} 
\end{small}
where $\mathbf{ \mathcal{BB} }_{\psi_{\uptau}} ( \lambda ) : = B_{\psi_{\uptau}}(\lambda) - \lambda B_{\psi_{\uptau}}(1)$ and $\displaystyle \mathbf{ \mathcal{JB} }_{\psi_{\uptau}} ( \lambda ) : = \int_0^{\lambda} \boldsymbol{J}_c(r) dB_{ \psi_{ \uptau} } - \lambda \int_0^{1} \boldsymbol{J}_c(r) dB_{ \psi_{ \uptau} }$. Thus, under the null hypothesis the OLS based functional with local unit root regressors converges into a nonstandard and nonpivotal limiting distribution. 

\medskip

\underline{ \textbf{Part (\textit{ii})} }
\

The limit expression of the fluctuation type test based on the IVX estimator is given as 
\begin{small}
\begin{align*}
\mathcal{L} 
&:=
\hat{\mathcal{J}}_n \left( \lambda, \uptau_0, \widehat{ \boldsymbol{\beta} }^{ivx}_n( \uptau_0 ) \right) - \lambda \hat{\mathcal{J}}_n \left( 1 , \uptau_0, \widehat{ \boldsymbol{\beta} }^{ivx}_n( \uptau_0 ) \right)
\\
&= 
\left( \sum_{t=1}^n \tilde{\boldsymbol{Z}}_{t-1,n} \boldsymbol{X}^{\prime}_{t-1,n} \right)^{- 1 / 2} \left\{ \sum_{t=1}^{ \floor{\lambda n} } \tilde{\boldsymbol{Z}}_{t-1,n} \psi_{\uptau} \big( u_t (\uptau_0) \big) - \lambda \sum_{t=1}^{n} \tilde{\boldsymbol{Z}}_{t-1,n} \psi_{\uptau} \big( u_t (\uptau_0) \big) \right\}
\\
&=
\left( \tilde{\boldsymbol{D}}_n^{-1} \left[ \sum_{t=1}^n \tilde{\boldsymbol{z}}_{t-1} \boldsymbol{x}^{\prime}_{t-1} \right] \tilde{\boldsymbol{D}}_n^{-1} \right)^{- 1 / 2} \left\{ \tilde{\boldsymbol{D}}_n^{-1} \sum_{t=1}^{ \floor{\lambda n} } \tilde{\boldsymbol{z}}_{t-1} \psi_{\uptau} \big( u_t (\uptau_0) \big) - \lambda \tilde{\boldsymbol{D}}_n^{-1} \sum_{t=1}^{n} \tilde{\boldsymbol{z}}_{t-1} \psi_{\uptau} \big( u_t (\uptau_0) \big) \right\}
\\
&\Rightarrow
\boldsymbol{\Gamma}_{cxz}^{-1 / 2} 
\times
\bigg\{ \mathcal{N} \displaystyle \bigg( \boldsymbol{0}, \uptau_0 (1 - \uptau_0) \lambda \boldsymbol{V}_{cxz} \bigg) 
- \lambda 
\mathcal{N} \bigg( \boldsymbol{0}, \uptau_0 (1 - \uptau_0) \boldsymbol{V}_{cxz} \bigg) \bigg\} 
\\
&=
\sqrt{\uptau_0 (1 - \uptau_0)} \times \boldsymbol{\Gamma}_{cxz}^{-1 / 2} \times
\bigg\{ \mathcal{N} \displaystyle \bigg( \boldsymbol{0}, \lambda \boldsymbol{V}_{cxz} \bigg) 
- \lambda 
\mathcal{N} \bigg( \boldsymbol{0},  \boldsymbol{V}_{cxz} \bigg) \bigg\} 
\equiv  \sqrt{\uptau_0 (1 - \uptau_0)} \big[ \boldsymbol{W}_p ( \lambda ) - \lambda \boldsymbol{W}_p ( 1 ) \big]. 
\end{align*}
\end{small}
provided that $\boldsymbol{\Gamma}_{cxz} \equiv \boldsymbol{V}_{xx}$, which applies when $\upgamma_x \in (0,  \upgamma_z)$. Proposition \ref{proposition1A} \textbf{(\textit{ii})} shows that the limiting distribution of the fluctuation type test is nonstandard in general when the IVX estimator is employed since we employ limit results which hold for both LUR and MI regressors. However, when the coefficient of persistence of regressors has an exponent rate with an absolute value less than the exponent rate of the mildly integrated instruments, then the asymptotic covariance matrix of the Gaussian variant has a simpified form due to the stochastic dominance property of these covariance matrices. 

\newpage

\underline{ \textbf{Part (\textit{iii})} }
\

We obtain the following limit result which holds for both LUR and MI regressors
\begin{small}
\begin{align*}
\mathcal{L} 
&:= \hat{\mathcal{J}}^{ivz}_n \left( \lambda, \uptau_0, \widehat{ \boldsymbol{ \beta} }^{ivz}_n( \uptau_0 ) \right) - \lambda \hat{\mathcal{J}}^{ivz}_n \left( 1, \uptau_0,  \widehat{ \boldsymbol{ \beta} }^{ivz}_n( \uptau_0 ) \right) 
\\
&= 
\big( \tilde{\boldsymbol{Z}}^{\prime} \tilde{\boldsymbol{Z}} \big)^{- 1 / 2}  \left\{ \tilde{\boldsymbol{D}}_n^{-1} \sum_{t=1}^{\floor{ \lambda n} } \tilde{\boldsymbol{z}}_{t-1} \psi_{ \uptau } \left( y_t - \tilde{\boldsymbol{z}}_{t-1}^{\prime} \widehat{\boldsymbol{\beta}}^{ivz}_n(\uptau_0) \right) - \lambda  \tilde{\boldsymbol{D}}_n^{-1}  \sum_{t=1}^{n} \tilde{\boldsymbol{z}}_{t-1} \psi_{ \uptau } \left( y_t -  \tilde{\boldsymbol{z}}_{t-1}^{\prime} \widehat{\boldsymbol{\beta}}^{ivz}_n(\uptau_0) \right) \right\}
\\
&=
\left( \tilde{\boldsymbol{D}}_n^{-1} \left[ \sum_{t=1}^{ n }   \tilde{\boldsymbol{z}}_{t-1} \tilde{\boldsymbol{z}}_{t-1}^{\prime} \right] \tilde{\boldsymbol{D}}_n^{-1} \right)^{- 1 / 2} \left\{  \tilde{\boldsymbol{D}}_n^{-1}  \sum_{t=1}^{\floor{ \lambda n} } \tilde{\boldsymbol{z}}_{t-1} \psi_{ \uptau } \big( y_t - \tilde{\boldsymbol{z}}_{t-1}^{\prime}\boldsymbol{\beta}_0(\uptau_0)  \big) - \lambda  \tilde{\boldsymbol{D}}_n^{-1}  \sum_{t=1}^{n} \boldsymbol{x}_{t-1} \psi_{ \uptau } \big( y_t - \tilde{\boldsymbol{z}}_{t-1}^{\prime}\boldsymbol{\beta}_0(\uptau_0) \big) \right\} 
\\
&\Rightarrow 
\boldsymbol{V} _{cxz}^{- 1 / 2} \times \bigg\{ \mathcal{N} \bigg(  \boldsymbol{0},   \uptau_0 ( 1 - \uptau_0 ) \lambda \boldsymbol{V}_{cxz} \bigg) - \lambda \mathcal{N} \bigg(  \boldsymbol{0}, \uptau_0 ( 1 - \uptau_0 ) \boldsymbol{V}_{cxz} \bigg) \bigg\}
\\
&=
\boldsymbol{V} _{cxz}^{- 1 / 2} \times \sqrt{ \uptau_0 ( 1 - \uptau_0 ) } \times \bigg\{ \mathcal{N} \bigg(  \boldsymbol{0}, \lambda \boldsymbol{V}_{cxz} \bigg) - \lambda \mathcal{N} \bigg(  \boldsymbol{0}, \boldsymbol{V}_{cxz} \bigg) \bigg\}
\\
&=
\sqrt{ \uptau_0 ( 1 - \uptau_0 ) } \ \boldsymbol{V}_{cxz}^{- 1/ 2} \times  \boldsymbol{V}_{cxz}^{ 1/ 2} \times \bigg\{ \mathcal{N} \big(  \boldsymbol{0}, \lambda \boldsymbol{I}_p  \big) - \lambda \mathcal{N} \big(  \boldsymbol{0}, \boldsymbol{I}_p \big) \bigg\}
\\
&=
\sqrt{ \uptau_0 ( 1 - \uptau_0 ) } \bigg[ \boldsymbol{W}_p ( \lambda ) - \lambda \boldsymbol{W}_p ( 1 ) \bigg], \ \ \text{since} \ \boldsymbol{V}_{cxz}^{- 1/ 2} \times  \boldsymbol{V}_{cxz}^{ 1/ 2} = \boldsymbol{I}_p.
\end{align*}
\end{small}
which implies that
\begin{align*}
\frac{1}{ \sqrt{\uptau_0 ( 1- \uptau_0) } } \bigg[ \mathcal{J}_n \big( \lambda, \uptau_0, \widehat{ \boldsymbol{ \beta} }^{ivz}_n( \uptau_0 ) \big) - \lambda \mathcal{J}_n \big( 1 , \uptau_0, \widehat{ \boldsymbol{ \beta} }^{ivz}_n( \uptau_0 ) \big) \bigg]
\Rightarrow
\frac{\sqrt{ \uptau_0 ( 1 - \uptau_0 ) }}{ \sqrt{\uptau_0 ( 1- \uptau_0) } }  \bigg[ \boldsymbol{W}_p ( \lambda ) - \lambda \boldsymbol{W}_p ( 1 ) \bigg]
\end{align*}
Thus, for some $0 < \lambda < 1$ and $\uptau_0 \in (0,1)$ it holds that
\begin{align}
\mathcal{SQ}^{ivz}_n ( \lambda ; \uptau_0 ) \Rightarrow 
\underset{ \lambda \in [0,1] }{ \mathsf{sup} } \ \big\| \boldsymbol{ \mathcal{BB} }_p( \lambda) \big\|_{\infty}
\end{align}
Overall, it seems that the fluctuation type statistics are non-pivotal, at the first glance, for all estimators and across the two persistence classes, due to the dependence of their limiting distributions on the nuisance coefficient of persistence; appearing in the estimation of the covariance matrix $\boldsymbol{V}_{cxz}$ of the Gaussian variant that the corresponding partial sum processes converge to. However, the IVZ based statistic for both types of persistence  induce an approximation which weakly converges into a Brownian bridge type limit. In practise, when we utilize the IVZ estimator (see also Theorem 3.2 and Proposition 3.2 in \cite{lee2016predictive}) then the limiting distribution of the  moment matrix between the covariates and the residuals of the model simplifies, and thus the overall limit is nuisance-parameter free. Furthermore, a similar limit for the IVX based test hold in the case of mildly integrated regressors. A nonstandard limit distribution appears in the case of high persistence (e.g., LUR) for both the OLS and IVX based test statistics. 
\end{proof}

\newpage 

\subsubsection{Proof of Proposition \ref{proposition1}}

We need to prove the following limit result
\begin{align}
\mathcal{ SW }^{ols}_n \big(  \lambda ; \uptau_0 \big) 
&\Rightarrow 
\underset{ \lambda \in [0,1] }{ \mathsf{sup} } \ \frac{ \ \big\| \boldsymbol{ \mathcal{BB} }_{p+1}( \lambda) \ \big\|^2 }{ \lambda( 1 - \lambda)} 
\end{align}
where the exponent rate that captures the degree of persistence for the original regressors can be $\upgamma_x = 1$ or $\upgamma_x \in (0,1)$. In particular to prove the Brownian Bridge limit, we need to derive the limiting distribution of the term $\mathcal{J}_{nx} \big( \lambda, \uptau_0, \widehat{ \boldsymbol{\theta} }^{ols}_n( \uptau_0 ) \big) - \lambda \mathcal{J}_{nx} \big( 1, \uptau_0,  \widehat{ \boldsymbol{\theta} }^{ols}_n( \uptau_0 ) \big)$.

Using Definition \ref{definition3}, for the case of mildly integrated regressors, $\upgamma_x \in (0,1)$, it holds that 
\begin{align}
\mathcal{J}^{ols}_{nx} \big( \lambda, \uptau_0, \widehat{ \boldsymbol{\theta} }^{ols}_n( \uptau_0 ) \big) 
\nonumber
&:= 
\big( \boldsymbol{X}^{\prime} \boldsymbol{X} \big)^{- 1 / 2} \sum_{t=1}^{\floor{ \lambda n} } \boldsymbol{X}_{t-1} \psi_{ \uptau } \left( y_t - \boldsymbol{X}_{t-1}^{\prime} \widehat{ \boldsymbol{\theta} }^{ols}_n( \uptau_0 ) \right)
\\
&\Rightarrow
\sqrt{ \uptau_0 (1 - \uptau_0) } \times \bigg[ \boldsymbol{W}_{p+1} ( \lambda ) - \lambda \boldsymbol{W}_{p+1} ( 1 ) \bigg].
\end{align}
where $\boldsymbol{W}_{p+1} ( . )$ is a $p-$vector of independent Wiener processes and the convergence holds because $\big\{ \boldsymbol{x}_{t-1} \psi_{ \uptau } \big( y_t - \boldsymbol{x}_{t-1}^{\prime} \boldsymbol{\beta}_0(\uptau_0 ) \big) \big\}$ is a sequence of martingale differences under the null. The particular limit is derived in the proof of Proposition \ref{proposition1A}.

\underline{\textbf{Part \textit{(i)}}}

\begin{proof}
Let $\boldsymbol{X}_{t-1} = \big( \boldsymbol{1}, \boldsymbol{x}^{\prime}_{t-1} \big)^{\prime}$ the design matrix and $\boldsymbol{\theta} (\uptau_0)= \big( \alpha(\uptau_0), \boldsymbol{\beta}^{\prime}(\uptau_0) \big)^{\prime}$ to be the parameter vector. Then, to derive the asymptotic distribution of the OLS-Wald test we employ the following functional
\begin{small}
\begin{align}
\hat{S}_{nx}^{ols} \big( \lambda, \uptau_0 , \widehat{\boldsymbol{\theta}}^{ols}_n ( \uptau_0 ) \big) 
:=  
\boldsymbol{D}_n^{-1} \sum_{t=1}^{\floor{ \lambda n} } \boldsymbol{X}_{t-1} \psi_{ \uptau } \big( u_t ( \uptau_0 ) \big), \ \ \text{for some} \ \ 0 < \lambda < 1,
\end{align}
\end{small}
where $u_t ( \uptau_0 ) = \big( y_t -  \boldsymbol{X}_{t-1}^{\prime} \widehat{\boldsymbol{\theta}}^{ols}_n( \uptau_0 ) \big)$ with $\uptau_0 \in (0,1)$. Then, the OLS based estimator for the subsample $1 \leq t \leq \floor{ \lambda n }$ denoted with $ \widehat{ \boldsymbol{\theta}}^{ols}_{1} (\lambda ; \uptau_0)$ is given by
\begin{small}
\begin{align}
\widehat{ \boldsymbol{\theta}}^{ols}_{1} (\lambda ; \uptau_0) 
&= 
\left( \frac{1}{ k } \sum_{t=1}^{\floor{\lambda n}} \textcolor{red}{ f_{ u_t(\uptau), t-1}(0) } \boldsymbol{X}_{t-1} \boldsymbol{X}_{t-1}^{\prime} \right)^{-1} \left( \frac{1}{k} \sum_{t=1}^{\floor{\lambda n}} \boldsymbol{X}_{t-1} y_{t}  \right) 
\end{align}
\end{small}
Therefore, it holds that 
\begin{small}
\begin{align}
\bigg\| \hat{S}_{nx}^{ols} \left( \lambda, \uptau_0, \widehat{ \boldsymbol{\theta}}^{ols}_{1} (\lambda ; \uptau_0) \right) \bigg\| 
&\leq 
\bigg\| \ \boldsymbol{D}_n^{-1} \sum_{t=1}^{ \floor{ \lambda n } } \boldsymbol{X}_{t-1} \bigg[ \mathds{1} \big\{ y_t = \boldsymbol{X}_{t-1}^{\prime} \widehat{\boldsymbol{\theta}}^{ols}_{1}( \lambda ; \uptau_0 ) \big\} \bigg] \ \bigg\|
\nonumber
\\
&\leq
(p+1) D_n^{- 1} \underset{ 1 \leq i \leq n }{ \mathsf{max} } \norm{ \boldsymbol{X}_{t-1} } \overset{ \mathbb{P} }{ \to } 0. 
\end{align}
\end{small}
which implies that $\hat{S}_{nx}^{ols} \left( \lambda, \uptau_0, \widehat{ \boldsymbol{\theta} }^{ols}_{1} (\lambda ; \uptau_0) \right) = o_{ \mathbb{P} }(1)$.

\newpage 

Moreover, for the estimator of the first subsample we obtain the following expression 
\begin{small}
\begin{align}
\label{term11}
\boldsymbol{D}_n \left( \widehat{\boldsymbol{\theta}}_{1}^{ols} ( \lambda ; \uptau_0 ) - \boldsymbol{\theta}_0 ( \uptau_0 )  \right)
&\overset{ p }{ \to } 
\bigg\{ \textcolor{red}{ f_{ u_t(\uptau)}(0) }  \lambda \mathbb{V}_{xx} \bigg\}^{-1} \times S_{nx}^{ols}  \big( \lambda, \uptau_0, \boldsymbol{\theta}_0 ( \uptau_0 ) \big) + o_{ \mathbb{P} }(1)
\nonumber
\\
&\equiv
\textcolor{red}{  \frac{ 1 }{ f_{ u_t(\uptau) }(0) } }
\frac{1}{\lambda} \mathbb{V}_{xx}^{-1} \times \mathcal{N} \displaystyle  \left( \boldsymbol{0}, \uptau_0 (1 - \uptau_0) 
 \lambda  
  \begin{bmatrix}
  1  &  \boldsymbol{0}^{\prime} \\
  \boldsymbol{0}  & \boldsymbol{V}_{xx}
  \end{bmatrix} 
\right)
\nonumber
\\
&=
\textcolor{red}{  \frac{ 1 }{ f_{ u_t(\uptau) }(0) } }
\frac{1}{\lambda} \sqrt{ \uptau_0 (1 - \uptau_0) } \mathbb{V}_{xx}^{-1} \times
\begin{bmatrix}
  1  &  \boldsymbol{0}^{\prime} \\
  \boldsymbol{0}  & \boldsymbol{V}_{xx}
\end{bmatrix}^{1 / 2}  \times  
\mathcal{N} \displaystyle \big( \boldsymbol{0}, \lambda  \big)
\nonumber
\\
&=
\textcolor{red}{  \frac{ 1 }{ f_{ u_t(\uptau) }(0) } } \frac{1}{\lambda} \sqrt{ \uptau_0 (1 - \uptau_0) }  \times \mathbb{V}_{xx}^{- 1/2} \times \boldsymbol{W}_{p+1} ( \lambda ).
\end{align}
\end{small}
since it holds that $S_{x}^{ols} \big( \lambda, \uptau_0 , \boldsymbol{\theta} ( \uptau_0 ) \big) 
\equiv
\mathcal{N} \displaystyle  \left( \boldsymbol{0}, \uptau_0 (1 - \uptau_0) 
 \lambda  
  \begin{bmatrix}
  1  &  \boldsymbol{0}^{\prime} \\
  \boldsymbol{0}  & \boldsymbol{V}_{xx}
  \end{bmatrix} 
\right)$. 
Similarly, 
\begin{small}
\begin{align}
\label{term12}
\boldsymbol{D}_n \left( \widehat{\boldsymbol{\theta}}^{ols}_{2} ( \lambda ; \uptau_0 ) - \boldsymbol{\theta}_0 (\uptau_0) \right)
&\overset{ p }{ \to }  
\bigg\{ \textcolor{red}{ f_{ u_t(\uptau)}(0) }  ( 1 - \lambda ) \mathbb{V}_{xx} \bigg\}^{-1} \times S_{nx}^{ols}  \big( \lambda, \uptau_0 , \boldsymbol{\theta}_0 (\uptau_0) \big) + o_{ \mathbb{P} }(1)
\nonumber
\\
&\equiv 
\textcolor{red}{  \frac{ 1 }{ f_{ u_t(\uptau) }(0) } } \frac{1}{1 - \lambda} \sqrt{ \uptau_0 (1 - \uptau_0) } \times \mathbb{V}_{xx}^{- 1/2} \times \bigg[ \boldsymbol{W}_{p+1}(1 ) - \boldsymbol{W}_{p+1}( \lambda ) \bigg]. 
\end{align}
\end{small}
Therefore, combining \eqref{term11} and \eqref{term12} we obtain the following expression 
\begin{small}
\begin{align*}
\boldsymbol{D}_n \left[ \Delta \widehat{\boldsymbol{\theta}}^{ols}_n \left( \lambda ; \uptau_0 \right) \right]
\Rightarrow 
- \textcolor{red}{  \frac{ 1 }{ f_{ u_t(\uptau) }(0) } } \frac{1}{ \lambda ( 1 - \lambda )} \sqrt{ \uptau_0 (1 - \uptau_0) } \times \mathbb{V}_{xx}^{- 1/2}  \times \big[ \boldsymbol{W}_{p+1}( \lambda ) - \lambda \boldsymbol{W}_{p+1}( 1 ) \big].
\end{align*}
\end{small}
Moreover, the convergence of the covariance matrix follows as below
\begin{small}
\begin{align}
\underset{ n \to \infty }{ \mathsf{plim} }  \widehat{\boldsymbol{V}}^{ols}_{n}( \lambda ; \uptau_0 ) 
\equiv 
\uptau_0 (1 - \uptau_0) \times \bigg\{ \underset{ n \to \infty }{ \mathsf{plim} } \widehat{\boldsymbol{V}}^{ols}_{1n}( \lambda; \uptau_0 ) 
+ 
\underset{ n \to \infty }{ \mathsf{plim} }  \widehat{\boldsymbol{V}}^{ols}_{2n}( \lambda; \uptau_0 ) \bigg\}
\end{align}
\end{small}
Therefore, it holds that
\begin{small}
\begin{align}
\underset{ n \to \infty }{ \mathsf{plim} }  &\widehat{\boldsymbol{V}}^{ols}_{1n}( \lambda ; \uptau_0 ) 
\\
&= 
\left\{ \underset{n \to \infty}{ \mathsf{plim} }  \textcolor{blue}{ \tilde{\boldsymbol{L}}_{nx} \left( \uptau_0, \boldsymbol{\theta}_n^{ols}(\uptau_0) \right)} \right\}^{-1} \left\{ \underset{ n \to \infty }{ \mathsf{plim} }  \boldsymbol{D}_n^{-1} \left[ \sum_{t=1}^{ \floor{\lambda n } } \boldsymbol{X}_{t-1} \boldsymbol{X}_{t-1}^{\prime} \right] \boldsymbol{D}_n^{-1}  \right\} \left\{ \underset{n \to \infty}{ \mathsf{plim} }  \textcolor{blue}{ \tilde{\boldsymbol{L}}_{nx} \left( \uptau_0, \boldsymbol{\theta}_n^{ols}(\uptau_0) \right)} \right\}^{-1}
\nonumber
\\
&=
\left\{ \underset{ n \to \infty }{ \mathsf{plim} }  \textcolor{blue}{\boldsymbol{D}_n^{-1} \left[ \sum_{t=1}^{ \floor{\lambda n } }   f_{u_t (\uptau), t-1} (0) \boldsymbol{X}_{t-1} \boldsymbol{X}_{t-1}^{\prime} \right] \boldsymbol{D}_n^{-1}  } \right\}^{-1} 
\left\{ \underset{ n \to \infty }{ \mathsf{plim} }  \boldsymbol{D}_n^{-1} \left[ \sum_{t=1}^{ \floor{\lambda n } }  \boldsymbol{X}_{t-1} \boldsymbol{X}_{t-1}^{\prime} \right] \boldsymbol{D}_n^{-1}  \right\} 
\nonumber
\\
&\times
\left\{ \underset{ n \to \infty }{ \mathsf{plim} }  \textcolor{blue}{\boldsymbol{D}_n^{-1} \left[ \sum_{t=1}^{ \floor{\lambda n } }   f_{u_t (\uptau), t-1} (0) \boldsymbol{X}_{t-1} \boldsymbol{X}_{t-1}^{\prime} \right] \boldsymbol{D}_n^{-1}  } \right\}^{-1} 
\nonumber
\\
&\equiv 
\bigg\{  \textcolor{red}{ f_{u_t (\uptau)}(0)} \lambda \mathbb{V}_{xx} \bigg\}^{-1} \times \bigg\{ \lambda \mathbb{V}_{xx} \bigg\}  \times \bigg\{ \textcolor{red}{ f_{u_t (\uptau)}(0)} \mathbb{V}_{xx} \bigg\}^{-1} 
\nonumber
\\
&= 
\textcolor{red}{ \frac{1}{ f_{u_t (\uptau)}(0)^2} } \frac{1}{ \lambda}  \mathbb{V}_{xx}^{-1}
\end{align}
\end{small}

\newpage 

Similarly, we obtain that 
\begin{align}
\underset{ n \to \infty }{ \mathsf{plim} }  &\widehat{\boldsymbol{V}}^{ols}_{2n}( \lambda ; \uptau_0 ) \Rightarrow \textcolor{red}{ \frac{1}{ f_{u_t (\uptau)}(0)^2} } \frac{1}{ 1 - \lambda}  \mathbb{V}_{xx}^{-1}
\end{align}
Thus, 
\begin{align}
\underset{ n \to \infty }{ \mathsf{plim} }  \widehat{\boldsymbol{V}}^{ols}_{n}( \lambda ; \uptau_0 ) 
\equiv  
\textcolor{red}{ \frac{\uptau_0( 1- \uptau_0) }{ f_{u_t (\uptau)}(0)^2} } \mathbb{V}_{xx}^{-1} \left\{ \frac{1}{\lambda} + \frac{1}{1- \lambda} \right\}
=
\textcolor{red}{ \frac{1}{ f_{u_t (\uptau)}(0)^2} } \frac{ \uptau_0( 1- \uptau_0) }{\lambda(1- \lambda)} \mathbb{V}_{xx}^{-1}
\end{align}
which implies that 
\begin{align}
\mathcal{W}_n^{ols} \left( \lambda; \uptau_0 \right) 
&:=   
\boldsymbol{D}_n  
\bigg\{ \Delta \widehat{ \boldsymbol{\theta} }^{ols}_n ( \lambda ; \uptau_0 ) \bigg\}^{\prime} \bigg[ \widehat{\boldsymbol{V}}_n( \lambda ; \uptau_0 ) \bigg]^{-1} \bigg\{ \Delta \widehat{ \boldsymbol{\theta} }^{ols}_n ( \lambda ; \uptau_0 ) \bigg\} 
\nonumber
\\
&\Rightarrow
\textcolor{red}{ \frac{1}{ f_{u_t (\uptau)}(0)^2} } 
\frac{ \uptau_0 ( 1 - \uptau_0 )  }{ \left[ \lambda ( 1 - \lambda ) \right]^2 } \big[ \boldsymbol{W}_{p+1}( \lambda ) - \lambda \boldsymbol{W}_{p+1}( 1 ) \big]^{\prime} \left( \mathbb{V}_{xx}^{- 1/2} \right)^{\prime} \times \left\{  \textcolor{red}{ f_{u_t (\uptau)}(0)^2 }  \frac{ \lambda ( 1 - \lambda )  }{\uptau_0 ( 1 - \uptau_0 )  } \mathbb{V}_{xx} \right\} 
\nonumber
\\
& \ \ \ \ \ \ \ \ \times \mathbb{V}_{xx}^{- 1/2}  \big[ \boldsymbol{W}_{p+1}( \lambda ) - \lambda \boldsymbol{W}_{p+1}( 1 ) \big]
\nonumber
\\
&\equiv
\frac{1}{ \lambda ( 1 - \lambda )  } \big[ \boldsymbol{W}_{p+1}( \lambda ) - \lambda \boldsymbol{W}_{p+1}( 1 ) \big]^{\prime} \big[ \boldsymbol{W}_{p+1}( \lambda ) - \lambda \boldsymbol{W}_{p+1}( 1 ) \big].
\end{align}
Hence, we have that 
\begin{align}
\mathcal{SW}^{ols}_n ( \lambda ; \uptau_0 ) 
&:= 
\underset{ \lambda \in \Lambda_{\eta} }{ \mathsf{sup} } \ \boldsymbol{D}_n  \left\{ \Delta \widehat{ \boldsymbol{\theta} }^{ols}_n ( \lambda ; \uptau_0 ) \right\}^{\prime} \times \bigg[ \widehat{\boldsymbol{V}}_n( \lambda ; \uptau_0 ) \bigg]^{-1} \times \left\{ \Delta \widehat{ \boldsymbol{\theta} }^{ols}_n ( \lambda ; \uptau_0 ) \right\}
\nonumber
\\
&\Rightarrow
\underset{ \lambda \in \Lambda_{\eta} }{ \mathsf{sup} } \
\frac{ \big[ \boldsymbol{W}_{p+1}( \lambda ) - \lambda \boldsymbol{W}_{p+1}( 1 ) \big]^{\prime} \big[ \boldsymbol{W}_{p+1}( \lambda ) - \lambda \boldsymbol{W}_{p+1}( 1 ) \big] }{ \lambda ( 1 - \lambda ) } 
\nonumber
\\
&\equiv 
\underset{ \lambda \in \Lambda_{\eta} }{ \mathsf{sup} } \
\frac{ \big\| \boldsymbol{ \mathcal{BB} }_{p+1}( \lambda ) \big\|^2 }{ \lambda ( 1 - \lambda )  }, \ \text{for} \ \upgamma_x \in (0,1),
\end{align}  
where $\boldsymbol{ \mathcal{BB} }_{p+1}( \lambda)$ is a Brownian Bridge process, which holds for the case of mildly integrated regressors, that is, $\upgamma_x \in (0,1)$ and holds even under the presence of model intercept. Next, we provide of some auxiliary derivations employed for Part \textbf{\textit{(i)}}.
\end{proof}

\begin{proof}
\begin{align*}
\widehat{ \boldsymbol{\theta}}^{ols}_{1} (\lambda ; \uptau_0) 
&= 
\left( \frac{1}{\kappa} \sum_{t=1}^{\floor{\lambda n}} \boldsymbol{X}_{t-1} \boldsymbol{X}_{t-1}^{\prime} \right)^{-1} \left( \frac{1}{\kappa} \sum_{t=1}^{\floor{\lambda n}} \boldsymbol{X}_{t-1} y_{t} \right)
\\
&= 
\left( \frac{1}{\kappa} \sum_{t=1}^{\floor{\lambda n}} \textcolor{red}{ f_{ u_t(\uptau), t-1}(0) } \boldsymbol{X}_{t-1} \boldsymbol{X}_{t-1}^{\prime} \right)^{-1} \left( \frac{1}{\kappa} \sum_{t=1}^{\floor{\lambda n}}  \boldsymbol{X}_{t-1} \big[  \boldsymbol{X}_{t-1}^{\prime} \boldsymbol{\theta}_0( \uptau_0 ) + \psi_{ \uptau } \big( u_t ( \uptau_0 ) \big) \big] \right) 
\\
&=
\boldsymbol{\theta}_0( \uptau_0 ) + \left( \frac{1}{\kappa} \sum_{t=1}^{\floor{\lambda n}} \textcolor{red}{ f_{ u_t(\uptau), t-1}(0) } \boldsymbol{X}_{t-1} \boldsymbol{X}_{t-1}^{\prime} \right)^{-1}
\left( \frac{1}{\kappa} \sum_{t=1}^{\floor{\lambda n}} \boldsymbol{X}_{t-1} \psi_{ \uptau } \big( u_t ( \uptau_0 ) \big) \right) + o_{ \mathbb{P} }(1).
\end{align*}

\newpage 

Thus, 
\begin{align*}
\boldsymbol{D}_n \left( \widehat{ \boldsymbol{\theta}}^{ols}_{1} (\lambda ; \uptau_0)  -  \boldsymbol{\theta}_0( \uptau_0 ) \right)
= 
\left( \boldsymbol{D}_n^{-1}  \left[ \sum_{t=1}^{\floor{\lambda n}} \boldsymbol{X}_{t-1} \boldsymbol{X}_{t-1}^{\prime} \right] \boldsymbol{D}_n^{-1}  \right)^{-1} \left( \boldsymbol{D}_n^{-1} \sum_{t=1}^{\floor{\lambda n}} \boldsymbol{X}_{t-1}  \psi_{ \uptau } \big( u_t ( \uptau_0 ) \big) \right) + o_{ \mathbb{P} }(1).
\end{align*} 
which implies that
\begin{align}
\boldsymbol{D}_n \left( \widehat{ \boldsymbol{\theta}}^{ols}_{1} (\lambda ; \uptau_0) - \boldsymbol{\theta}_0( \uptau_0 ) \right) 
&\Rightarrow 
\big\{ \lambda \mathbb{V}_{xx} \big\}^{-1} \times \mathcal{N} \displaystyle \big( \boldsymbol{0}, \uptau_0 (1 - \uptau_0) \lambda \mathbb{V}_{xx} \big) 
\nonumber
\\
&\equiv 
\frac{1}{\lambda} \sqrt{\uptau_0 ( 1 - \uptau_0)} \mathbb{V}_{xx}^{-1} \times \mathbb{V}_{xx}^{1/2} \times \boldsymbol{W}_{p+1} ( \lambda )
\nonumber
\\
&= 
\frac{1}{\lambda} \sqrt{\uptau_0 ( 1 - \uptau_0)} \times \mathbb{V}_{xx}^{- 1/2} \times \boldsymbol{W}_{p+1}( \lambda ).
\end{align}
Recall that
\begin{align*}
S_{nx}^{ols} \big( \lambda, \uptau_0 , \boldsymbol{\theta}^{ols}_n ( \uptau_0 ) \big) 
:=  
\boldsymbol{D}_n^{-1} \sum_{t=1}^{\floor{ \lambda n} } \boldsymbol{X}_{t-1} \psi_{ \uptau } \big( u_t ( \uptau_0 ) \big), \ \ \text{for} \ ( \lambda, \uptau_0 )\in (0,1),
\end{align*}
where $u_t ( \uptau_0 ) = \big( y_t -  \boldsymbol{X}_{t-1}^{\prime} \boldsymbol{\theta}_n( \uptau_0 ) \big)$. Thus, $
S_{nx}^{ols} \big( \lambda, \uptau_0 , \boldsymbol{\theta}^{ols}_n ( \uptau_0 ) \big) 
\Rightarrow
\mathcal{N} \displaystyle  \left( \boldsymbol{0}, \uptau_0 (1 - \uptau_0) 
 \lambda  
  \begin{bmatrix}
  1  &  \boldsymbol{0}^{\prime} \\
  \boldsymbol{0}  & \boldsymbol{V}_{xx}
  \end{bmatrix} 
\right)$.
Denote with $\mathbb{V}_{xx} := \begin{bmatrix}
  1  &  \boldsymbol{0}^{\prime} \\
  \boldsymbol{0}  & \boldsymbol{V}_{xx}
  \end{bmatrix}$, which implies that 
\begin{align*}
S_{nx}^{ols} \big( \lambda, \uptau_0 , \boldsymbol{\theta}^{ols}_n ( \uptau_0 ) \big) 
\Rightarrow
\mathcal{N} \displaystyle \big( \boldsymbol{0}, \uptau_0 (1 - \uptau_0) \lambda \mathbb{V}_{xx} \big) 
\equiv 
\sqrt{\uptau_0 (1 - \uptau_0)} \times \mathbb{V}_{xx}^{- 1/2} \times\boldsymbol{W}_{p+1} ( \lambda ).
\end{align*}  
Similarly, we can prove that 
\begin{align*}
\boldsymbol{D}_n \left( \widehat{ \boldsymbol{\theta}}^{ols}_{2} (\lambda ; \uptau_0) - \boldsymbol{\theta}_0( \uptau_0 ) \right) 
&\Rightarrow 
\frac{1}{1 - \lambda} \sqrt{\uptau_0 ( 1 - \uptau_0)} \times \mathbb{V}_{xx}^{- 1/2} \times \big[ \boldsymbol{W}_{p+1} (1) -\boldsymbol{W}_{p+1} ( \lambda ) \big].
\end{align*}
Practically, the above results can be deduced from Assumption \ref{assumption2} $(b)$ such that 
\begin{align}
\underset{ r \in [0, \lambda] }{ \mathsf{sup} } \left| \frac{1}{ n^{ 1 + \upgamma_x } } \sum_{t=1}^{ \floor{ \lambda n } } \boldsymbol{x}_{t-1} \boldsymbol{x}_{t-1}^{\prime} - r \mathbb{E} \big[ \boldsymbol{x}_{t-1} \boldsymbol{x}_{t-1}^{\prime} \big] \right| = o_{ \mathbb{P} }(1), \ \text{as} \ n \to \infty.
\end{align}
where $\mathbb{E} \big[ \boldsymbol{x}_{t-1} \boldsymbol{x}_{t-1}^{\prime} \big] \equiv \boldsymbol{V}_{xx}$, which implies that 
$\displaystyle \frac{1}{ n^{ 1 + \upgamma_x } } \sum_{t=1}^{ \floor{ \lambda n } } \boldsymbol{x}_{t-1} \boldsymbol{x}_{t-1}^{\prime} \overset{ p }{\to} \lambda \boldsymbol{V}_{xx}$. Similarly, 
\begin{align}
\underset{ r \in (\lambda, 1 ] }{ \mathsf{sup} } \left| \frac{1}{ n^{ 1 + \upgamma_x } } \sum_{t = \floor{ \lambda n } + 1}^{n} \boldsymbol{x}_{t-1} \boldsymbol{x}_{t-1}^{\prime} - r \mathbb{E} \big[ \boldsymbol{x}_{t-1} \boldsymbol{x}_{t-1}^{\prime} \big] \right| = o_{ \mathbb{P} }(1), \ \text{as} \ n \to \infty.
\end{align}
which implies that $\displaystyle \frac{1}{ n^{ 1 + \upgamma_x } } \sum_{t = \floor{ \lambda n } + 1 }^{n} \boldsymbol{x}_{t-1} \boldsymbol{x}_{t-1}^{\prime} \overset{ p }{\to} ( 1 - \lambda) \boldsymbol{V}_{xx}$.

\end{proof}

\newpage

\underline{\textbf{Part \textit{(ii)}}}

Next, we investigate the asymptotic behaviour of the OLS-Wald test statistic in the case of local unit root regressors (i.e., high persistent). To minimize complexity of notation for the derivations of this proof we denote with
\begin{small}
\begin{align}
\mathbb{S}_{xx} :=
\begin{bmatrix}
1  &  \displaystyle \int_0^1 \boldsymbol{J}_c(r)^{\prime} 
\\
\displaystyle \int_0^1 \boldsymbol{J}_c(r)  &      \displaystyle \int_0^1 \boldsymbol{J}_c(r) \boldsymbol{J}_c(r)^{\prime}
\end{bmatrix}  \ \ \text{and} \ \ 
\mathbb{S}_{xx}(\lambda) :=
\begin{bmatrix}
\lambda  &  \displaystyle \int_0^{\lambda} \boldsymbol{J}_c(r)^{\prime} 
\\
\displaystyle \int_0^{\lambda} \boldsymbol{J}_c(r)  &      \displaystyle \int_0^{\lambda} \boldsymbol{J}_c(r) \boldsymbol{J}_c(r)^{\prime}
\end{bmatrix}
\end{align}
\end{small}
Moreover, it holds that $\displaystyle \left( \boldsymbol{D}_n^{-1}  \left[ \sum_{t=1}^{n} \boldsymbol{X}_{t-1} \boldsymbol{X}_{t-1}^{\prime} \right] \boldsymbol{D}_n^{-1}  \right) 
\Rightarrow \mathbb{S}_{xx}$ and
\begin{align}
\left( \boldsymbol{D}_n^{-1}  \left[ \sum_{t=1}^{ \floor{\lambda n }  } \boldsymbol{X}_{t-1} \boldsymbol{X}_{t-1}^{\prime} \right] \boldsymbol{D}_n^{-1}  \right) 
\Rightarrow
\begin{bmatrix}
\lambda  &  \displaystyle \int_0^{\lambda} \boldsymbol{J}_c(r)^{\prime} 
\\
\displaystyle \int_0^{\lambda} \boldsymbol{J}_c(r)  &      \displaystyle \int_0^{\lambda} \boldsymbol{J}_c(r) \boldsymbol{J}_c(r)^{\prime}
\end{bmatrix}, \text{when} \ \upgamma_x = 1.
\end{align}

\begin{proof}
Therefore, for LUR regressors, $\upgamma_x = 1$, we have that 
\begin{small}
\begin{align*}
\boldsymbol{D}_n \left( \widehat{ \boldsymbol{\theta}}^{ols}_{1} (\lambda ; \uptau_0)  -  \boldsymbol{\theta}_0( \uptau_0 ) \right)
= 
\left( \boldsymbol{D}_n^{-1}  \left[ \sum_{t=1}^{\floor{\lambda n}} \boldsymbol{X}_{t-1} \boldsymbol{X}_{t-1}^{\prime} \right] \boldsymbol{D}_n^{-1}  \right)^{-1} \left( \boldsymbol{D}_n^{-1} \sum_{t=1}^{\floor{\lambda n}} \boldsymbol{X}_{t-1}  \psi_{ \uptau } \big( u_t ( \uptau_0 ) \big) \right) + o_{ \mathbb{P} }(1)
\end{align*}
\end{small} 
which implies that
\begin{align}
\label{term111}
\boldsymbol{D}_n \left( \widehat{ \boldsymbol{\theta}}^{ols}_{1} (\lambda ; \uptau_0) - \boldsymbol{\theta}_0( \uptau_0 ) \right) 
&\Rightarrow 
\displaystyle \mathbb{S}_{xx}^{-1}  (\lambda) \times 
\begin{bmatrix}
B_{\psi_{\uptau}}(\lambda) 
\\
\displaystyle \int_0^{\lambda} \boldsymbol{J}_c(r) dB_{ \psi_{ \uptau} } 
\end{bmatrix}
\\
\nonumber
\\
\label{term112}
\boldsymbol{D}_n \left( \widehat{ \boldsymbol{\theta}}^{ols}_{2} (\lambda ; \uptau_0) - \boldsymbol{\theta}_0( \uptau_0 ) \right) 
&\Rightarrow 
\big( \displaystyle \mathbb{S}_{xx}(1) - \mathbb{S}_{xx}(\lambda)  \big)^{-1}
\times
\begin{bmatrix}
B_{\psi_{\uptau}}(1)  -  B_{\psi_{\uptau}}(\lambda) 
\\
\displaystyle \int_0^{1} \boldsymbol{J}_c(r) dB_{ \psi_{ \uptau} }  - \displaystyle  \int_0^{\lambda} \boldsymbol{J}_c(r) dB_{ \psi_{ \uptau} }
\end{bmatrix}
\end{align}

\newpage 

Therefore, combining \eqref{term111} and \eqref{term112} we obtain the following expression 
\begin{align}
&\boldsymbol{D}_n \left[ \Delta \widehat{\boldsymbol{\theta}}^{ols}_n \left( \lambda; \uptau_0 \right) \right]
\nonumber
\\
&\Rightarrow 
\begin{bmatrix}
\displaystyle \big[ \mathbb{S}_{xx}(1) - \mathbb{S}_{xx}(\lambda) \big]^{-1} \times \bigg\{ B_{\psi_{\uptau}}(1) -  B_{\psi_{\uptau}}(\lambda) \bigg\} - \mathbb{S}^{-1}_{xx}(\lambda) \times  B_{\psi_{\uptau}}(\lambda)
\\
\displaystyle \big[ \mathbb{S}_{xx}(1) - \mathbb{S}_{xx}(\lambda)  \big]^{-1} \times \bigg\{ \int_0^{1} \boldsymbol{J}_c(r) dB_{ \psi_{ \uptau} } -  \int_0^{\lambda} \boldsymbol{J}_c(r) dB_{ \psi_{ \uptau} } \bigg\} - \mathbb{S}^{-1}_{xx}(\lambda) \times \int_0^{\lambda} \boldsymbol{J}_c(r) dB_{ \psi_{ \uptau} }
\end{bmatrix}
\nonumber
\\
&\equiv
\big[ \mathbb{S}_{xx}(1) - \mathbb{S}_{xx}(\lambda) \big]^{-1}
\times
\begin{bmatrix}
\displaystyle B_{\psi_{\uptau}}(1) - B_{\psi_{\uptau}}(\lambda) 
\\
\displaystyle \int_0^{1} \boldsymbol{J}_c(r) dB_{ \psi_{ \uptau} } -  \int_0^{\lambda} \boldsymbol{J}_c(r) dB_{ \psi_{ \uptau} } 
\end{bmatrix}
- 
\mathbb{S}^{-1}_{xx}(\lambda)
\times
\begin{bmatrix}
\displaystyle B_{\psi_{\uptau}}(\lambda) 
\\
\displaystyle \int_0^{\lambda} \boldsymbol{J}_c(r) dB_{ \psi_{ \uptau} } 
\end{bmatrix}
\end{align}
Then, the OLS-Wald test for testing the null hypothesis of no parameter instability in the nonstationary quantile predictive regression model at an unknown break-point location $\kappa = \floor{ \lambda n }$ is given by the following expression 
\begin{align}
\mathcal{W}_n^{ols} \left( \lambda; \uptau_0 \right) :=   \uptau_0 ( 1 - \uptau_0) \boldsymbol{D}_n  
\bigg\{ \Delta \widehat{ \boldsymbol{\theta} }^{ols}_n ( \lambda ; \uptau_0 ) \bigg\}^{\prime} \bigg[ \widehat{\boldsymbol{V}}_n( \lambda ; \uptau_0 ) \bigg]^{-1} \bigg\{ \Delta \widehat{ \boldsymbol{\theta} }^{ols}_n ( \lambda ; \uptau_0 ) \bigg\} 
\end{align}
Furthermore, we study the limiting variance of the OLS-Wald test under the null hypothesis. In particular, the convergence of the covariance matrix follows as below
\begin{align}
\underset{ n \to \infty }{ \mathsf{plim} } \ &\bigg\{ \widehat{\boldsymbol{V}}^{ols}_{1n}( \lambda ; \uptau_0 ) \bigg\}
\nonumber
\\
&\equiv 
\left\{ \underset{ n \to \infty }{ \mathsf{plim} } \ \textcolor{blue}{ \boldsymbol{L}_{nx} \left( \uptau_0, \boldsymbol{\theta}_n^{ols} (\uptau_0) \right)} \right\}^{-1} \left\{ \underset{ n \to \infty }{ \mathsf{plim} } \ \boldsymbol{D}_n^{-1} \left[ \sum_{t=1}^{ \floor{\lambda n } } \boldsymbol{X}_{t-1} \boldsymbol{X}_{t-1}^{\prime} \right] \boldsymbol{D}_n^{-1}  \right\} \left\{ \underset{ n \to \infty }{ \mathsf{plim} } \ \textcolor{blue}{ \boldsymbol{L}_{nx} \left( \uptau_0, \boldsymbol{\theta}_n^{ols} (\uptau_0) \right)} \right\}^{-1}
\nonumber
\\
&=
\left\{ \underset{ n \to \infty }{ \mathsf{plim} } \ \textcolor{blue}{\boldsymbol{D}_n^{-1} \left[ \sum_{t=1}^{ \floor{\lambda n } }   f_{u_t (\uptau), t-1} (0) \boldsymbol{X}_{t-1} \boldsymbol{X}_{t-1}^{\prime} \right] \boldsymbol{D}_n^{-1}  } \right\}^{-1} 
\times
\left\{ \underset{ n \to \infty }{ \mathsf{plim} } \ \boldsymbol{D}_n^{-1} \left[ \sum_{t=1}^{ \floor{\lambda n } }  \boldsymbol{X}_{t-1} \boldsymbol{X}_{t-1}^{\prime} \right] \boldsymbol{D}_n^{-1}  \right\} 
\nonumber
\\
&\times 
\left\{ \underset{ n \to \infty }{ \mathsf{plim} } \ \textcolor{blue}{  \boldsymbol{D}_n^{-1} \left[ \sum_{t=1}^{ \floor{\lambda n } }  f_{u_t (\uptau), t-1} (0) \boldsymbol{X}_{t-1} \boldsymbol{X}_{t-1}^{\prime} \right] \boldsymbol{D}_n^{-1}  } \right\}^{-1}
\nonumber
\\
&\equiv 
\bigg\{ \textcolor{red}{ f_{ u_t(\uptau)}(0) } \mathbb{S}_{xx}(\lambda) \bigg\}^{-1} \times \mathbb{S}_{xx}(\lambda) \times \bigg\{ \textcolor{red}{ f_{ u_t(\uptau)}(0) }  \mathbb{S}_{xx}(\lambda) \bigg\}^{-1} 
\nonumber
\\
&= 
\textcolor{red}{ \frac{1}{ f_{ u_t(\uptau)}(0)^2 } }   \mathbb{S}_{xx}^{-1}(\lambda).
\end{align}
and
\begin{align}
\underset{ n \to \infty }{ \mathsf{plim} } \ \bigg\{ \widehat{\boldsymbol{V}}^{ols}_{2n}( \lambda ; \uptau_0 ) \bigg\}
\Rightarrow \textcolor{red}{ \frac{1}{ f_{ u_t(\uptau)}(0)^2 } }   \bigg[ \mathbb{S}_{xx}(1) - \mathbb{S}_{xx}(\lambda) \bigg]^{-1}. 
\end{align}



\newpage

Therefore, 
\begin{align}
\underset{ n \to \infty }{ \mathsf{plim} } \ \bigg\{ \widehat{\boldsymbol{V}}^{ols}_{n}( \lambda ; \uptau_0 ) \bigg\}
\Rightarrow 
\textcolor{red}{ \frac{1}{ f_{ u_t(\uptau)}(0)^2 } } 
\bigg\{  \mathbb{S}_{xx}^{-1}(\lambda) + \bigg[ \mathbb{S}_{xx}(1) - \mathbb{S}_{xx}(\lambda) \bigg]^{-1} \bigg\}. 
\end{align}
We can also simplify further the term $\bigg\{ \underset{ n \to \infty }{ \mathsf{plim} } \  \widehat{\boldsymbol{V}}^{ols}_{n}( \lambda ; \uptau_0 ) \bigg\}^{-1} 
= 
\textcolor{red}{ f_{ u_t(\uptau)}(0)^2 } \bigg[ \mathbb{S}_{xx}(\lambda) - \mathbb{S}_{xx}(\lambda) \mathbb{S}^{-1}_{xx}(1)  \mathbb{S}_{xx}(\lambda) \bigg]$.
Denote with 
\color{blue}
\begin{align*}
\boldsymbol{\Delta}^{ols}_0 \big( \lambda; \uptau_0 \big)  
:=
\big[ \mathbb{S}_{xx}(1) - \mathbb{S}_{xx}(\lambda) \big]^{-1}
\times
\begin{bmatrix}
\displaystyle B_{\psi_{\uptau}}(1) - B_{\psi_{\uptau}}(\lambda) 
\\
\displaystyle \int_0^{1} \boldsymbol{J}_c(r) dB_{ \psi_{ \uptau} } -  \int_0^{\lambda} \boldsymbol{J}_c(r) dB_{ \psi_{ \uptau} } 
\end{bmatrix}
- 
\mathbb{S}^{-1}_{xx}(\lambda)
\times
\begin{bmatrix}
\displaystyle B_{\psi_{\uptau}}(\lambda) 
\\
\displaystyle \int_0^{\lambda} \boldsymbol{J}_c(r) dB_{ \psi_{ \uptau} } 
\end{bmatrix}
\end{align*}
\color{black}
Then, it follows that
\begin{align*}
\mathcal{W}_n^{ols} \left( \lambda; \uptau_0 \right) 
&:=   
\boldsymbol{D}_n  
\bigg\{ \Delta \widehat{ \boldsymbol{\theta} }^{ols}_n ( \lambda ; \uptau_0 ) \bigg\}^{\prime} \bigg[ \widehat{\boldsymbol{V}}^{ols}_n( \lambda ; \uptau_0 ) \bigg]^{-1} \bigg\{ \Delta \widehat{ \boldsymbol{\theta} }^{ols}_n ( \lambda ; \uptau_0 ) \bigg\} 
\\
&\Rightarrow
\boldsymbol{\Delta}^{ols}_0 \big( \lambda; \uptau_0 \big) ^{\prime} \big[ \boldsymbol{\Sigma}^{-1} _0 \big( \lambda; \uptau_0 \big)  \big]\boldsymbol{\Delta}^{ols}_0 \big( \lambda; \uptau_0 \big) 
\\ 
&\equiv 
\textcolor{red}{ f_{ u_t(\uptau)}(0)^2 } 
\bigg\{
\boldsymbol{\Delta}^{ols}_0 \big( \lambda; \uptau_0 \big) \bigg\}^{\prime} 
 \bigg[ \mathbb{S}_{xx}(\lambda) - \mathbb{S}_{xx}(\lambda) \mathbb{S}^{-1}_{xx}(1)  \mathbb{S}_{xx}(\lambda) \bigg]
\bigg\{
\boldsymbol{\Delta}^{ols}_0 \big( \lambda; \uptau_0 \big) \bigg\}.
\end{align*}
\end{proof}

Therefore, the asymptotic distribution of the sup OLS-Wald test statistic is nonpivotal and nonstandard and has the above analytical expression. Furthermore, we verify that a trivial aspect such as the inclusion of a model intercept can complicate the asymptotic theory of the structural break test since the model intercept and the slopes are known to have different rates of convergence.  

\newpage 

\subsubsection{Proof of Theorem \ref{Theorem1}}

\begin{small}
\begin{align}
\mathcal{SW}_n^{ivx}\big( \lambda; \uptau_0 \big)  \Rightarrow \underset{ \lambda \in \Lambda_{\eta} }{ \mathsf{sup} } \bigg\{ \boldsymbol{\Delta}_0 \big( \lambda; \uptau_0 \big) ^{\prime} \big[ \boldsymbol{\Sigma}_0 \big( \lambda; \uptau_0 \big)  \big]^{-1} \boldsymbol{\Delta}_0 \big( \lambda; \uptau_0 \big)  \bigg\}
\end{align}
where $\Lambda_{\eta} := [ \eta,  1 - \eta ]$ with $0 < \eta < 1/2$ and
\begin{align}
\boldsymbol{\Delta}_0 \big( \lambda; \uptau_0 \big)   
&:= 
\boldsymbol{W}_p (\lambda) - \boldsymbol{\Psi}_c(\lambda) \boldsymbol{W}_p(1)
\\
\boldsymbol{\Sigma}_0 \big( \lambda; \uptau_0 \big)  
&:= 
\lambda \big( \boldsymbol{I}_p - \boldsymbol{\Psi}_c(\lambda)  \big) \big( \boldsymbol{I}_p - \boldsymbol{\Psi}_c(\lambda)  \big)^{\prime} + (1 - \lambda ) \boldsymbol{\Psi}_c(\lambda) \boldsymbol{\Psi}_c( \lambda)^{\prime} 
\end{align}
such that
\begin{equation*}
\boldsymbol{\Psi}_c (\lambda)
=
\begin{cases}
\displaystyle \left( \lambda  \boldsymbol{\Omega}_{xx} +  \int_0^{\lambda } \boldsymbol{J}_c^{\mu} (r) d\boldsymbol{J}_c^{\prime} \right) \left( \boldsymbol{\Omega}_{xx} + \int_0^{1} \boldsymbol{J}^{\mu}_c (r) d\boldsymbol{J}_c^{\prime} \right)^{-1}  & ,\text{for} \ \upgamma_x = 1 
\\
\\
\displaystyle \lambda \boldsymbol{I}_p  & , \text{for} \ \upgamma_x \in (0,1)
\end{cases}
\end{equation*}
where $\boldsymbol{W}_p(.)$ is a $p-$dimensional standard Brownian motion, $\boldsymbol{J}_c ( \lambda ) = \int_0^{\lambda} e^{(\lambda - s) \boldsymbol{C}_p} d \boldsymbol{B}(s)$ is an \textit{Ornstein-Uhkenbeck} process and we denote with $\boldsymbol{J}^{\mu}_c (\lambda) = \boldsymbol{J}_c (\lambda) - \int_0^1 \boldsymbol{J}_c(s) ds$ and $\boldsymbol{W}_p^{\mu} (\lambda ) = \boldsymbol{W}_p(\lambda) - \int_0^1 \boldsymbol{W}(s) ds$ the demeaned processes of $\boldsymbol{J}_c(\lambda)$ and $\boldsymbol{W}_p(\lambda)$ respectively.
\end{small}

\medskip

\begin{proof}
We consider the limiting distribution of the sup IVX-Wald test for the nonstationary quantile predictive regression model. An analytic expression for the limiting variance of the IVX-Wald test statistic, $\widehat{\boldsymbol{V}}_n^{ivx} \big( \lambda; \uptau_0 \big)  := \tilde{\mathbf{Q}}_1 \big( \lambda; \uptau_0 \big)  + \tilde{\mathbf{Q}}_2 \big( \lambda; \uptau_0 \big) $, can be obtained by formulating the test statistic with respect to the linear restrictions matrix, under the null hypothesis of no structural break for a fixed quantile $\uptau_0 \in (0,1)$. Furthermore, by applying weakly convergence arguments to the particular expression (with zero off-diagonal terms) holds due to the orthogonality property of the regressors of the two regimes with respect to the unknown break-location $\kappa = \floor{\lambda n}$, $0 < \lambda < 1$. 
\begin{small}
\begin{align}
\label{cov.ivx.wald}
\widehat{\boldsymbol{V}}_n^{ivx} \big( \lambda; \uptau_0 \big)   
&:= 
\bigg[ \boldsymbol{I}_p - \boldsymbol{I}_p \bigg] \begin{bmatrix}
\left(\boldsymbol{Z}_1^{\prime} \boldsymbol{X}_1 \right)^{-1} & \boldsymbol{0}                  \\
\boldsymbol{0}                & \left( \boldsymbol{Z}_2^{\prime} \boldsymbol{X}_2 \right)^{-1}   \\
\end{bmatrix} 
\begin{bmatrix}
\boldsymbol{Z}_1^{\prime} \boldsymbol{Z}_1 & \boldsymbol{0}                  \\
\boldsymbol{0}                 & \boldsymbol{Z}_2^{\prime} \boldsymbol{Z}_2   \\
\end{bmatrix}
\begin{bmatrix}
\left(\boldsymbol{X}_1^{\prime} \boldsymbol{Z}_1 \right)^{-1} & \boldsymbol{0}  \\
\boldsymbol{0}                 & \left( \boldsymbol{X}_2^{\prime} \boldsymbol{Z}_2 \right)^{-1}   \\
\end{bmatrix}
\begin{bmatrix}
\boldsymbol{I}_p \\
-\boldsymbol{I}_p
\end{bmatrix}
\nonumber
\\
&\equiv
\bigg[ \tilde{\mathbf{Q}}_1 \big( \lambda; \uptau_0 \big)  + \tilde{\mathbf{Q}}_2 \big( \lambda; \uptau_0 \big)  \bigg]
\end{align}
\end{small}
where 
\begin{small}
\begin{align}
\tilde{\mathbf{Q}}_1 \big( \lambda; \uptau_0 \big) 
&:= 
\left( \tilde{ \boldsymbol{Z} }_1^{\prime} \boldsymbol{X}_1 \right)^{-1} \left( \tilde{\boldsymbol{Z}}_1^{\prime} \tilde{\boldsymbol{Z}}_1 \right) \left(\boldsymbol{X}_1^{\prime} \tilde{\boldsymbol{Z}}_1 \right)^{-1} 
\\
\tilde{\mathbf{Q}}_2 \big( \lambda; \uptau_0 \big) 
&:=
\left(\tilde{\boldsymbol{Z}}_2^{\prime} \boldsymbol{X}_2 \right)^{-1} \left( \tilde{\boldsymbol{Z}}_2^{\prime} \tilde{\boldsymbol{Z}}_2 \right) \left(\boldsymbol{X}_2^{\prime} \tilde{\boldsymbol{Z}}_2 \right)^{-1}  
\\
\tilde{\boldsymbol{D}}_n \widehat{\boldsymbol{V}}_n^{ivx} \big( \lambda; \uptau_0 \big)\tilde{\boldsymbol{D}}_n
&=
\tilde{\boldsymbol{D}}_n \left[ \tilde{\mathbf{Q}}_1 \big( \lambda; \uptau_0 \big)  + \tilde{\mathbf{Q}}_2 \big( \lambda; \uptau_0 \big) \right] \tilde{\boldsymbol{D}}_n
\end{align}
\end{small}

\newpage 

Therefore, by employing the corresponding normalized versions it holds that  
\begin{small}
\begin{align}
\tilde{\mathbf{Q}}_1 \big( \lambda; \uptau_0 \big) 
&:= 
\left(  \sum_{t=1}^{\floor{ \lambda n } } \boldsymbol{X}_{t-1,n} \tilde{\boldsymbol{Z}}_{t-1,n}^{\prime} \right)^{-1} 
\left( \sum_{t=1}^{\floor{ \lambda n } } \tilde{\boldsymbol{Z}}_{t-1,n} \tilde{\boldsymbol{Z}}_{t-1,n}^{\prime}  \right) 
\left( \sum_{t=1}^{\floor{ \lambda n } } \tilde{\boldsymbol{Z}}_{t-1,n}\boldsymbol{X}_{t-1,n}^{\prime} \right)^{-1}
\nonumber
\\
&=
\left( \tilde{\boldsymbol{D}}_n^{-1} \sum_{t=1}^{\floor{ \lambda n } } \boldsymbol{x}_{t-1} \tilde{\boldsymbol{z}}_{t-1}^{\prime} \tilde{\boldsymbol{D}}_n^{-1} \right)^{-1} 
\left( \tilde{\boldsymbol{D}}_n^{-1} \sum_{t=1}^{\floor{ \lambda n } } \tilde{\boldsymbol{z}}_{t-1} \tilde{\boldsymbol{z}}_{t-1}^{\prime} \tilde{\boldsymbol{D}}_n^{-1} \right) 
\left( \tilde{\boldsymbol{D}}_n^{-1} \sum_{t=1}^{\floor{ \lambda n } } \tilde{\boldsymbol{z}}_{t-1}\boldsymbol{x}_{t-1}^{\prime} \tilde{\boldsymbol{D}}_n^{-1} \right)^{-1}
\nonumber
\\
&\equiv
\tilde{\boldsymbol{D}}_n \left( \sum_{t=1}^{\floor{ \lambda n } } \boldsymbol{x}_{t-1} \tilde{\boldsymbol{z}}_{t-1}^{\prime} \right)^{-1} 
\left( \sum_{t=1}^{\floor{ \lambda n } } \tilde{\boldsymbol{z}}_{t-1} \tilde{\boldsymbol{z}}_{t-1}^{\prime} \right) 
\left( \sum_{t=1}^{\floor{ \lambda n } } \tilde{\boldsymbol{z}}_{t-1}\boldsymbol{x}_{t-1}^{\prime} \right)^{-1} \tilde{\boldsymbol{D}}_n 
\end{align}
\end{small}
Moreover, we denote the weakly convergence of the following  moment matrices as below 
\begin{align}
\tilde{\boldsymbol{D}}_n^{-1} \sum_{t=1}^{\floor{ \lambda n } } \boldsymbol{x}_{t-1} \tilde{\boldsymbol{z}}_{t-1}^{\prime} \tilde{\boldsymbol{D}}_n^{-1} \Rightarrow \boldsymbol{\Gamma}_{cxz} ( \lambda) \ \ \ \ \text{and} \ \ \ \ \tilde{\boldsymbol{D}}_n^{-1} \sum_{t=1}^{n } \boldsymbol{x}_{t-1} \tilde{\boldsymbol{z}}_{t-1}^{\prime}  \tilde{\boldsymbol{D}}_n^{-1}\Rightarrow \boldsymbol{\Gamma}_{cxz} (1) 
\end{align}
By applying weak convergence arguments we obtain that 
\begin{align}
\tilde{\boldsymbol{D}}_n \times \tilde{\mathbf{Q}}_1 \big( \lambda; \uptau_0 \big) \times \tilde{\boldsymbol{D}}_n 
&\Rightarrow 
\boldsymbol{\Gamma}_{cxz} ( \lambda)^{-1} \big( \lambda \boldsymbol{V}_{cxz} \big) \big( \boldsymbol{\Gamma}_{cxz} ( \lambda)^{\prime} \big)^{-1}
\nonumber
\\
&\equiv
\lambda \left( \boldsymbol{\Gamma}_{cxz} ( \lambda) \boldsymbol{V}_{cxz}^{-1} \boldsymbol{\Gamma}_{cxz} ( \lambda)^{\prime} \right)^{-1}
\end{align}
Similarly, it holds that 
\begin{small}
\begin{align*}
\tilde{\mathbf{Q}}_2 \big( \lambda; \uptau_0 \big) 
&:= 
\left(  \sum_{t=\floor{ \lambda n } + 1}^{n } \boldsymbol{X}_{t-1,n} \tilde{\boldsymbol{Z}}_{t-1,n}^{\prime} \right)^{-1} 
\left( \sum_{t=\floor{ \lambda n } + 1}^{n } \tilde{\boldsymbol{Z}}_{t-1,n} \tilde{\boldsymbol{Z}}_{t-1,n}^{\prime}  \right) 
\left( \sum_{t=\floor{ \lambda n } + 1}^{n } \tilde{\boldsymbol{Z}}_{t-1,n}\boldsymbol{X}_{t-1,n}^{\prime} \right)^{-1}
\\
&=
\left( \tilde{\boldsymbol{D}}_n^{-1} \sum_{t=\floor{ \lambda n } + 1}^{n } \boldsymbol{x}_{t-1} \tilde{\boldsymbol{z}}_{t-1}^{\prime} \tilde{\boldsymbol{D}}_n^{-1} \right)^{-1} 
\left( \tilde{\boldsymbol{D}}_n^{-1} \sum_{t=\floor{ \lambda n } + 1}^{n } \tilde{\boldsymbol{z}}_{t-1} \tilde{\boldsymbol{z}}_{t-1}^{\prime} \tilde{\boldsymbol{D}}_n^{-1} \right) 
\left( \tilde{\boldsymbol{D}}_n^{-1} \sum_{t=\floor{ \lambda n } + 1}^{n } \tilde{\boldsymbol{z}}_{t-1}\boldsymbol{x}_{t-1}^{\prime} \tilde{\boldsymbol{D}}_n^{-1} \right)^{-1}
\\
&\equiv
\tilde{\boldsymbol{D}}_n \left( \sum_{t=\floor{ \lambda n } + 1}^{n } \boldsymbol{x}_{t-1} \tilde{\boldsymbol{z}}_{t-1}^{\prime} \right)^{-1} 
\left( \sum_{t=\floor{ \lambda n } + 1}^{n } \tilde{\boldsymbol{z}}_{t-1} \tilde{\boldsymbol{z}}_{t-1}^{\prime} \right) 
\left( \sum_{t=\floor{ \lambda n } + 1}^{n } \tilde{\boldsymbol{z}}_{t-1}\boldsymbol{x}_{t-1}^{\prime} \right)^{-1} \tilde{\boldsymbol{D}}_n 
\end{align*}
\end{small}
Thus, it follows that
\begin{small} 
\begin{align}
\tilde{\boldsymbol{D}}_n \times \tilde{\mathbf{Q}}_2 \big( \lambda; \uptau_0 \big) \times \tilde{\boldsymbol{D}}_n 
&\Rightarrow 
\big( \boldsymbol{\Gamma}_{cxz} (1) - \boldsymbol{\Gamma}_{cxz} ( \lambda) \big)^{-1} \big( (1 - \lambda) \boldsymbol{V}_{cxz} \big) \left[ \big( \boldsymbol{\Gamma}_{cxz} (1) - \boldsymbol{\Gamma}_{cxz} ( \lambda) \big)^{\prime} \right]^{-1}
\nonumber
\\
&\equiv
(1 - \lambda) \bigg\{ \big( \boldsymbol{\Gamma}_{cxz} (1) - \boldsymbol{\Gamma}_{cxz} ( \lambda) \big) \boldsymbol{V}_{cxz}^{-1} \big( \boldsymbol{\Gamma}_{cxz} (1) - \boldsymbol{\Gamma}_{cxz} ( \lambda) \big)^{\prime} \bigg\}^{-1}
\end{align}
\end{small}
Therefore, for the limiting variance of the sup IVX-Wald test we have that 
\begin{align*}
&\tilde{\boldsymbol{D}}_n \widehat{\boldsymbol{V}}_n^{ivx} \big( \lambda; \uptau_0 \big)\tilde{\boldsymbol{D}}_n
\\
&=
\tilde{\boldsymbol{D}}_n \left[ \tilde{\mathbf{Q}}_1 \big( \lambda; \uptau_0 \big)  + \tilde{\mathbf{Q}}_2 \big( \lambda; \uptau_0 \big) \right] \tilde{\boldsymbol{D}}_n
\\
&=
\bigg\{ \lambda \boldsymbol{\Gamma}_{cxz} ( \lambda)^{-1} \boldsymbol{V}_{cxz} \boldsymbol{\Gamma}_{cxz}^{\prime} ( \lambda)^{-1} + (1 - \lambda)  \big( \boldsymbol{\Gamma}_{cxz} (1) - \boldsymbol{\Gamma}_{cxz} ( \lambda) \big)^{-1} \boldsymbol{V}_{cxz} \big( \boldsymbol{\Gamma}^{\prime}_{cxz} (1) - \boldsymbol{\Gamma}_{cxz}^{\prime} ( \lambda) \big)^{-1}  \bigg\}
\end{align*}

\newpage 

Furthermore, it holds that
\begin{small}
\begin{align}
\label{ExpressionA}
\tilde{\boldsymbol{D}}_n \left( \widehat{ \boldsymbol{\beta}}^{ivx}_{1} (\lambda ; \uptau_0) - \boldsymbol{\beta}_0( \uptau_0 ) \right) 
&\Rightarrow 
\boldsymbol{\Gamma}^{-1}_{cxz} (\lambda) \times \mathcal{N} \displaystyle \big( \boldsymbol{0}, \uptau_0 (1 - \uptau_0) \lambda \boldsymbol{V}_{cxz} \big) 
\nonumber
\\
&\equiv 
\sqrt{\uptau_0 ( 1 - \uptau_0)} \times \boldsymbol{\Gamma}^{-1}_{cxz} (\lambda) \times \boldsymbol{V}_{cxz}^{1/2} \times \boldsymbol{W}_{p} ( \lambda ).
\\
\nonumber
\\
\label{ExpressionB}
\tilde{\boldsymbol{D}}_n \left( \widehat{\boldsymbol{\beta}}_{2}^{ivx} ( \lambda; \uptau_0 ) - \boldsymbol{\beta}_0 (\uptau) \right)
&\Rightarrow 
\bigg( \boldsymbol{\Gamma}_{cxz}(1) - \boldsymbol{\Gamma}_{cxz}( \lambda) \bigg)^{-1} \times \mathcal{N} \displaystyle \big( \boldsymbol{0}, \uptau_0 (1 - \uptau_0) (1 - \lambda) \boldsymbol{V}_{cxz} \big) 
\nonumber
\\
&\equiv 
\sqrt{\uptau_0 ( 1 - \uptau_0)} \times \bigg( \boldsymbol{\Gamma}_{cxz}(1) - \boldsymbol{\Gamma}_{cxz}( \lambda)  \bigg)^{-1} \times \boldsymbol{V}_{cxz}^{1/2} \times \bigg[ \boldsymbol{W}_p (1) - \boldsymbol{W}_p (\lambda) \bigg].
\end{align}
\end{small}
which implies that
\begin{align}
\tilde{\boldsymbol{D}}_n \left( \widehat{\boldsymbol{\beta}}_{2}^{ivx} ( \lambda; \uptau_0 ) - \widehat{\boldsymbol{\beta}}_{1}^{ivx} ( \lambda; \uptau_0 ) \right)
\Rightarrow
- \sqrt{\uptau_0 ( 1 - \uptau_0)} \boldsymbol{V}_{cxz}^{1/2} \bigg\{ \textcolor{red}{\mathcal{A}} \boldsymbol{W}_p (\lambda) - \textcolor{red}{\mathcal{B}} \boldsymbol{W}_p (1) \bigg\}
\end{align}
where
\begin{small}
\begin{align}
\textcolor{red}{\mathcal{A}} 
= 
\left\{ \bigg( \boldsymbol{\Gamma}_{cxz}(1) - \boldsymbol{\Gamma}_{cxz}( \lambda) \bigg)^{-1} + \boldsymbol{\Gamma}^{-1}_{cxz}( \lambda) \right\} \ \ \ \text{and} \ \ \ 
\textcolor{red}{\mathcal{B}} 
= 
\bigg( \boldsymbol{\Gamma}_{cxz}(1) - \boldsymbol{\Gamma}_{cxz}( \lambda) \bigg)^{-1}
\end{align}
\end{small}
Therefore, putting all the above together the expression can be simplified as below
\begin{small}
\begin{align*}
&\mathcal{SW}_n^{ivx}\big( \lambda; \uptau_0 \big) \Rightarrow 
\uptau_0 ( 1 - \uptau_0)
\\
&\times
\bigg( \boldsymbol{W}_p(\lambda) - \boldsymbol{\Gamma}_{cxz} (\lambda) \boldsymbol{\Gamma}^{-1}_{cxz} (1) \boldsymbol{W}_p(1) \bigg)^{\prime} 
\\  
&\times \bigg\{ \lambda \big( \boldsymbol{I}_p - \boldsymbol{\Gamma}_{cxz} (\lambda) \boldsymbol{\Gamma}^{-1}_{cxz} (1)  \big) \big( \boldsymbol{I}_p - \boldsymbol{\Gamma}_{cxz} (\lambda) \boldsymbol{\Gamma}^{-1}_{cxz} (1) \big)^{\prime} +  (1 - \lambda) \left( \boldsymbol{\Gamma}_{cxz} (\lambda) \boldsymbol{\Gamma}^{-1}_{cxz} (1) \right) \left( \boldsymbol{\Gamma}_{cxz} (\lambda) \boldsymbol{\Gamma}^{-1}_{cxz} (1)\right)^{\prime} \bigg\}^{-1} 
\\
&\times 
\bigg( \boldsymbol{W}_p(\lambda) - \boldsymbol{\Gamma}_{cxz} (\lambda) \boldsymbol{\Gamma}^{-1}_{cxz}(1) \boldsymbol{W}_p(1) \bigg).   
\end{align*} 
\end{small} 
Denote with 
\begin{small}
\begin{equation*}
\boldsymbol{\Psi}_c(\lambda) 
:= 
\boldsymbol{\Gamma}_{cxz} (\lambda) \boldsymbol{\Gamma}^{-1}_{cxz} (1)
=
\begin{cases}
\displaystyle \left( \lambda \mathbf{\Omega}_{xx} +  \int_0^{\lambda} \boldsymbol{J}^{\mu}_c(r) d\boldsymbol{J}_c^{\prime} \right) \left( \mathbf{\Omega}_{xx} +  \int_0^{1} \boldsymbol{J}^{\mu}_c(r) d\boldsymbol{J}_c^{\prime} \right)^{-1}  
& ,\text{for} \ \gamma_x = 1
\\
\\
\displaystyle \lambda \boldsymbol{I}_p
& ,\text{for} \ 0 < \gamma_x < 1. 
\end{cases}
\end{equation*}
\end{small}
Thus, in summary, the sup IVX-Wald statistic can be formulated as below
\begin{small}
\begin{align*}
\mathcal{SW}_n^{ivx}\big( \lambda; \uptau_0 \big)  
&\Rightarrow 
\underset{ \lambda \in \Lambda_{\eta} }{ \mathsf{sup} } \bigg\{ \boldsymbol{\Delta}_0 \big( \lambda; \uptau_0 \big) ^{\prime} \big[ \boldsymbol{\Sigma}_0 \big( \lambda; \uptau_0 \big)  \big]^{-1} \boldsymbol{\Delta}_0 \big( \lambda; \uptau_0 \big)  \bigg\}
\\
\boldsymbol{\Delta}_0 \big( \lambda; \uptau_0 \big)   
&:= 
\boldsymbol{W}_p (\lambda) - \boldsymbol{\Psi}_c(\lambda) \boldsymbol{W}_p(1)
\\
\boldsymbol{\Sigma}_0 \big( \lambda; \uptau_0 \big)  
&:= 
\lambda \big( \boldsymbol{I}_p - \boldsymbol{\Psi}_c(\lambda)  \big) \big( \boldsymbol{I}_p - \boldsymbol{\Psi}_c(\lambda)  \big)^{\prime} + (1 - \lambda ) \boldsymbol{\Psi}_c(\lambda) \boldsymbol{\Psi}_c( \lambda)^{\prime} 
\end{align*}
\end{small}
\end{proof}

\newpage 

\paragraph{Proof of Expression \eqref{ExpressionA}:}

\begin{proof}
We consider the dequantiled model with $p$ regressors which can be either mildly integrated or high persistent. Then, by employing the  embedded normalization matrices we can derive analytic expressions for the IVX estimators of the two subsamples. 
\begin{align*}
\widehat{ \boldsymbol{\beta} }_{1}^{ivx} ( \lambda; \uptau_0 ) 
&= 
\left( \frac{1}{ \kappa } \sum_{t=1}^{ \floor{\lambda n} }  \tilde{\boldsymbol{Z}}_{t-1,n} \boldsymbol{X}_{t-1,n}^{\prime} \right)^{-1} \left( \frac{1}{ \kappa } \sum_{t=1}^{ \floor{\lambda n} }  \tilde{\boldsymbol{Z}}_{t-1,n}y_t (\uptau) \right) 
\\
&= 
\left( \frac{1}{ \kappa } \sum_{t=1}^{\floor{\lambda n}} \tilde{\boldsymbol{Z}}_{t-1,n} \boldsymbol{X}_{t-1,n}^{\prime} \right)^{-1} \left( \frac{1}{ \kappa } \sum_{t=1}^{\floor{\lambda n}} \tilde{\boldsymbol{Z}}_{t-1,n} \left[  \boldsymbol{X}_{t-1,n}^{\prime} \boldsymbol{\beta}_0( \uptau_0 ) \ + \psi_{ \uptau } \big( u_t ( \uptau_0 ) \big) \right] \right) 
\\
&=
\boldsymbol{\beta}_0( \uptau_0 ) + \left( \frac{1}{ \kappa } \sum_{t=1}^{\floor{\lambda n}} \tilde{\boldsymbol{Z}}_{t-1,n} \boldsymbol{X}_{t-1,n}^{\prime} \right)^{-1}
\left( \frac{1}{ \kappa } \sum_{t=1}^{\floor{\lambda n}} \tilde{\boldsymbol{Z}}_{t-1,n} \psi_{ \uptau } \big( u_t ( \uptau_0 ) \big) \right) + o_{ \mathbb{P} }(1)
\end{align*}
Therefore, it holds that 
\begin{align*}
\textcolor{red}{\tilde{\boldsymbol{D}}_n}  \left( \widehat{ \boldsymbol{\beta} }_{1}^{ivx} ( \lambda; \uptau_0 ) - \boldsymbol{\beta}_0 (\uptau) \right)
= 
\left( \textcolor{red}{\tilde{\boldsymbol{D}}_n} \sum_{t=1}^{\floor{\lambda n}} \tilde{\boldsymbol{Z}}_{t-1,n} \boldsymbol{X}_{t-1,n}^{\prime} \right)^{-1} \left( \textcolor{red}{\tilde{\boldsymbol{D}}_n} \sum_{t=1}^{\floor{\lambda n}} \tilde{\boldsymbol{Z}}_{t-1,n}  \psi_{ \uptau } \big( u_t ( \uptau_0 ) \big) \right) + o_{ \mathbb{P} }(1)
\end{align*} 
Notice that it holds that 
\begin{small}
\begin{align}
\left( \sum_{t=1}^{\floor{\lambda n}} \tilde{\boldsymbol{Z}}_{t-1,n} \boldsymbol{X}_{t-1,n}^{\prime} \right)^{-1} 
&= 
\left( \sum_{t=1}^{\floor{\lambda n}} \textcolor{blue}{\tilde{\boldsymbol{D}}^{-1}_n} \tilde{\boldsymbol{z}}_{t-1} \left( \textcolor{blue}{\tilde{\boldsymbol{D}}^{-1}_n} \boldsymbol{x}_{t-1} \right)^{\prime} \right)^{-1} 
\nonumber
\\
&=
\left( \textcolor{blue}{\tilde{\boldsymbol{D}}^{-1}_n} \left[ \sum_{t=1}^{\floor{\lambda n}}  \tilde{\boldsymbol{z}}_{t-1}  \boldsymbol{x}_{t-1}^{\prime} \right] \textcolor{blue}{\tilde{\boldsymbol{D}}^{-1}_n}  \right)^{-1} 
\nonumber
\\
&=
\left( \textcolor{blue}{ \frac{1}{ n^{ 1 + ( \upgamma_x \wedge \upgamma_z ) } } \otimes \boldsymbol{I}_p  } \sum_{t=1}^{\floor{\lambda n}}   \tilde{\boldsymbol{z}}_{t-1}  \boldsymbol{x}_{t-1}^{\prime} \right)^{-1}
\Rightarrow
\boldsymbol{\Gamma}^{-1}_{cxz} (\lambda)  
\end{align}
For the second term of the expression for $\textcolor{red}{\tilde{\boldsymbol{D}}_n}  \left( \widehat{ \boldsymbol{\beta} }_{1}^{ivx} ( \lambda; \uptau_0 ) - \boldsymbol{\beta}_0 (\uptau) \right)$ it holds that 
\begin{align}
\left(  \sum_{t=1}^{\floor{\lambda n}} \tilde{\boldsymbol{Z}}_{t-1,n}  \psi_{ \uptau } \big( u_t ( \uptau_0 ) \big) \right)
&= 
\left( \textcolor{blue}{\tilde{\boldsymbol{D}}_n^{-1}} \sum_{t=1}^{\floor{\lambda n}} \tilde{\boldsymbol{z}}_{t-1}  \psi_{ \uptau } \big( u_t ( \uptau_0 ) \big) \right)
\nonumber
\\
&=
\left( \textcolor{blue}{ n^{ - \frac{1 + ( \upgamma_x \wedge \upgamma_z ) }{2} } \otimes \boldsymbol{I}_p  } \sum_{t=1}^{\floor{\lambda n}} \tilde{\boldsymbol{z}}_{t-1}  \psi_{ \uptau } \big( u_t ( \uptau_0 ) \big) \right)
\nonumber
\\
&\Rightarrow
\mathcal{N} \bigg( \boldsymbol{0}, \uptau_0 ( 1 - \uptau_0)  \lambda \boldsymbol{V}_{cxz}  \bigg)
\end{align}
\end{small}
which implies that
\begin{align}
\textcolor{red}{\tilde{\boldsymbol{D}}_n} \left( \widehat{ \boldsymbol{\beta}}^{ivx}_{1} (\lambda ; \uptau_0) - \boldsymbol{\beta}_0( \uptau_0 ) \right) 
&\Rightarrow 
\boldsymbol{\Gamma}^{-1}_{cxz} (\lambda)   \times \mathcal{N} \displaystyle \big( \boldsymbol{0}, \uptau_0 (1 - \uptau_0) \lambda \boldsymbol{V}_{cxz} \big) 
\nonumber
\\
&\equiv 
\sqrt{\uptau_0 ( 1 - \uptau_0)} \times \boldsymbol{\Gamma}^{-1}_{cxz} (\lambda) \times \boldsymbol{V}_{cxz}^{1/2} \times \boldsymbol{W}_{p} ( \lambda ).
\end{align}

\newpage 

where $\boldsymbol{\Gamma}_{cxz} (\lambda)$ corresponds to the weakly convergence argument of the partial sum process expression such that
\begin{align}
\boldsymbol{\Gamma}_{cxz} (\lambda)  
\equiv
\begin{cases}
\displaystyle  - \boldsymbol{C}_z^{-1} \left( \lambda \boldsymbol{\Omega}_{xx} + \int_0^{\lambda} \boldsymbol{J}_c (r) d\boldsymbol{J}_c^{\prime} \right) & \ \text{if} \ \upgamma_x = 1,
\\
- \lambda \boldsymbol{C}_z^{-1} \bigg( \boldsymbol{\Omega}_{xx} +    \boldsymbol{C}_p \boldsymbol{V}_{xx} \bigg) &  \ \text{if} \ \upgamma_x \in ( \upgamma_z, 1 ),
\\
\lambda \boldsymbol{V}_{xx} &  \ \text{if} \ \upgamma_x \in ( 0, \upgamma_z ).
\end{cases}
\end{align}
Following similar arguments expression \eqref{ExpressionB} follows, and the rest of the arguments for the proof of Theorem \ref{Theorem1}. 
\end{proof}

\underline{\textbf{Known break-point:}}

\begin{proof} For a known-break point the asymptotic term that corresponds to the case of high persistence, $\upgamma_x = 1$,   can simplify further, when $\lambda$ is fixed say $\lambda \equiv \lambda_0$, such that 
\begin{align}
\boldsymbol{\Gamma}_{cxz} (\lambda_0)  
&\equiv 
\displaystyle  - \boldsymbol{C}_z^{-1} \left( \lambda_0 \boldsymbol{\Omega}_{xx} + \int_0^{\lambda_0} \boldsymbol{J}_c (r) d\boldsymbol{J}_c^{\prime} \right)
\nonumber
\\
&=
\displaystyle  - \lambda_0 \boldsymbol{C}_z^{-1} \left( \boldsymbol{\Omega}_{xx} +  \int_0^{1} \boldsymbol{J}_c (r) d\boldsymbol{J}_c^{\prime} \right) \equiv \lambda_0 \boldsymbol{\Gamma}_{cxz}(1).
\end{align}
Thus,  further generalizations occur based on the formulation of Theorem \ref{Theorem1}, which cover both the case of mildly integrated and persistent regressors, demonstrating the convergence to a nuisance-free limiting distribution. In particular,
\begin{align}
\tilde{\mathbf{Q}}_1 \big( \lambda_0 ; \uptau_0 \big) 
&\Rightarrow
\big\{ \lambda_0  \boldsymbol{\Gamma}_{cxz} \big\}^{-1} 
\times 
\big\{ \lambda_0  \boldsymbol{V}_{cxz} \big\} 
\times 
\big\{ \lambda_0  \boldsymbol{\Gamma}_{cxz}^{\prime} \big\}^{-1} 
\nonumber
\\
&=
\left( \frac{1}{\lambda_0 } \boldsymbol{\Gamma}_{cxz}^{-1} \boldsymbol{V}_{cxz} \left( \boldsymbol{\Gamma}_{cxz}^{\prime} \right)^{-1} \right) 
\equiv 
\bigg( \lambda_0  \boldsymbol{\Gamma}_{cxz} \boldsymbol{V}_{cxz}^{-1} \boldsymbol{\Gamma}_{cxz}^{\prime} \bigg)^{-1} 
\\
\tilde{\mathbf{Q}}_2 \big( \lambda_0 ; \uptau_0 \big) 
&\Rightarrow
\big\{ (1 - \lambda_0 ) \boldsymbol{\Gamma}_{cxz} \big\}^{-1} 
\times \big\{ (1 - \lambda_0 ) \boldsymbol{V}_{cxz} \big\} 
\times 
\big\{ (1 - \lambda_0 ) \boldsymbol{\Gamma}_{cxz}^{\prime} \big\}^{-1} 
\nonumber
\\
&=
\left( \frac{1}{1 - \lambda_0 } \boldsymbol{\Gamma}_{cxz}^{-1} \boldsymbol{V}_{cxz} \left( \boldsymbol{\Gamma}_{cxz}^{\prime} \right)^{-1} \right) 
\equiv 
\bigg( (1 - \lambda_0 ) \boldsymbol{\Gamma}_{cxz} \boldsymbol{V}_{cxz}^{-1} \boldsymbol{\Gamma}_{cxz}^{\prime} \bigg)^{-1} 
\end{align}
Therefore, it follows that 
\begin{align}
\widehat{\boldsymbol{V}}_n^{ivx} \big( \lambda_0 ; \uptau_0 \big)  
\Rightarrow
\left\{ \bigg( \lambda_0  \boldsymbol{\Gamma}_{cxz} \boldsymbol{V}_{cxz}^{-1} \boldsymbol{\Gamma}_{cxz}^{\prime} \bigg)^{-1} + \bigg( (1 - \lambda_0 ) \boldsymbol{\Gamma}_{cxz} \boldsymbol{V}_{cxz}^{-1} \boldsymbol{\Gamma}_{cxz}^{\prime} \bigg)^{-1} \right\}
\end{align}
and by denoting with $\mathbb{S}_{cxz} := \big( \boldsymbol{\Gamma}_{cxz} \boldsymbol{V}_{cxz}^{-1} \boldsymbol{\Gamma}_{cxz}^{\prime} \big)$ we obtain
\begin{align}
\widehat{\boldsymbol{V}}_n^{ivx} \big( \lambda_0 ; \uptau_0 \big) 
\Rightarrow
\bigg\{ \big( \lambda_0   \mathbb{S}_{cxz} \big)^{-1} +  \big( (1 - \lambda_0  ) \mathbb{S}_{cxz} \big)^{-1} \bigg\} 
\equiv 
\bigg\{ \frac{1}{\lambda_0 } + \frac{1}{1 - \lambda_0 } \bigg\} 
\times 
\mathbb{S}_{cxz}^{-1}
\end{align}

\newpage 

Then, the expression for the IVX-Wald test statistic given by
\begin{align}
\mathcal{W}_n^{ivx} \big( \lambda_0 ; \uptau_0 \big)  
\Rightarrow 
 \bigg\{ \boldsymbol{\Delta}_0 \big( \lambda_0; \uptau_0 \big) ^{\prime} \big[ \boldsymbol{\Sigma}_0 \big( \lambda_0; \uptau_0 \big)  \big]^{-1} \boldsymbol{\Delta}_0 \big( \lambda_0; \uptau_0 \big)  \bigg\}
\end{align}
can be simplified further by combining the asymptotic convergence results above 
\begin{align*}
\Delta \widehat{ \boldsymbol{\beta} }^{ivx}_n \big( \lambda_0; \uptau_0 \big)  
&\Rightarrow 
\boldsymbol{\Delta}_0 \big( \lambda_0; \uptau_0 \big)  
:=
\frac{1}{\lambda_0(1 - \lambda_0)} \sqrt{\uptau_0 ( 1 - \uptau_0)} \boldsymbol{\Gamma}_{cxz}^{-1} \times \boldsymbol{V}_{cxz}^{1/2} \times \big[ \boldsymbol{W}_p(\lambda_0) - \lambda \boldsymbol{W}_p(1) \big].
\\
\widehat{\boldsymbol{V}}_n^{ivx} \big( \lambda_0; \uptau_0 \big) 
&\Rightarrow
\boldsymbol{\Sigma}_0 \big( \lambda_0; \uptau_0 \big) 
:=
\frac{1}{ \lambda_0 ( 1 - \lambda_0)} \times \mathbb{S}_{cxz}^{-1}.
\end{align*}
Therefore, we obtain 
\begin{align}
\label{wald.ivx.limit}
&\mathcal{W}_n^{ivx}\big( \lambda_0; \uptau_0 \big) 
\Rightarrow 
\nonumber
\\
&\bigg\{ \frac{\uptau_0 ( 1 - \uptau_0)}{\lambda_0 (1 - \lambda_0)} 
\big[ \boldsymbol{W}_p(\lambda_0) - \lambda \boldsymbol{W}_p(1) \big] ^{\prime} \left( \boldsymbol{V}_{cxz}^{1/2} \right) \left( \boldsymbol{\Gamma}_{cxz}^{-1} \right)^{\prime} \mathbb{S}_{cxz}
\boldsymbol{\Gamma}_{cxz}^{-1}  \boldsymbol{V}_{cxz}^{1/2}
\big[ \boldsymbol{W}_p(\lambda_0) - \lambda \boldsymbol{W}_p(1) \big] \bigg\}
\end{align}
Consider the formulation for the covariance matrix such that
\begin{align*}
\left( \boldsymbol{V}_{cxz}^{1/2} \right) \left( \boldsymbol{\Gamma}_{cxz}^{-1} \right)^{\prime}
\mathbb{S}_{cxz} 
\boldsymbol{\Gamma}_{cxz}^{-1}  \boldsymbol{V}_{cxz}^{1/2} 
\equiv
\left( \boldsymbol{V}_{cxz}^{1/2} \right) \left( \boldsymbol{\Gamma}_{cxz}^{-1} \right)^{\prime}
\big( \boldsymbol{\Gamma}_{cxz} \boldsymbol{V}_{cxz}^{-1} \boldsymbol{\Gamma}_{cxz}^{\prime} \big) 
\boldsymbol{\Gamma}_{cxz}^{-1}  \boldsymbol{V}_{cxz}^{1/2} 
=
\boldsymbol{I}_p.
\end{align*}
In summary, we prove that the asymptotic distribution of the IVX-Wald test for a known break-point $\lambda_0$ is nuisance-parameter free as below
\begin{align}
\label{final.wald.ivx.limit}
\frac{1}{\uptau_0 ( 1 - \uptau_0)} \times \mathcal{W}_n^{ivx} \big( \lambda_0; \uptau_0 \big)  
\Rightarrow 
\frac{ \big[ \boldsymbol{W}_p(\lambda_0) - \lambda \boldsymbol{W}_p(1) \big] ^{\prime} \big[ \boldsymbol{W}_p(\lambda_0) - \lambda \boldsymbol{W}_p(1) \big]}{ \lambda_0 (1 - \lambda_0) } 
\Rightarrow 
\chi^2_p.
\end{align}
which holds regardless of the degree of persistence driving the regressors of the model (at least with respect to high persistence or mildly integrated regressors).

\end{proof}




\newpage 


\subsubsection{Proof of Corollary \ref{Corollary55AA}}

\begin{proof}

\begin{align*}
\widehat{ \boldsymbol{\beta} }_{1}^{ivz} ( \lambda; \uptau_0 ) 
&= 
\left( \frac{1}{\kappa} \sum_{t=1}^{ \floor{\lambda n} } \tilde{\boldsymbol{Z}}_{t-1,n} \tilde{\boldsymbol{Z}}_{t-1,n}^{\prime} \right)^{-1} 
\left( \frac{1}{\kappa} \sum_{t=1}^{ \floor{\lambda n} } \tilde{\boldsymbol{Z}}_{t-1,n} y_t (\uptau) \right) 
\\
&= 
\left( \frac{1}{ \kappa } \sum_{t=1}^{\floor{\lambda n}} \tilde{\boldsymbol{Z}}_{t-1,n} \tilde{\boldsymbol{Z}}_{t-1,n}^{\prime} \right)^{-1} 
\left( \frac{1}{ \kappa } \sum_{t=1}^{\floor{\lambda n}} \tilde{\boldsymbol{Z}}_{t-1,n} \bigg[ \tilde{\boldsymbol{Z}}_{t-1,n}^{\prime} \boldsymbol{\beta}_0( \uptau_0 ) + \psi_{ \uptau } \big( u_t ( \uptau_0 ) \big) \bigg] \right) 
\\
&=
\boldsymbol{\beta}_0( \uptau_0 ) + \left( \frac{1}{ \kappa } \sum_{t=1}^{\floor{\lambda n}} \tilde{\boldsymbol{Z}}_{t-1,n} \tilde{\boldsymbol{Z}}_{t-1,n}^{\prime} \right)^{-1}
\left( \frac{1}{\kappa} \sum_{t=1}^{\floor{\lambda n}} \tilde{\boldsymbol{Z}}_{t-1,n} \psi_{ \uptau } \big( u_t ( \uptau_0 ) \big) \right) + o_{ \mathbb{P} }(1)
\end{align*}
Therefore, it holds that
\begin{align*}
\tilde{\boldsymbol{D}}_n  \left( \widehat{ \boldsymbol{\beta} }_{1}^{ivz} ( \lambda; \uptau_0 ) - \boldsymbol{\beta}_0 (\uptau) \right)
= 
\left( \sum_{t=1}^{\floor{\lambda n}} \tilde{\boldsymbol{Z}}_{t-1,n} \tilde{\boldsymbol{Z}}_{t-1,n}^{\prime} \right)^{-1} \left( \sum_{t=1}^{\floor{\lambda n}} \tilde{\boldsymbol{Z}}_{t-1,n}  \psi_{ \uptau } \big( u_t ( \uptau_0 ) \big) \right) + o_{ \mathbb{P} }(1)
\end{align*} 
which implies that
\begin{align}
\tilde{\boldsymbol{D}}_n \left( \widehat{ \boldsymbol{\beta}}^{ivz}_{1} (\lambda ; \uptau_0) - \boldsymbol{\beta}_0( \uptau_0 ) \right) 
&\Rightarrow 
\big( \lambda \boldsymbol{V}_{cxz} \big)^{-1} \times \mathcal{N} \displaystyle \big( \boldsymbol{0}, \uptau_0 (1 - \uptau_0) \lambda \boldsymbol{V}_{cxz} \big) 
\nonumber
\\
&\equiv 
\frac{1}{\lambda} \sqrt{\uptau_0 ( 1 - \uptau_0)} \times \boldsymbol{V}_{cxz}^{-1} \times \boldsymbol{V}_{cxz}^{1/2} \times \boldsymbol{W}_{p} ( \lambda )
\nonumber
\\
&=
\frac{1}{\lambda} \sqrt{\uptau_0 ( 1 - \uptau_0)} \times \boldsymbol{V}_{cxz}^{-1/2} \times \boldsymbol{W}_{p} ( \lambda ).
\end{align}
Similarly, it can be proved that 
\begin{align}
\tilde{\boldsymbol{D}}_n \left( \widehat{\boldsymbol{\beta}}_{2}^{ivz} ( \lambda; \uptau_0 ) - \boldsymbol{\beta}_0 (\uptau) \right)
&\Rightarrow 
\big( ( 1 - \lambda) \boldsymbol{V}_{cxz} \big)^{-1} \times \mathcal{N} \displaystyle \big( \boldsymbol{0}, \uptau_0 (1 - \uptau_0) (1 - \lambda) \boldsymbol{V}_{cxz} \big) 
\nonumber
\\
&\equiv 
\frac{1}{1 - \lambda} \sqrt{\uptau_0 ( 1 - \uptau_0)} \times  \boldsymbol{V}_{cxz}^{- 1/2} \times \bigg[ \boldsymbol{W}_{p} ( \lambda ) - \boldsymbol{W}_{p} (1) \bigg].
\end{align}
Furthermore, it holds that 
\begin{align}
\tilde{\Delta} \widehat{ \boldsymbol{\beta} }^{ivz}_n \big( \lambda ; \uptau_0 \big)
&:=  
\tilde{\boldsymbol{D}}_n \bigg[ \widehat{ \boldsymbol{\beta} }_{2}^{ivz} ( \lambda ; \uptau_0 ) - \widehat{ \boldsymbol{\beta} }_{1}^{ivz} ( \lambda ; \uptau_0 ) \bigg]
\nonumber
\\
&\Rightarrow
- \frac{1}{\lambda(1 - \lambda)} \sqrt{\uptau_0 ( 1 - \uptau_0)} \times  \boldsymbol{V}_{cxz}^{- 1/2} \times 
\bigg[ \boldsymbol{W}_{p} (\lambda) - \lambda \boldsymbol{W}_{p} (1) \bigg].
\end{align}
Moreover, 
\begin{align*}
\tilde{\boldsymbol{D}}_n \tilde{\mathbf{Q}}_1 \big( \lambda; \uptau_0 \big) \tilde{\boldsymbol{D}}_n 
&\Rightarrow 
\boldsymbol{\Gamma}_{cxz} ( \lambda)^{-1} \big( \lambda \boldsymbol{V}_{cxz} \big) \big( \boldsymbol{\Gamma}_{cxz} ( \lambda)^{\prime} \big)^{-1}
\equiv
\lambda \left(  \lambda \boldsymbol{V}_{cxz} \boldsymbol{V}_{cxz}^{-1} \lambda \boldsymbol{V}_{cxz}^{\prime} \right)^{-1}
=
\frac{1}{\lambda} \boldsymbol{V}_{cxz}^{-1} 
\\
\tilde{\boldsymbol{D}}_n \tilde{\mathbf{Q}}_2 \big( \lambda; \uptau_0 \big) \tilde{\boldsymbol{D}}_n 
&\Rightarrow 
\frac{1}{1- \lambda} \boldsymbol{V}_{cxz}^{-1} 
\end{align*}

\newpage 

Therefore, the limiting variance for the IVZ estimator is given by 
\begin{align}
\tilde{\boldsymbol{D}}_n \widehat{\boldsymbol{V}}_n^{ivz} \big( \lambda; \uptau_0 \big)\tilde{\boldsymbol{D}}_n
\nonumber
&=
\tilde{\boldsymbol{D}}_n \bigg[ \tilde{\mathbf{Q}}_1 \big( \lambda; \uptau_0 \big)  + \tilde{\mathbf{Q}}_2 \big( \lambda; \uptau_0 \big) \bigg] \tilde{\boldsymbol{D}}_n
\\
&\Rightarrow
\left\{ \frac{1}{\lambda} +  \frac{1}{1 - \lambda}  \right\} \boldsymbol{V}_{cxz}^{-1} 
\end{align} 
and
\begin{align}
\left[ \tilde{\boldsymbol{D}}_n \widehat{\boldsymbol{V}}_n^{ivz} \big( \lambda; \uptau_0 \big)\tilde{\boldsymbol{D}}_n \right]^{-1}
&\Rightarrow
\lambda ( 1 - \lambda) \boldsymbol{V}_{cxz}
\end{align} 
Then, the limiting distribution of the IVZ-Wald test is given by  
\color{black}
\begin{align}
&\left\{ \tilde{\boldsymbol{D}}_n \left( \widehat{ \boldsymbol{\beta}}^{ivz}_{2} (\lambda ; \uptau_0) - \widehat{ \boldsymbol{\beta}}^{ivz}_{1} (\lambda ; \uptau_0) \right) \right\}^{\prime} 
\left[ \tilde{\boldsymbol{D}}_n \widehat{\boldsymbol{V}}_n^{ivz} \big( \lambda; \uptau_0 \big)\tilde{\boldsymbol{D}}_n \right]^{-1}
\left\{ \tilde{\boldsymbol{D}}_n \left( \widehat{ \boldsymbol{\beta}}^{ivz}_{2} (\lambda ; \uptau_0) - \widehat{ \boldsymbol{\beta}}^{ivz}_{1} (\lambda ; \uptau_0) \right) \right\}
\nonumber
\\
&\Rightarrow
\frac{ \uptau_0 ( 1 - \uptau_0) }{ \big[ \lambda ( 1 - \lambda) \big]^2 } \bigg[ \boldsymbol{W}_{p} (\lambda) -   \lambda \boldsymbol{W}_{p} (1) \bigg]^{\prime} \boldsymbol{V}_{cxz}^{-1/2} \times \lambda ( 1 - \lambda) \boldsymbol{V}_{cxz} \times \boldsymbol{V}_{cxz}^{-1/2} \bigg[ \boldsymbol{W}_{p} (\lambda) -   \lambda \boldsymbol{W}_{p} (1) \bigg]
\nonumber
\\
&\equiv
\frac{ \uptau_0 ( 1 - \uptau_0) }{ \lambda ( 1 - \lambda)} \bigg[ \boldsymbol{W}_{p} (\lambda) -   \lambda \boldsymbol{W}_{p} (1) \bigg]^{\prime} \bigg[ \boldsymbol{W}_{p} (\lambda) -   \lambda \boldsymbol{W}_{p} (1) \bigg]
\end{align}
Specifically, the IVZ-based test statistic shows an equivalent asymptotic behaviour with the corresponding IVX-based test when testing for structural breaks in nonstationary quantile predictive regressions. 
\end{proof}

\medskip

\begin{remark}
The formulation of the IVZ-Wald statistic shows the equivalence of  an OLS-Wald based test to the IVZ-Wald statistic, since the IVZ estimation is equivalent to when original regressors are converted into stationary time series and then constructing an OLS based test; in which case we avoid any possible computational complexity of the second stage regression which requires the estimation of long-run covariance matrices to filter out potential long-run bias effects in the limiting variance of the test.  
\end{remark}

\newpage 

\subsection{Asymptotic results on stochastic integrals}
\label{Appendix8.2}

In this section we summarize main invariance principles employed for deriving some of the theoretical results of the paper. Extensive details on these results can be found in the framework proposed by \cite{PhillipsMagdal2009econometric}.  

\medskip

\begin{lemma}
\label{LemmaB1}
Let $\mathbf{V}_{xx} := \displaystyle \int_0^{\infty} e^{r \boldsymbol{C}_p } \mathbf{\Omega}_{xx} e^{r \boldsymbol{C}_p } dr$ where $\mathbf{\Omega}_{xx}$ is the long-run covariance of the error term $\boldsymbol{v}_{t}$. Then, under the null hypothesis of no structural break in the predictive regression model the following large sample theory holds: 
\begin{enumerate}

\item[\textit{(i)}] the sample covariance weakly convergence to the following limit (see Corollary \ref{corollary4A})
\begin{align}
\frac{1}{ \sqrt{ \uptau_0 (1 - \uptau_0)} } \tilde{\boldsymbol{D}}_n^{-1}
 \sum_{t=1}^{ \floor{ \lambda n } } \tilde{ \boldsymbol{z} }_{t-1}  \psi_{\uptau} \big( u_t ( \uptau_0 ) \big) \Rightarrow  \boldsymbol{\mathcal{U}}_p \left( \lambda \right)
\end{align}
where $\boldsymbol{\mathcal{U}}_p \left( . \right)$ is a Brownian motion with variance $\boldsymbol{V}_{cxz} $ as defined below
\begin{small}
\begin{equation}
\boldsymbol{V}_{cxz} 
\equiv
\begin{cases}
\mathbf{V}_{zz} = \displaystyle \int_0^{\infty} e^{r \boldsymbol{C}_z} \mathbf{\Omega}_{xx} e^{r \boldsymbol{C}_z} dr 
& ,\ \text{when} \ 0 < \upgamma_z < \upgamma_x  < 1,
\\
\\
\displaystyle \mathbf{V}_{xx} = \int_0^{\infty} e^{r \boldsymbol{C}_p } \mathbf{\Omega}_{xx} e^{ r \boldsymbol{C}_p } dr   
& , \ \text{when} \ 0 < \upgamma_x < \upgamma_z < 1.  
\end{cases}
\end{equation}
\end{small}
\end{enumerate} 

\item[\textit{(ii)}] the sample covariance weakly convergence to the following limit  
\begin{align}
\tilde{\boldsymbol{D}}_n^{-1} \left[ \sum_{t=1}^{ \floor{ \lambda n } } \boldsymbol{x}_{t-1} \tilde{ \boldsymbol{z} }_{t-1}^{\prime} \right] \tilde{\boldsymbol{D}}_n^{-1}
\Rightarrow \textcolor{blue}{\equiv \boldsymbol{\Gamma}_{cxz}(\lambda) }
\end{align}
where the exact analytic form of $\boldsymbol{\varPsi} ( \lambda )$ depends on which of the two exponents rates of persistence stochastically dominates such as  
\begin{equation*}
\boldsymbol{\Gamma}_{cxz}(\lambda) 
:=
\begin{cases}
\displaystyle - \boldsymbol{C}_z^{-1} \left( \lambda \mathbf{\Omega}_{xx} +  \int_0^{ \lambda } \boldsymbol{J}^{\mu}_c(r) d\boldsymbol{J}_c^{\prime} \right)    & ,\ \text{when} \ \upgamma_x = 1 
\\
\\
\displaystyle - \lambda \boldsymbol{C}_z^{-1} \bigg( \mathbf{\Omega}_{xx} + \boldsymbol{C}_p \boldsymbol{V}_{xx} \bigg)   
& ,\ \text{when} \ 0 < \upgamma_z < \upgamma_x < 1  
\\
\\
\displaystyle \lambda \int_0^{ \infty } e^{r \boldsymbol{C}_p } \boldsymbol{\Omega}_{xx} e^{r\boldsymbol{C}_p} dr \ \textcolor{blue}{\equiv \lambda \boldsymbol{V}_{xx} }
& ,\ \text{when} \ 0 < \upgamma_x < \upgamma_z  < 1   
\end{cases}
\end{equation*}
where $\boldsymbol{B}_p(.)$ is a $p-$dimensional standard Brownian motion, $\boldsymbol{J}_c ( \lambda ) = \int_0^{\lambda} e^{(\lambda - s) \boldsymbol{C}_p } d \boldsymbol{B}(s)$ is an \textit{Ornstein-Uhkenbeck} process and we denote with $\boldsymbol{J}^{\mu}_c (\lambda) = \boldsymbol{J}_c (\lambda) - \int_0^1 \boldsymbol{J}_c(s) ds$ and $\boldsymbol{B}_p^{\mu} (\lambda ) = \boldsymbol{B}(\lambda) - \int_0^1 \boldsymbol{B}(s) ds$ the demeaned processes of $\boldsymbol{J}_c(\lambda)$ and $\boldsymbol{B}_p(\lambda)$ respectively.

\item[\textit{(iii)}] The weakly joint convergence result applies and the asymptotic terms given by expressions in \textit{(i)} and \textit{(ii)} are stochastically independent.  

\end{lemma}

\newpage 

\begin{proof}
We present the main conjectures for deriving the invariance principles presented by Lemma \ref{LemmaB1} (see, \cite{PhillipsMagdal2009econometric} and \cite{kostakis2015Robust} for details)
\begin{align}
\label{limitA}
n^{ - ( 1 + \upgamma_z ) }  \sum_{t=1}^n \tilde{\boldsymbol{z}}_{t-1} \tilde{\boldsymbol{z}}_{t-1}^{\prime}  \overset{ \mathbb{P} }{ \to } \boldsymbol{V}_{zz} :=  \int_0^{\infty} e^{ r\boldsymbol{C}_z } \boldsymbol{ \Omega }_{xx} e^{r\boldsymbol{C}_z} dr  
\end{align}
Moreover, we have the weakly convergence result from \cite{PhillipsMagdal2009econometric}:
\begin{align}
\label{mixed.gaussian}
n^{ - \frac{ 1 + \upgamma_z }{ 2 } } \sum_{t=1}^n \left( \tilde{\boldsymbol{z}}_{t-1} \otimes \boldsymbol{v}_t  \right)  
\Rightarrow \mathcal{N} \big( \boldsymbol{0}, \boldsymbol{V}_{zz} \otimes \mathbf{\Sigma}_{ vv } \big)   
\end{align}
Expression \eqref{mixed.gaussian} shows weakly convergence into a mixed Gaussian limit distribution. In particular, this implies that the limit distribution of $n^{ - (1 + \upgamma_z )/2} \sum_{t=1}^n \left( \tilde{\boldsymbol{z}}_{t-1} \otimes \boldsymbol{v}_t  \right)$ is Gaussian with mean zero and covariance matrix equal to the probability limit of $n^{ - (1 + \upgamma_z )/2} \sum_{t=1}^n \left( \tilde{\boldsymbol{z}}_{t-1} \otimes \boldsymbol{v}_t  \right)$, which is equal to $\boldsymbol{V}_{zz} \otimes \boldsymbol{\Sigma}_{ vv }$, where $\boldsymbol{V}_{zz}$ is defined in \eqref{limitA}. Specifically, the above Mixed Gaussianity convergence argument, is a powerful property of the IVX filtration and ensures the robustness of the instrumental variable based procedure for abstract persistence. The dependence of the covariance matrix on the degree of persistence of the IVX instrument, induces exactly the Mixed Gaussianity. Similarly, the limit distribution below follows from Lemma 3.3 of PM.
\begin{align*}
n^{ - (1 + \upgamma_z )/2} \sum_{t=1}^n \big( \boldsymbol{x}_{t-1} \otimes \boldsymbol{v}_t  \big)  
\Rightarrow \mathcal{N} \big( 0 , \boldsymbol{V}_{xx} \otimes \boldsymbol{\Sigma}_{ vv } \big), \ \text{where} \ \boldsymbol{V}_{xx} := \int_0^{\infty} e^{r \boldsymbol{C}_p } \boldsymbol{\Omega}_{xx} e^{r \boldsymbol{C}_p } dr
\end{align*}
\end{proof}


\bibliographystyle{apalike}

{\small  
\bibliography{myreferences1}}

\begin{thebibliography}{}

\bibitem[Andersen and Varneskov, 2021]{andersen2021consistent}
Andersen, T.~G. and Varneskov, R.~T. (2021).
\newblock Consistent inference for predictive regressions in persistent
  economic systems.
\newblock {\em Journal of Econometrics}, 224(1):215--244.

\bibitem[Andrews, 1993]{andrews1993tests}
Andrews, D.~W. (1993).
\newblock Tests for parameter instability and structural change with unknown
  change point.
\newblock {\em \href{https://doi.org/10.2307/2951764}{Econometrica: Journal of
  the Econometric Society}}, pages 821--856.

\bibitem[Andrews and Ploberger, 1994]{andrews1994optimal}
Andrews, D.~W. and Ploberger, W. (1994).
\newblock Optimal tests when a nuisance parameter is present only under the
  alternative.
\newblock {\em Econometrica: Journal of the Econometric Society}, pages
  1383--1414.

\bibitem[Atanasov et~al., 2020]{atanasov2020consumption}
Atanasov, V., M{\o}ller, S.~V., and Priestley, R. (2020).
\newblock Consumption fluctuations and expected returns.
\newblock {\em The Journal of Finance}, 75(3):1677--1713.

\bibitem[Aue et~al., 2017]{aue2017piecewise}
Aue, A., Cheung, R.~C., Lee, T.~C., and Zhong, M. (2017).
\newblock Piecewise quantile autoregressive modeling for nonstationary time
  series.
\newblock {\em Bernoulli}, 23(1):1--22.

\bibitem[Bai, 1996]{bai1996testing}
Bai, J. (1996).
\newblock Testing for parameter constancy in linear regressions: an empirical
  distribution function approach.
\newblock {\em Econometrica: Journal of the Econometric Society}, pages
  597--622.

\bibitem[Banerjee et~al., 1993]{banerjee1993co}
Banerjee, A., Dolado, J.~J., Galbraith, J.~W., Hendry, D., et~al. (1993).
\newblock Co-integration, error correction, and the econometric analysis of
  non-stationary data.
\newblock {\em OUP Catalogue}.

\bibitem[Bickel, 1975]{bickel1975one}
Bickel, P.~J. (1975).
\newblock One-step huber estimates in the linear model.
\newblock {\em Journal of the American Statistical Association},
  70(350):428--434.

\bibitem[Billingsley, 1968]{billingsley1968convergence}
Billingsley, P. (1968).
\newblock {\em Convergence of probability measures}.
\newblock John Wiley \& Sons.

\bibitem[Cai and Wang, 2014]{cai2014testing}
Cai, Z. and Wang, Y. (2014).
\newblock Testing predictive regression models with nonstationary regressors.
\newblock {\em Journal of Econometrics}, 178:4--14.

\bibitem[Cai et~al., 2015]{cai2015testing}
Cai, Z., Wang, Y., and Wang, Y. (2015).
\newblock Testing instability in a predictive regression model with
  nonstationary regressors.
\newblock {\em Econometric Theory}, 31(5):953.

\bibitem[Campbell and Yogo, 2006]{campbell2006efficient}
Campbell, J.~Y. and Yogo, M. (2006).
\newblock Efficient tests of stock return predictability.
\newblock {\em Journal of financial economics}, 81(1):27--60.

\bibitem[Canarella et~al., 2012]{canarella2012unit}
Canarella, G., Miller, S., and Pollard, S. (2012).
\newblock Unit roots and structural change: an application to us house price
  indices.
\newblock {\em Urban Studies}, 49(4):757--776.

\bibitem[Caner and Hansen, 2001]{caner2001threshold}
Caner, M. and Hansen, B.~E. (2001).
\newblock Threshold autoregression with a unit root.
\newblock {\em Econometrica}, 69(6):1555--1596.

\bibitem[Chernozhukov, 2005]{chernozhukov2005extremal}
Chernozhukov, V. (2005).
\newblock Extremal quantile regression.
\newblock {\em The Annals of Statistics}, 33(2):806--839.

\bibitem[Cho et~al., 2015]{cho2015quantile}
Cho, J.~S., Kim, T.-h., and Shin, Y. (2015).
\newblock Quantile cointegration in the autoregressive distributed-lag modeling
  framework.
\newblock {\em Journal of econometrics}, 188(1):281--300.

\bibitem[Chow, 1960]{chow1960tests}
Chow, G.~C. (1960).
\newblock Tests of equality between sets of coefficients in two linear
  regressions.
\newblock {\em Econometrica: Journal of the Econometric Society}, pages
  591--605.

\bibitem[Chu et~al., 1996]{chu1996monitoring}
Chu, C.-S.~J., Stinchcombe, M., and White, H. (1996).
\newblock Monitoring structural change.
\newblock {\em Econometrica: Journal of the Econometric Society}, pages
  1045--1065.

\bibitem[Davidson, 2000]{davidson2000econometric}
Davidson, J. (2000).
\newblock {\em Econometric theory}.
\newblock Wiley-Blackwell.

\bibitem[Davies, 1977]{davies1977hypothesis}
Davies, R.~B. (1977).
\newblock Hypothesis testing when a nuisance parameter is present only under
  the alternative.
\newblock {\em Biometrika}, 64(2):247--254.

\bibitem[De~Haan and Ferreira, 2006]{De2006extreme}
De~Haan, L. and Ferreira, A. (2006).
\newblock {\em Extreme value theory: an introduction}, volume~21.
\newblock Springer.

\bibitem[Demetrescu et~al., 2020]{demetrescu2020testing}
Demetrescu, M., Georgiev, I., Rodrigues, P.~M., and Taylor, A.~R. (2020).
\newblock Testing for episodic predictability in stock returns.
\newblock {\em Journal of Econometrics}.

\bibitem[Dou and M{\"u}ller, 2021]{dou2021generalized}
Dou, L. and M{\"u}ller, U.~K. (2021).
\newblock Generalized local-to-unity models.
\newblock {\em Econometrica}, 89(4):1825--1854.

\bibitem[Elliott, 2011]{elliott2011control}
Elliott, G. (2011).
\newblock A control function approach for testing the usefulness of trending
  variables in forecast models and linear regression.
\newblock {\em Journal of econometrics}, 164(1):79--91.

\bibitem[Elliott et~al., 2015]{elliott2015nearly}
Elliott, G., M{\"u}ller, U.~K., and Watson, M.~W. (2015).
\newblock Nearly optimal tests when a nuisance parameter is present under the
  null hypothesis.
\newblock {\em Econometrica}, 83(2):771--811.

\bibitem[Elliott and Stock, 1994]{elliott1994inference}
Elliott, G. and Stock, J.~H. (1994).
\newblock Inference in time series regression when the order of integration of
  a regressor is unknown.
\newblock {\em Econometric theory}, 10(3-4):672--700.

\bibitem[Escanciano and Goh, 2018]{escanciano2018quantile}
Escanciano, J.~C. and Goh, S. (2018).
\newblock Quantile-regression inference with adaptive control of size.
\newblock {\em Journal of the American Statistical Association}.

\bibitem[Fan and Lee, 2019]{fan2019predictive}
Fan, R. and Lee, J.~H. (2019).
\newblock Predictive quantile regressions under persistence and conditional
  heteroskedasticity.
\newblock {\em Journal of Econometrics}, 213(1):261--280.

\bibitem[Furno, 2014]{furno2014quantile}
Furno, M. (2014).
\newblock Quantile regression estimates and the analysis of structural breaks.
\newblock {\em Quantitative Finance}, 14(12):2185--2192.

\bibitem[Galvao et~al., 2011]{Galvao2011threshold}
Galvao, A., Montes-Rojas, G., and Olmo, J. (2011).
\newblock Threshold quantile autoregressive models.
\newblock {\em Journal of Time Series Analysis}, 32(3):253--267.

\bibitem[Galvao et~al., 2014]{galvao2014testing}
Galvao, A.~F., Kato, K., Montes-Rojas, G., and Olmo, J. (2014).
\newblock Testing linearity against threshold effects: uniform inference in
  quantile regression.
\newblock {\em Annals of the Institute of Statistical Mathematics},
  66(2):413--439.

\bibitem[Georgiev et~al., 2021]{georgiev2021extensions}
Georgiev, I., Demetrescu, M., Rodrigues, P.~M., and Taylor, A. (2021).
\newblock Extensions to ivx methods of inference for return predictability.

\bibitem[Georgiev et~al., 2018]{georgiev2018testing}
Georgiev, I., Harvey, D.~I., Leybourne, S.~J., and Taylor, A.~R. (2018).
\newblock Testing for parameter instability in predictive regression models.
\newblock {\em Journal of Econometrics}, 204(1):101--118.

\bibitem[Goh and Knight, 2009]{goh2009nonstandard}
Goh, S.~C. and Knight, K. (2009).
\newblock Nonstandard quantile-regression inference.
\newblock {\em Econometric Theory}, 25(5):1415--1432.

\bibitem[Gonzalo and Pitarakis, 2012]{gonzalo2012regime}
Gonzalo, J. and Pitarakis, J.-Y. (2012).
\newblock Regime-specific predictability in predictive regressions.
\newblock {\em Journal of Business \& Economic Statistics}, 30(2):229--241.

\bibitem[Gonzalo and Pitarakis, 2017]{gonzalo2017inferring}
Gonzalo, J. and Pitarakis, J.-Y. (2017).
\newblock Inferring the predictability induced by a persistent regressor in a
  predictive threshold model.
\newblock {\em Journal of Business \& Economic Statistics}, 35(2):202--217.

\bibitem[Hansen, 1996]{hansen1996inference}
Hansen, B.~E. (1996).
\newblock Inference when a nuisance parameter is not identified under the null
  hypothesis.
\newblock {\em Econometrica: Journal of the econometric society}, pages
  413--430.

\bibitem[Hansen, 2000]{hansen2000testing}
Hansen, B.~E. (2000).
\newblock Testing for structural change in conditional models.
\newblock {\em Journal of Econometrics}, 97(1):93--115.

\bibitem[Hanson, 2002]{hanson2002tests}
Hanson, B.~E. (2002).
\newblock Tests for parameter instability in regressions with i (1) processes.
\newblock {\em Journal of Business \& Economic Statistics}, 20(1):45--59.

\bibitem[Harvey et~al., 2021]{harvey2021simpleA}
Harvey, D.~I., Leybourne, S.~J., and Taylor, A.~R. (2021).
\newblock Simple tests for stock return predictability with good size and power
  properties.
\newblock {\em Journal of Econometrics}.

\bibitem[Hoga, 2017]{hoga2017change}
Hoga, Y. (2017).
\newblock Change point tests for the tail index of $\beta$-mixing random
  variables.
\newblock {\em Econometric Theory}, 33(4):915--954.

\bibitem[Hoga, 2018]{hoga2018structural}
Hoga, Y. (2018).
\newblock A structural break test for extremal dependence in $\beta$-mixing
  random vectors.
\newblock {\em Biometrika}, 105(3):627--643.

\bibitem[Jansson and Moreira, 2006]{jansson2006optimal}
Jansson, M. and Moreira, M.~J. (2006).
\newblock Optimal inference in regression models with nearly integrated
  regressors.
\newblock {\em Econometrica}, 74(3):681--714.

\bibitem[Kasparis et~al., 2015]{Kasparis2015nonparametric}
Kasparis, I., Andreou, E., and Phillips, P. C.~B. (2015).
\newblock Nonparametric predictive regression.
\newblock {\em Journal of Econometrics}, 185(2):468--494.

\bibitem[Kato, 2009]{kato2009asymptotics}
Kato, K. (2009).
\newblock Asymptotics for argmin processes: Convexity arguments.
\newblock {\em Journal of Multivariate Analysis}, 100(8):1816--1829.

\bibitem[Katsouris, 2021]{Katsouris2021breaks}
Katsouris, C.~G. (2021).
\newblock Robust structural break tests in predictive regressions with
  persistent predictors.
\newblock {\em University of Southampton, Working paper}.

\bibitem[Kiefer, 1967]{kiefer1967bahadur}
Kiefer, J. (1967).
\newblock On bahadur's representation of sample quantiles.
\newblock {\em The Annals of Mathematical Statistics}, 38(5):1323--1342.

\bibitem[Knight, 1989]{knight1989limit}
Knight, K. (1989).
\newblock Limit theory for autoregressive-parameter estimates in an
  infinite-variance random walk.
\newblock {\em The Canadian Journal of Statistics/La Revue Canadienne de
  Statistique}, pages 261--278.

\bibitem[Koenker and Bassett, 1978]{Koenker1978regression}
Koenker, R. and Bassett, G. (1978).
\newblock Regression quantiles.
\newblock {\em Econometrica: journal of the Econometric Society}, pages 33--50.

\bibitem[Koenker and Bassett, 1982]{Koenker1982Robust}
Koenker, R. and Bassett, G. (1982).
\newblock Robust tests for heteroscedasticity based on regression quantiles.
\newblock {\em Econometrica: Journal of the Econometric Society}, pages 43--61.

\bibitem[Koenker and Hallock, 2001]{koenker2001quantile}
Koenker, R. and Hallock, K.~F. (2001).
\newblock Quantile regression.
\newblock {\em Journal of economic perspectives}, 15(4):143--156.

\bibitem[Koenker and Machado, 1999]{koenker1999goodness}
Koenker, R. and Machado, J.~A. (1999).
\newblock Goodness of fit and related inference processes for quantile
  regression.
\newblock {\em Journal of the american statistical association},
  94(448):1296--1310.

\bibitem[Koenker and Portnoy, 1987]{koenker1987estimation}
Koenker, R. and Portnoy, S. (1987).
\newblock L-estimation for linear models.
\newblock {\em Journal of the American statistical Association},
  82(399):851--857.

\bibitem[Koenker and Xiao, 2002]{Koenker2002inference}
Koenker, R. and Xiao, Z. (2002).
\newblock Inference on the quantile regression process.
\newblock {\em Econometrica}, 70(4):1583--1612.

\bibitem[Koenker and Xiao, 2004]{Koenker2004unit}
Koenker, R. and Xiao, Z. (2004).
\newblock Unit root quantile autoregression inference.
\newblock {\em Journal of the American Statistical Association},
  99(467):775--787.

\bibitem[Koenker and Xiao, 2006]{koenker2006quantile}
Koenker, R. and Xiao, Z. (2006).
\newblock Quantile autoregression.
\newblock {\em Journal of the American statistical association},
  101(475):980--990.

\bibitem[Koltchinskii, 1997]{koltchinskii1997m}
Koltchinskii, V.~I. (1997).
\newblock M-estimation, convexity and quantiles.
\newblock {\em The annals of Statistics}, pages 435--477.

\bibitem[Kostakis et~al., 2015]{kostakis2015Robust}
Kostakis, A., Magdalinos, T., and Stamatogiannis, M.~P. (2015).
\newblock Robust econometric inference for stock return predictability.
\newblock {\em The Review of Financial Studies}, 28(5):1506--1553.

\bibitem[Koul and Saleh, 1995]{koul1995autoregression}
Koul, H.~L. and Saleh, A. M.~E. (1995).
\newblock Autoregression quantiles and related rank-scores processes.
\newblock {\em The Annals of Statistics}, pages 670--689.

\bibitem[Kuan and Chen, 1994]{kuan1994implementing}
Kuan, C.-M. and Chen, M.-Y. (1994).
\newblock Implementing the fluctuation and moving-estimates tests in dynamic
  econometric models.
\newblock {\em Economics Letters}, 44(3):235--239.

\bibitem[Kulperger et~al., 2005]{kulperger2005high}
Kulperger, R., Yu, H., et~al. (2005).
\newblock High moment partial sum processes of residuals in garch models and
  their applications.
\newblock {\em The Annals of Statistics}, 33(5):2395--2422.

\bibitem[Kwiatkowski et~al., 1992]{Kwiatkowski1992testing}
Kwiatkowski, D., Phillips, P. C.~B., Schmidt, P., and Shin, Y. (1992).
\newblock Testing the null hypothesis of stationarity against the alternative
  of a unit root: How sure are we that economic time series have a unit root?
\newblock {\em Journal of econometrics}, 54(1-3):159--178.

\bibitem[Lee, 2016]{lee2016predictive}
Lee, J.~H. (2016).
\newblock Predictive quantile regression with persistent covariates: Ivx-qr
  approach.
\newblock {\em Journal of Econometrics}, 192(1):105--118.

\bibitem[Leisch et~al., 2000]{leisch2000monitoring}
Leisch, F., Hornik, K., and Kuan, C.-M. (2000).
\newblock Monitoring structural changes with the generalized fluctuation test.
\newblock {\em Econometric Theory}, 16(6):835--854.

\bibitem[Li et~al., 2016]{li2016estimation}
Li, H., Zheng, C., and Guo, Y. (2016).
\newblock Estimation and test for quantile nonlinear cointegrating regression.
\newblock {\em Economics Letters}, 148:27--32.

\bibitem[Magdalinos and Phillips, 2009]{Magdalinos2009limit}
Magdalinos, T. and Phillips, P. C.~B. (2009).
\newblock Limit theory for cointegrated systems with moderately integrated and
  moderately explosive regressors.
\newblock {\em Econometric Theory}, 25(2):482--526.

\bibitem[McCloskey, 2017]{mccloskey2017bonferroni}
McCloskey, A. (2017).
\newblock Bonferroni-based size-correction for nonstandard testing problems.
\newblock {\em Journal of Econometrics}, 200(1):17--35.

\bibitem[Mikusheva, 2007]{mikusheva2007uniform}
Mikusheva, A. (2007).
\newblock Uniform inference in autoregressive models.
\newblock {\em Econometrica}, 75(5):1411--1452.

\bibitem[Neocleous and Portnoy, 2008]{neocleous2008monotonicity}
Neocleous, T. and Portnoy, S. (2008).
\newblock On monotonicity of regression quantile functions.
\newblock {\em Statistics \& probability letters}, 78(10):1226--1229.

\bibitem[Newey, 1991]{newey1991uniform}
Newey, W.~K. (1991).
\newblock Uniform convergence in probability and stochastic equicontinuity.
\newblock {\em Econometrica: Journal of the Econometric Society}, pages
  1161--1167.

\bibitem[Paye, 2012]{paye2012deja}
Paye, B.~S. (2012).
\newblock Deja vol: Predictive regressions for aggregate stock market
  volatility using macroeconomic variables.
\newblock {\em Journal of Financial Economics}, 106(3):527--546.

\bibitem[Perron, 1991]{perron1991continuous}
Perron, P. (1991).
\newblock A continuous time approximation to the unstable first-order
  autoregressive process: the case without an intercept.
\newblock {\em Econometrica: Journal of the Econometric Society}, pages
  211--236.

\bibitem[Phillips, 1987a]{Phillips1987time}
Phillips, P. C.~B. (1987a).
\newblock Time series regression with a unit root.
\newblock {\em Econometrica: Journal of the Econometric Society}, pages
  277--301.

\bibitem[Phillips, 1987b]{Phillips1987towards}
Phillips, P. C.~B. (1987b).
\newblock Towards a unified asymptotic theory for autoregression.
\newblock {\em Biometrika}, 74(3):535--547.

\bibitem[Phillips, 1988]{Phillips1988Regression}
Phillips, P. C.~B. (1988).
\newblock Regression theory for near-integrated time series.
\newblock {\em Econometrica: Journal of the Econometric Society}, pages
  1021--1043.

\bibitem[Phillips, 2014]{Phillips2014Confidence}
Phillips, P. C.~B. (2014).
\newblock On confidence intervals for autoregressive roots and predictive
  regression.
\newblock {\em Econometrica}, 82(3):1177--1195.

\bibitem[Phillips and Lee, 2013]{Phillips2013predictive}
Phillips, P. C.~B. and Lee, J.~H. (2013).
\newblock Predictive regression under various degrees of persistence and robust
  long-horizon regression.
\newblock {\em Journal of Econometrics}, 177(2):250--264.

\bibitem[Phillips and Lee, 2016]{Phillips2016robust}
Phillips, P. C.~B. and Lee, J.~H. (2016).
\newblock Robust econometric inference with mixed integrated and mildly
  explosive regressors.
\newblock {\em Journal of Econometrics}, 192(2):433--450.

\bibitem[Phillips and Magdalinos, 2007]{Phillips2007limit}
Phillips, P. C.~B. and Magdalinos, T. (2007).
\newblock Limit theory for moderate deviations from a unit root.
\newblock {\em Journal of Econometrics}, 136(1):115--130.

\bibitem[Phillips and Magdalinos, 2009]{PhillipsMagdal2009econometric}
Phillips, P. C.~B. and Magdalinos, T. (2009).
\newblock Econometric inference in the vicinity of unity.
\newblock {\em Singapore Management University, CoFie Working Paper}, 7.

\bibitem[Phillips and Park, 1988]{Phillips1988asymptotic}
Phillips, P. C.~B. and Park, J.~Y. (1988).
\newblock Asymptotic equivalence of ordinary least squares and generalized
  least squares in regressions with integrated regressors.
\newblock {\em Journal of the American Statistical Association},
  83(401):111--115.

\bibitem[Phillips and Perron, 1988]{Phillips1988testing}
Phillips, P. C.~B. and Perron, P. (1988).
\newblock Testing for a unit root in time series regression.
\newblock {\em Biometrika}, 75(2):335--346.

\bibitem[Phillips and Solo, 1992]{Phillips1992asymptotics}
Phillips, P. C.~B. and Solo, V. (1992).
\newblock Asymptotics for linear processes.
\newblock {\em The Annals of Statistics}, pages 971--1001.

\bibitem[Pitarakis, 2004]{pitarakis2004least}
Pitarakis, J.-Y. (2004).
\newblock Least squares estimation and tests of breaks in mean and variance
  under misspecification.
\newblock {\em The Econometrics Journal}, 7(1):32--54.

\bibitem[Pitarakis, 2014]{pitarakis2014joint}
Pitarakis, J.-Y. (2014).
\newblock A joint test for structural stability and a unit root in
  autoregressions.
\newblock {\em Computational Statistics \& Data Analysis}, 76:577--587.

\bibitem[Pitarakis, 2017]{pitarakis2017simple}
Pitarakis, J.-Y. (2017).
\newblock A simple approach for diagnosing instabilities in predictive
  regressions.
\newblock {\em Oxford Bulletin of Economics and Statistics}, 79(5):851--874.

\bibitem[Pollard, 1991]{pollard1991asymptotics}
Pollard, D. (1991).
\newblock Asymptotics for least absolute deviation regression estimators.
\newblock {\em Econometric Theory}, 7(2):186--199.

\bibitem[Portnoy, 1991]{portnoy1991asymptotic}
Portnoy, S. (1991).
\newblock Asymptotic behavior of regression quantiles in non-stationary,
  dependent cases.
\newblock {\em Journal of Multivariate analysis}, 38(1):100--113.

\bibitem[Portnoy, 2012]{portnoy2012nearly}
Portnoy, S. (2012).
\newblock Nearly root-$ n $ approximation for regression quantile processes.
\newblock {\em The Annals of Statistics}, 40(3):1714--1736.

\bibitem[Qu, 2008]{qu2008testing}
Qu, Z. (2008).
\newblock Testing for structural change in regression quantiles.
\newblock {\em Journal of Econometrics}, 146(1):170--184.

\bibitem[Qu and Perron, 2007]{qu2007estimating}
Qu, Z. and Perron, P. (2007).
\newblock Estimating and testing structural changes in multivariate
  regressions.
\newblock {\em Econometrica}, 75(2):459--502.

\bibitem[Qu and Yoon, 2015]{qu2015nonparametric}
Qu, Z. and Yoon, J. (2015).
\newblock Nonparametric estimation and inference on conditional quantile
  processes.
\newblock {\em Journal of Econometrics}, 185(1):1--19.

\bibitem[Ren et~al., 2019]{ren2019balanced}
Ren, Y., Tu, Y., and Yi, Y. (2019).
\newblock Balanced predictive regressions.
\newblock {\em Journal of Empirical Finance}, 54:118--142.

\bibitem[R{\'e}v{\'e}sz, 1982]{revesz1982increments}
R{\'e}v{\'e}sz, P. (1982).
\newblock On the increments of wiener and related processes.
\newblock {\em The Annals of Probability}, pages 613--622.

\bibitem[Seo, 1998]{seo1998tests}
Seo, B. (1998).
\newblock Tests for structural change in cointegrated systems.
\newblock {\em Econometric Theory}, 14(2):222--259.

\bibitem[Shao, 2010]{shao2010self}
Shao, X. (2010).
\newblock A self-normalized approach to confidence interval construction in
  time series.
\newblock {\em Journal of the Royal Statistical Society: Series B (Statistical
  Methodology)}, 72(3):343--366.

\bibitem[Stock, 1994]{stock1994unit}
Stock, J.~H. (1994).
\newblock Unit roots, structural breaks and trends.
\newblock {\em Handbook of econometrics}, 4:2739--2841.

\bibitem[Su and Xiao, 2008]{su2008testing}
Su, L. and Xiao, Z. (2008).
\newblock Testing for parameter stability in quantile regression models.
\newblock {\em Statistics \& Probability Letters}, 78(16):2768--2775.

\bibitem[Uematsu, 2019]{uematsu2019nonstationary}
Uematsu, Y. (2019).
\newblock Nonstationary nonlinear quantile regression.
\newblock {\em Econometric Reviews}, 38(4):386--416.

\bibitem[Van Der~Vaart and Wellner, 1996]{van1996WeakConergence}
Van Der~Vaart, A.~W. and Wellner, J.~A. (1996).
\newblock {\em Weak convergence and empirical processes: with applications to
  statistics}.
\newblock Springer Series in Statistics.

\bibitem[Wang and Phillips, 2012]{WangPhillips2012specification}
Wang, Q. and Phillips, P. C.~B. (2012).
\newblock A specification test for nonlinear nonstationary models.
\newblock {\em The Annals of Statistics}, 40(2):727--758.

\bibitem[Wellner and van~der Vaart, 2007]{wellner2007empirical}
Wellner, J. A.~W. and van~der Vaart, A.~W. (2007).
\newblock Empirical processes indexed by estimated functions.
\newblock In {\em Asymptotics: particles, processes and inverse problems},
  pages 234--252. Institute of Mathematical Statistics.

\bibitem[Xiao, 2009]{xiao2009quantile}
Xiao, Z. (2009).
\newblock Quantile cointegrating regression.
\newblock {\em Journal of econometrics}, 150(2):248--260.

\bibitem[Yang et~al., 2020]{yang2020testing}
Yang, B., Long, W., Peng, L., and Cai, Z. (2020).
\newblock Testing the predictability of us housing price index returns based on
  an ivx-ar model.
\newblock {\em Journal of the American Statistical Association}, pages 1--22.

\bibitem[Zhou and Portnoy, 1998]{zhou1998statistical}
Zhou, K.~Q. and Portnoy, S.~L. (1998).
\newblock Statistical inference on heteroscedastic models based on regression
  quantiles.
\newblock {\em Journal of Nonparametric Statistics}, 9(3):239--260.

\end{thebibliography}

\newpage 

\end{document}